\documentclass[longauth]{aa}  

\usepackage{graphicx}
\usepackage{txfonts}
\usepackage{booktabs}
\usepackage{natbib}
\usepackage[colorlinks=true,citecolor=blue]{hyperref}
\usepackage[normalem]{ulem}
\usepackage{siunitx}
\usepackage[utf8]{inputenc}
\usepackage{longtable}
\usepackage{fancyhdr}
\usepackage[version=4]{mhchem}
\usepackage{gensymb}

\def\hd20{HD~209458\,b} 
\def\gj34{GJ~3470\,b} 
\def\hat11{HAT-P-11\,b}

\newcommand{\tstar}{GJ~806\xspace}

\newcommand{\pytransit}{\textsc{PyTransit}\xspace}
\newcommand{\ldtk}{\textsc{LDTk}\xspace}

\newcommand{\tess}{\textit{TESS}\xspace}

\begin{document} 

\title{GJ~806 (TOI-4481): A bright nearby multi-planetary system with a transiting hot, low-density super-Earth}

\titlerunning{Mass determinations of GJ~806\,b and c}

 \author{
 E.~Palle\inst{\ref{iiac},\ref{iull}},
 J.~Orell-Miquel\inst{\ref{iiac},\ref{iull}},
 M.~Brady\inst{\ref{uchicago}}, 
 J.~Bean\inst{\ref{uchicago}},
 A.\,P.~Hatzes\inst{\ref{tls}},
 G.~Morello\inst{\ref{iiac},\ref{iull}},
 J.\,C.~Morales\inst{\ref{ice},\ref{ieec}},
 F.~Murgas\inst{\ref{iiac},\ref{iull}},
 K.~Molaverdikhani\inst{\ref{lmu},\ref{origins},\ref{mpia}}, 
 H.~Parviainen\inst{\ref{iiac},\ref{iull}},
 J.~Sanz-Forcada\inst{\ref{cab}},
 V.\,J.\,S.~B\'ejar\inst{\ref{iiac},\ref{iull}},
 J.\,A.~Caballero\inst{\ref{cab}},
 K.~R.~Sreenivas\inst{\ref{uniariel}},
 M.~Schlecker\inst{\ref{arizona}},
 I.~Ribas\inst{\ref{ice},\ref{ieec}}, 
 V.~Perdelwitz\inst{\ref{uniariel},\ref{hamstern}},
 L.~Tal-Or\inst{\ref{uniariel},\ref{iag}},  
 M.~P\'erez-Torres\inst{\ref{iaa}},
 R.~Luque\inst{\ref{iaa},\ref{uchicago}},
 S.~Dreizler\inst{\ref{iag}},
 B.~Fuhrmeister\inst{\ref{inst:hs}}, 
 F.~Aceituno\inst{\ref{iaa}}, 
 P.\,J.~Amado\inst{\ref{iaa}}, 
 G.~Anglada-Escud\'e\inst{\ref{ice},\ref{ieec}},
 D.A.~Caldwell\inst{\ref{seti}}, 
 D.~Charbonneau\inst{\ref{cfa}}, 
 C.~Cifuentes\inst{\ref{cab}},
 J.P.~de Leon\inst{\ref{utokio}}, 
 K.A.~Collins\inst{\ref{cfa}},  
 S.~Dufoer\inst{\ref{Vereniging}}, 
 N.~Espinoza\inst{\ref{stsci}}, 
 Z.~Essack\inst{\ref{deap},\ref{kavlimit}}, 
 A.~Fukui\inst{\ref{komaba},\ref{iiac}}, 
 Y.~G\'omez~Maqueo~Chew\inst{\ref{ciudad}},  
 M.A.~G\'omez-Mu\~noz\inst{\ref{uname}},  
 Th.~Henning\inst{\ref{inst:mpia}}, 
 E.~Herrero\inst{\ref{ieec}},
 S.V.~Jeffers\inst{\ref{inst:mps}},  
 J.~Jenkins\inst{\ref{ames}}, 
 A.~Kaminski\inst{\ref{ZHA}},
 J.~Kasper\inst{\ref{uchicago}}, 
 M.~Kunimoto\inst{\ref{mit}}, 
 D.~Latham\inst{\ref{cfa}}, 
 J. Lillo-Box\inst{\ref{cab}}, 
 M.\,J.~L\'opez-Gonz\'alez\inst{\ref{iaa}}, 
 D.~Montes\inst{\ref{inst:ucm}}, 
 M.~Mori\inst{\ref{utokio}}, 
 N.~Narita\inst{\ref{komaba},\ref{abc},\ref{iiac}}, 
 A.~Quirrenbach\inst{\ref{inst:lsw}}, 
 S.~Pedraz\inst{\ref{iaa}}, 
 A.~Reiners\inst{\ref{iag}},
 E.~Rodr\'iguez\inst{\ref{iaa}}, 
 C.~Rodr\'iguez-L\'opez\inst{\ref{iaa}}, 
 L.~Sabin\inst{\ref{uname}},  
 N.~Schanche\inst{\ref{unibe}},  
 R-P.~Schwarz\inst{\ref{cfa}}, 
 A.~Schweitzer\inst{\ref{inst:hs}}, 
 A.~Seifahrt\inst{\ref{uchicago}},  
 G.~Stefansson\inst{\ref{princeton}},   
 J.~Sturmer\inst{\ref{inst:lsw}},  
 T.~Trifonov\inst{\ref{inst:mpia}},
 S.~Vanaverbeke\inst{\ref{Vereniging},\ref{leuvenmat},\ref{astrolab}}, 
 R.D.~Wells\inst{\ref{unibe}}, 
 M.R.~Zapatero-Osorio\inst{\ref{cab}},  
 M.~Zechmeister\inst{\ref{iag}} 
}

\institute{
\label{iiac} Instituto de Astrof\'isica de Canarias (IAC), 38200 La Laguna, Tenerife, Spain
\and 
\label{iull} Deptartamento de Astrof\'isica, Universidad de La Laguna (ULL), 38206 La Laguna, Tenerife, Spain
\and
\label{uchicago}Department of Astronomy and Astrophysics, University of Chicago, Chicago, IL 60637, USA 
\and
\label{tls} Th\"uringer Landessternwarte Tautenburg, 07778 Tautenburg, Germany    
\and
\label{ice} Institut de Ciències de l’Espai (CSIC), Campus UAB, c/ de Can Magrans s/n, E-08193 Bellaterra, Barcelona, Spain
\and
\label{ieec} Institut d’Estudis Espacials de Catalunya, E-08034 Barcelona, Spain Spain
\and
\label{lmu} Universitäts-Sternwarte, Ludwig-Maximilians-Universität München, Scheinerstrasse 1, D-81679 München, Germany
\and
\label{origins} Exzellenzcluster Origins, Boltzmannstraße 2, 85748 Garching, Germany
\and
\label{mpia} Max-Planck-Institut für Astronomie, Königstuhl 17, D-69117 Heidelberg, Germany 
\and 
\label{cab} Centro de Astrobiolog\'ia (CAB, CSIC-INTA), Depto. de Astrof\'isica, ESAC campus, 28692, Villanueva de la Ca\~nada (Madrid), Spain 
\and
\label{uniariel} Department of Physics, Ariel University, Israel
\and
\label{arizona}Department of Astronomy/Steward Observatory, The University of Arizona, 933 North Cherry Avenue, Tucson, AZ 85721, USA
\and 
\label{hamstern} Hamburger Sternwarte, Universit\"at Hamburg, Gojenbergsweg 112, 21029 Hamburg, Germany
\and
\label{iag} Institut f\"ur Astrophysik und Geophysik, Georg-August-Universit\"at, Friedrich-Hund-Platz 1, 37077 G\"ottingen, Germany
\and
\label{iaa} Instituto de Astrof\'isica de Andaluc\'ia (IAA-CSIC), Glorieta de la Astronom\'ia s/n, 18008 Granada, Spain
\and
\label{inst:hs} Hamburger Sternwarte, Universität Hamburg, Gojenbergsweg 112, 21029 Hamburg, Germany
\and
\label{seti} SETI Institute/NASA Ames Research Center, 339 Bernardo Ave, Suite 200 Mountain View, CA 94043, United States
\and
\label{cfa} Center for Astrophysics | Harvard \& Smithsonian, 60 Garden St., Cambridge MA 02138 USA
\and
\label{utokio} Department of Astronomy, Graduate School of Science, The University of
Tokyo, 7-3-1 Hongo, Bunkyo-ku, Tokyo 113-0033, Japan
\and
\label{Vereniging} Vereniging Voor Sterrenkunde (VVS), Oostmeers 122 C, 8000 Brugge, Belgium
\and 
\label{stsci}Space Telescope Science Institute, 3700 San Martin Drive, Baltimore, MD 21218, USA
\and
\label{deap} Department of Earth, Atmospheric and Planetary Sciences, Massachusetts Institute of Technology, Cambridge, MA 02139, USA
\and
\label{kavlimit} Kavli Institute for Astrophysics and Space Research, Massachusetts Institute of Technology, Cambridge, MA 02139, USA
\and
\label{komaba} Komaba Institute for Science, The University of Tokyo, 3-8-1 Komaba,
Meguro, Tokyo 153-8902, Japan
\and
\label{ciudad} Universidad Nacional Aut\'onoma de M\'exico, Instituto de Astronom\'ia, AP 70-264, CDMX  04510, M\'exico
\and
\label{uname} Universidad Nacional  Aut\'onoma de M\'exico, Instituto de Astronom\'ia, AP 106, Ensenada 22800, BC, M\'exico
\and 
\label{inst:mpia}Max-Planck-Institut f\"ur Astronomie, K\"onigstuhl 17, 69117 Heidelberg, Germany
\and
\label{inst:mps}Max Planck Institute for Solar System Research, Justus-von-Liebig-Weg 3, D-37077 Göttingen, Germany
\and
\label{ames} NASA Ames Research Center, Moffett Field, CA 94035, USA 
\and
\label{ZHA} Zentrum fur Astronomie der Universitat Heidelberg Landessternwarte Konigstuhl 12 69117 Heidelberg, Germany
\and
\label{mit} Department of Physics and Kavli Institute for Astrophysics and Space Research, Massachusetts Institute of Technology, Cambridge, MA 02139, USA
\and
\label{inst:ucm}Departamento de F{\'i}sica de la Tierra y Astrof{\'i}sica \& IPARCOS-UCM (Instituto de F\'{i}sica de Part\'{i}culas y del Cosmos de la UCM), Facultad de Ciencias F{\'i}sicas, Universidad Complutense de Madrid, E-28040 Madrid, Spain
\and
\label{abc} Astrobiology Center, 2-21-1 Osawa, Mitaka, Tokyo 181-8588, Japan
\and
\label{inst:lsw}Landessternwarte, Zentrum für Astronomie der Universität Heidelberg, Königstuhl 12, 69117 Heidelberg, Germany
\and
\label{unibe} Center for Space and Habitability, University of Bern, Gesellschaftsstrasse 6, 3012, Bern, Switzerland 
\and
\label{princeton} Department of Astrophysical Sciences, Princeton University, 4 Ivy Lane, Princeton, NJ 08540, USA
\and
\label{leuvenmat} Centre for Mathematical Plasma-Astrophysics, Department of Mathematics, KU Leuven, Celestijnenlaan 200B, 3001 Heverlee, Belgium
\and
\label{astrolab} Belgium ASTROLAB IRIS, Provinciaal Domein “De Palingbeek”, Verbrandemolenstraat 5, 8902 Zillebeke, Ieper, Belgium
}
 
\authorrunning{Palle et al.}
    
\date{Received dd June 2022 / Accepted dd Month 2022}

\abstract{
One of the main scientific goals of the {\em TESS} mission is the discovery of transiting small planets around the closest and brightest stars in the sky. 
Here, using data from the CARMENES, MAROON-X, and HIRES spectrographs, together with {\em TESS}, we report the discovery and mass determination of a planetary system around the M1.5\,V star GJ~806 (TOI-4481).
GJ~806 is a bright ($V \approx$ 10.8\,mag, $J \approx$ 7.3\,mag) and nearby ($d$ = 12\,pc) M dwarf that hosts at least two planets. 
The innermost planet, GJ~806\,b, is transiting and has an ultra-short orbital period of 0.93\,d, a radius of $1.331\pm0.023\,R_{\oplus}$, a mass of $1.90\pm 0.17\,M_{\oplus}$, a mean density of $4.40 \pm 0.45$\,g\,cm$^{-3}$, and an equilibrium temperature of $940 \pm 10$\,K. 
We detect a second, non-transiting, super-Earth planet in the system, GJ~806\,c, with an orbital period of 6.6\,d, a minimum mass of $5.80 \pm 0.30\,M_{\oplus}$, and an equilibrium temperature of $490 \pm 5$\,K. The radial velocity data also shows evidence for a third periodicity at 13.6\,d, although the current dataset does not provide sufficient evidence to unambiguously distinguish between a third super-Earth mass ($M \sin\,i = 8.50\pm 0.45\,M_{\oplus}$) planet or stellar activity. Additionally, we report one transit observation of GJ~806\,b taken with CARMENES in search for a possible extended atmosphere of H or He, but we can only place upper limits to its existence. This is not surprising as our evolutionary models support the idea that any possible primordial H/He atmosphere that GJ~806\,b might have had, would long have been lost. However, GJ~806\,b's bulk density makes it likely that the planet hosts some type of volatile atmosphere. 
In fact, with a transmission spectroscopy metrics (TSM) of 44 and an emission spectroscopy metrics (ESM) of 24, GJ~806\,b is to date the third-ranked terrestrial planet around an M dwarf suitable for transmission spectroscopy studies using {\em JWST}, and the most promising terrestrial planet for emission spectroscopy studies. GJ 806b is also an excellent target for the detection of
radio emission via star-planet interactions.
}


\maketitle

\section{Introduction}
\label{sec:introduction}

The TESS mission \citep{Ricker2015} is conducting an all-sky survey to find transiting planets around the brightest and closests stars to the Solar System. Given that the majority of stars in the solar neighbourhood are M dwarfs, and the bandpass in which it observes covers red-optical wavelengths, TESS is especially suited for the detection of short-period transiting planets around these type of stars. This is important, as planets found orbiting around M dwarfs offer a unique opportunity for the future exploration of the atmospheric composition of small rocky planets, with the recently launched 
JWST and the upcoming ELTs \citep{Snellen2013}. 

Since the start of operations in mid-2018, TESS has released over 5000 planet candidates, known as TESS Objects of Interest (TOI)\footnote{\url{https://tess.mit.edu/toi-releases/}}. The large majority of these TOIs require ground-based follow-up to confirm their planetary nature, and radial velocity measurements in particular to measure the planetary masses. This confirmation process is carried out by a international and coordinated effort, involving a large fleet of professional and amateur observatories, known as TFOP (TESS Official Follow-up Program). Among the many facilities that can carry out precise radial velocity measurements, here we make use of data from the HIRES \citep{Vogt1994}, CARMENES \citep{CARMENES} and MAROON-X spectrographs \citep{Seifahrt18}. HIRES (High Resolution Echelle Spectrometer) is a visible (0.3--1 micron; $\mathcal{R}$\,=\,85\,000) slit echelle spectrograph mounted on the Keck Telescope in Hawaii.
CARMENES at Calar Alto observatory is a visible (0.52--0.96 micron; $\mathcal{R}$\,=\,94\,600) and near-infrared (0.96--1.71 micron; $\mathcal{R}$\,=\,80\,400) fiber-fed spectrograph well-suited to measure radial velocities, and confirm planets, around M dwarf stellar hosts. In fact, the synergy between TESS and CARMENES has already lead to the discovery of several small transiting planets around M dwarfs \citep{GJ357, GJ3473, LTT3780-1, trifonov2021, LHS1478, Gonzalez2022, Luque2022}. Finally, MAROON-X is an optical (0.5--0.92 micron; $\mathcal{R}$\,=\,85\,000) fiber-fed echelle spectrograph mounted on the Gemini telescope in Hawaii. The instrument was designed to detect Earth-size planets in the habitable zones of mid- to late-M dwarfs and provides extremely precise radial velocity measurements \citep{trifonov2021}.  Here, we make use of all three instruments to confirm the planetary nature of GJ~806\,b (TOI-4481.01), an ultra-short period (USP) planet candidate around an M1.5V  dwarf star. 

USP planets are arbitrarily defined as those having periods shorter than 1\,d \citep{Sahu2006}. Given their short distance to their host stars, these planets are subject to strong stellar irradiation, a fact that translates into a lack of intermediate-mass planets at short orbital periods, the so-called Neptunian desert \citep{Szabo2011, Mazeh2016, McDonald2019}. Smaller USP planets are more common and typically rocky in nature. The Neptune desert can be explained by photo-evaporation mechanisms leading to the loss of primordial H/He atmospheres, leaving behind the rocky cores \citep{valsecchi2014, konigl2017, OwenLAi18}, but other mechanisms such as high-eccentricity migration, disk-driven migration or in-situ formation have also been proposed \citep{Sanchis-Ojeda14,Mazeh2016,Lundkvist16,Lopez17}. 

It is thus of particular interest to accurately measure the mass, radius and density of these small USP planets in order to understand the mechanism that govern their formation and evolution. This is especially important for USP planets around M dwarfs that are accessible to further atmospheric characterization of their atmospheres, which can in turn inform us in more detail about the formation and evolutionary history of the system.

\section{Observations}
\label{sec:Obs}

\subsection{\tess photometry}
\label{sec:tess}

\begin{figure}
\includegraphics[width=1\linewidth]{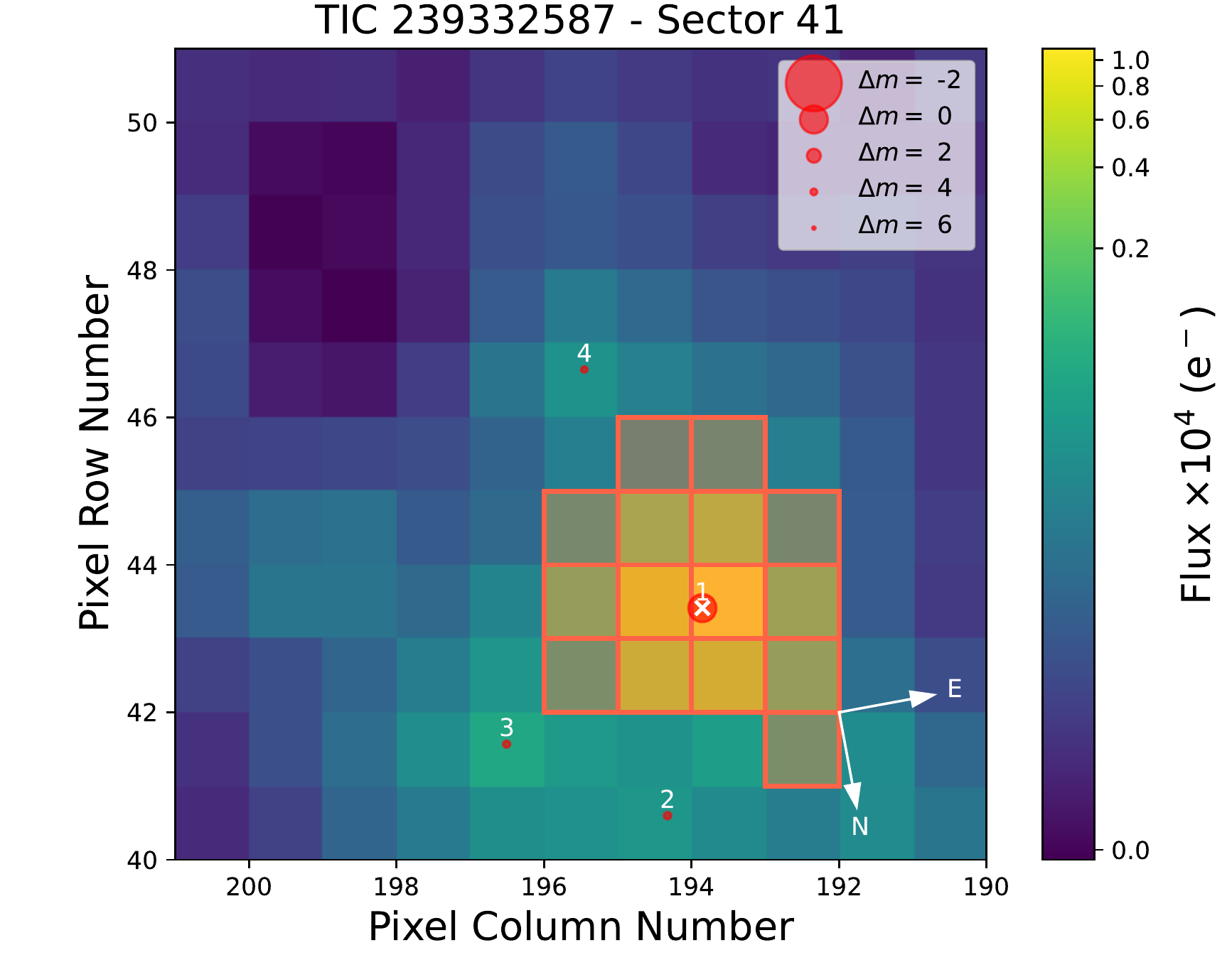}
\caption{\tess target pixel file for Sector 41 showing GJ~806 as a red circle with a white x, nearby stars as red circles, and the pixels included into the photometry aperture in red. The figure was created with \texttt{tpfplotter} \citep{2020A&A...635A.128A}. 
}
\label{fig:tpf_plot}
\end{figure}


The \tess satellite observed GJ~806 during its Primary mission sector 15, with a cadence of 30 minutes, and 
re-observed it again on sector 41, with a cadence of 2 minutes. While the planet candidate went originally unnoticed in 
Sector 15, it was announced as a \tess object of interest (TOI) on the public \tess data website of the 
Massachusetts Institute of Technology (MIT)\footnote{\url{https://tess.mit.edu/toi-releases/}} on October 7 
2021. TOI-4481.01 was announced as a potential 1.4~$R_{\oplus}$ planet candidate with an ultra-short orbital 
period of 0.93 days, around a bright ($T_{mag} = 8.73$) M dwarf.

We downloaded the official Sector 15 and 41 light curves 
generated by the Science Processing Operations Center (SPOC;  \citet{Jenkins2016}) at NASA Ames Research Center from the Mikulski Archive for Space 
Telescopes\footnote{\url{https://mast.stsci.edu}} (MAST) and used the systematic error-corrected Pre-search Data Conditioning 
photometry (PDC-SAP)  \citep{2012PASP..124.1000S,Stumpe2012PASP..124..985S,Stumpe2014PASP..126..100S} for sector 41, and the TESS-SPOC HLSP project light curve for sector 15 \citep{Caldwell2020}, in our photometric analyses. The SPOC pipeline determined low third light contamination levels of 3\% 
and 1\% for the two sectors (15 and 41, respectively). Figure \ref{fig:tpf_plot} shows a TESS image of 
the target star with its surroundings, which supports the low contamination levels. The field is not crowded 
and all the nearby stars are significantly fainter than the target star ($\Delta$mag > 4).

\subsection{Ground-based transit photometry with MuSCAT2}
\label{sec:lco}

Transits of GJ 806b were observed with the MuSCAT2 multi-imager instrument \citep{Narita2019} at the Telescopio Carlos S\'anchez (TCS) located at the Teide Observatory (Spain). MuSCAT2 observes simultaneously in 4 bands ($g'$, $r'$, $i'$, and $z'$).  Data reduction is performed using a custom Python pipeline developed specifically for MuSCAT2 \citep{Parviainen2019}. Each night a set of different aperture sizes is extracted for the target and the comparison stars, and the combinations that provide the most accurate photometry are selected to compute the light curves.


Three full transits were observed on the nights of October 10, 21 and 22 2021 with typical exposure times of 20, 14, 5 and 4 seconds for the $g'$, $r'$, $i'$, and $z'$ bands, respectively.



\subsection{Seeing-limited ground photometry}
\label{sec:groud_obs}

Ground-based seeing-limited observations were taken from a series of observatories in order to determine the rotational period of the host star. 




The T150 telescope \citep{Quirrenbach2022} is a 150 cm Ritchey-Chrètien telescope equipped with a CCD camera Andor Ikon-L DZ936N-BEX2-DD 2k$\times$2k, with a resulting field of view (FOV) of  7.92$\times$7.92 arcmin$^2$. Our set of observations, collected in Johnson V and R filters, consists of 27 epochs obtained during the  period October-December 2021. Each epoch typically consisted of 20 exposures of  50s and 30s in V and R filters, respectively. All CCD measurements were obtained by the method of fixed aperture  photometry using a 1 $\times$ 1 binning (no binning). Each CCD frame was corrected  in a standard way for bias and flat-fielding. Different aperture sizes were  also tested in order to choose the best one for our observations. 


GJ806 was monitored in the V-band filter with the 40cm telescopes of the Las Cumbres Observatory  Global Telescope (LCOGT) network \citep{brown13}. We obtained 48 epochs between October 8 and December 17 2021 
with the IAC2021B-002 program (IP: V. B\'ejar). The instrument mounted on the 40cm telescopes is a 3k$\times$2k SBIG CCD 
camera with a pixel scale of 0.571\,arcsec providing a field of view of 29.2$\times$19.5\,arcmin$^2$. 
Weather conditions at the observatories were mostly clear during our observations, and the average seeing varies between 1 and 2\,arcsecs. 
Raw data was processed using the BANZAI pipeline \citep{mccully18}, which includes bad pixel, bias, dark and 
flat field corrections for each individual night. 
Differential aperture photometry of our target with respect to several reference stars were done using AstroImageJ \citep{collins17}. 
We selected the optimal aperture that provides the lower dispersion of the light curves.


GJ806 was also observed from November 2021 to May 2022 with the 0.8\,m Joan Or\'o Telescope (TJO) at the Observatori Astron\`{o}mic del Montsec (OAdM), Sant Esteve de la Sarga, Catalonia, using the LAIA 4k$\times$4k CCD camera, which provides a field of view of 30\,arcmin, with a pixel scale of 0.4\,arcsec, and the Johnson R filter. The raw images were reduced with the icat pipeline of the TJO \citep{colome2006}, using darks, bias, and flat fields images for calibration. We performed differential photometry with AstroImageJ \citep{collins17}, using the aperture size that minimised the rms of the resulting relative fluxes, and a selection of the brightest reference stars in the field that did not show variability.


Finally, GJ806 was observed from e-EYE (shorthand for Entre Encinas y Estrellas\footnote{https://www.e-eye.es/}), in southern Spain. Observations in the B, V and R filters were taken between October 2021 and May 2022 using a 16" ODK Corrected-Dall-Kirkham reflector with 16803 CCD chip on a ASA DDM85 mount. The CCD camera is equipped with Astrodon filters. The effective pixel scale is 2.04 "/pixel with 3x3 binning. Reduction of images and differential aperture photometry of the target and several reference stars were performed using the Lesve photometry package\footnote{http://www.dppobservatory.net}.

\subsection{Spectroscopic observations}

\subsubsection{CARMENES}
\label{sec:carmrvs}

The CARMENES\footnote{Calar Alto high-Resolution search for M dwarfs with Exoearths with Near-infrared and optical Echelle Spectrographs: \url{http://carmenes.caha.es}} instrument at the 3.5\,m telescope at the Calar Alto Observatory in Almer\'ia, Spain is a dual channel spectrograph that operates at both the optical ($0.52-0.96~{\mu}$m) and near-infrared ($0.96-1.71~{\mu}$m) wavelengths. The average resolving power for the two wavelength regions is $\mathcal{R}~=~94\,600$ and $\mathcal{R}~=~80\,400$, respectively. 

\tstar was part of the original CARMENES survey of 300 M dwarfs in search for planetary companions, and thus observations started several years prior to the TESS candidate announcement. CARMENES obtained 67 spectra for GJ 806 between 23 April 2016 and 28 October 2021, with a total baseline spanning about 5.5 years. The exposure times were set to 900\,s. CARMENES data reduction was performed uniformly using the \texttt{CARACAL} \citep{Caballero2016} pipeline, and radial velocity measurements were extracted using the \texttt{SERVAL} \citep{SERVAL} pipeline. \texttt{SERVAL} RVs were further corrected using measured nightly zero point corrections as discussed in \citealt{Trifonov2020}. 

The \texttt{SERVAL} pipeline also produces a series of spectral activity indices that can be used to explore the stellar activity signals (see Table~\ref{tab:RV_table}). Here we use only the RV data from CARMENES's visible channel. The final CARMENES RV values, along with their uncertainties and BJD time stamp are given in Table~\ref{tab:RV_table}. The average signal-to-noise-ratio of the observations is 111 at 7370\,$\AA$ and the average radial velocity precision is 1.5\,m\,s$^{-1}$.

Additionally to the RV monitoring, a single transit of GJ~806b was observed with the CARMENES spectrograph on the night of 30 November 2021. We observed the target with both visible and near-infrared channels simultaneously, collecting a total of 28 high-resolution spectra in each, 10 of them between the first ($T_1$) and fourth ($T_4$) contacts, covering the full transit, and 8 of them before transit, and 10 after transit. The spectra were taken in good weather conditions with an exposure time of 300\,s, ensuring that any planetary absorption line was not spread over more than $\sim$\,2 pixels during any given exposure, and with a signal-to-noise ratio that varied from 34 to 52 (median value of 44) around 6560\,$\AA$ and from 73 to 101 (median value of 85) around 10\,830\,$\AA$.

Fibre A was used to observe GJ~806, while fibre B was placed on the sky in order to monitor the sky emission lines (fibres A and B are permanently separated by 88\,arcsec in the east-west direction). The observations were reduced using the CARMENES pipeline \texttt{caracal}, and both fibres were extracted with the flat-optimised extraction algorithm (\citealp{FOX_extraction}).

\subsubsection{MAROON-X}
\label{sec:maroonx}

\begin{figure*}[ht!]
\includegraphics[width=1\linewidth]{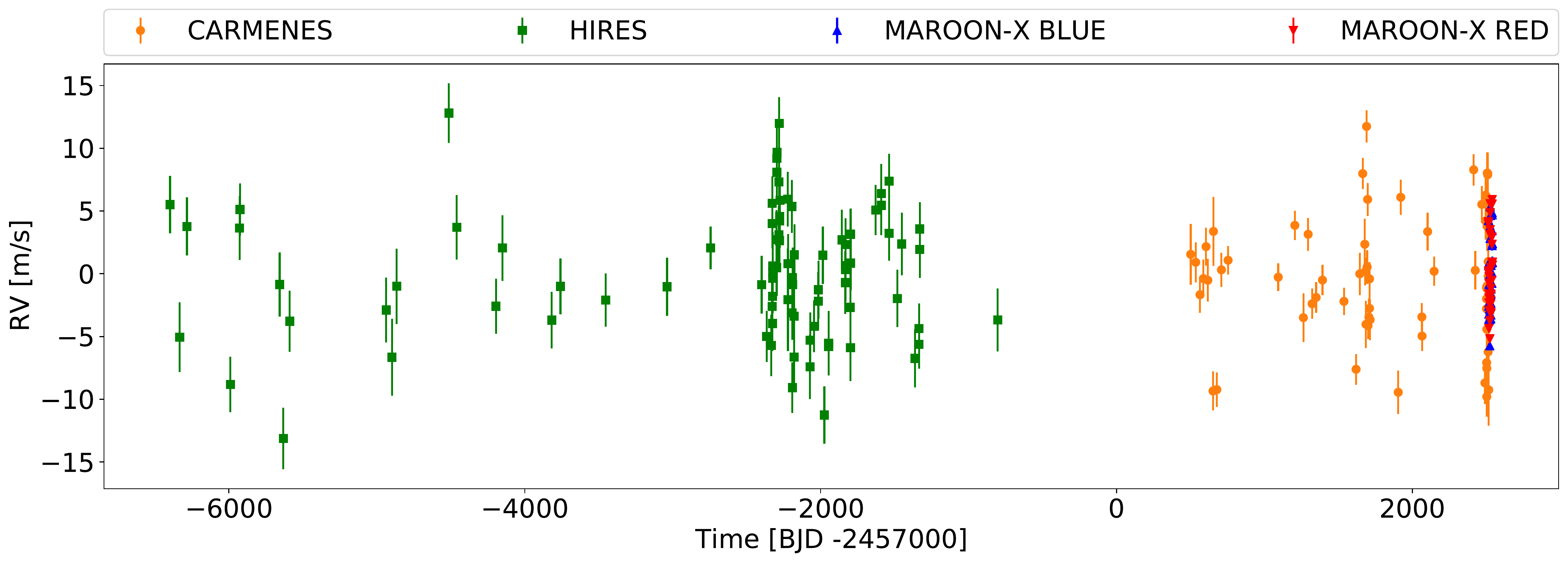}
\caption{Time series of RV measurements by HIRES, CARMENES VIS, and MAROON-X Red and Blue.}
\label{fig:RVs_vs_time}
\end{figure*}

MAROON-X\footnote{M dwarf Advanced Radial velocity Observer Of Neighboring eXoplanets: \url{https://www.gemini.edu/instrumentation/maroon-x}} is a stabilized, fiber-fed high-resolution ($\mathcal{R} \approx 85\,000$) spectrograph mounted at the 8.1-meter Gemini North telescope on Mauna Kea, Hawaii, USA \citep{Seifahrt2016,Seifahrt18,Seifahrt2020}.  MAROON-X has a blue and red arm, which encompass 500-678 and 654-920 nm, respectively.  During an observation, both arms are exposed simultaneously.

We observed GJ~806 with MAROON-X a total of 37 times between 27 October 2021 and 23 November 2021 as part of our ongoing survey of transiting planets identified by \textit{TESS} around M dwarfs within 30\,pc.  Exposure times were typically set to 1200 seconds.  When the weather permitted, the target was observed twice a night to allow for precise characterization of the 0.93-day transiting planet.

The data were reduced using a custom package, and radial velocities were extracted using a version of the \texttt{SERVAL} \citep{SERVAL} pipeline modified for use with MAROON-X data.  The red and blue arms were processed separately due to their different wavelength ranges.  Thus, both arms are analyzed as independent datasets.  The spectra had a peak SNR of around 450 in the red arm and 200 in the blue arm, with accompanying RV precisions of 0.3\,m\,s$^{-1}$ and 0.4\,m\,s$^{-1}$, respectively.  The higher precision in the red arm is an expected result given the cool host star.  As with CARMENES, \texttt{SERVAL} calculated a suite of line indices (H$\alpha$, NaD, and CaIRT) and spectral activity indicators (CRX and dLW) to describe the activity of the host star.

\subsubsection{HIRES}
\label{sec:hires}

HIRES obtained 86 spectra for GJ~806 between 2 June 1997 and 26 September 2012, with a total baseline spanning about 15.3 years. The HIRES RVs were originally published by \citet{Butler2017}, and were subsequently reprocessed by \citet{Lev2018}, including some nightly zero-point corrections. 
This last corrected dataset is the one used here (Tal-Or, priv. comm.). The average radial velocity precision is 2.4\,m\,s$^{-1}$.

The complete time series of the radial velocity measurements used in this work, including HIRES, CARMENES and MAROON-X, is plotted in Figure~\ref{fig:RVs_vs_time} and are given in Table~\ref{tab:RV_table}.

\section{The star}
\label{sec:properties}

\subsection{Stellar parameters}

\begin{table}
\centering
\small
\caption{Stellar parameters of GJ~806.} 
\label{tab:stellar_parameters}
\begin{tabular}{lcr}
\hline
\hline
\noalign{\smallskip}
Parameter & Value & Reference \\ 
\noalign{\smallskip}
\hline
\noalign{\smallskip}
\multicolumn{3}{c}{\em Basic identifiers and data}\\
\noalign{\smallskip}
GJ                                  & 806                   & Gli69 \\
BD                                  & +44 3567              & Arg1903 \\ 
Karmn                               & J20450+444            & AF15, Cab16a \\
TOI                                 & 4481                  & ExoFOP-{\em TESS} \\
TIC                                 & 239332587             & Sta18 \\  
Sp. type                            & M1.5\,V               & PMSU \\
$T$ [mag]                           & $8.7276 \pm 0.0073$   & ExoFOP-{\em TESS}$^a$ \\
\noalign{\smallskip}
\multicolumn{3}{c}{\em Astrometry and kinematics}\\
\noalign{\smallskip}
$\alpha$ (J2016.0)                  & 20:45:04.10  & {\it Gaia} EDR3 \\
$\delta$ (J2016.0)                  & +44:29:56.6  & {\it Gaia} EDR3 \\
$\mu_{\alpha}\cos\delta$ [$\mathrm{mas\,yr^{-1}}$]  & $+434.028 \pm 0.018$ & {\it Gaia} EDR3 \\
$\mu_{\delta}$ [$\mathrm{mas\,yr^{-1}}$] & $+271.022 \pm 0.020$ & {\it Gaia} EDR3 \\
$\varpi$ [mas]                      & $82.890 \pm 0.017$ & {\it Gaia} EDR3 \\
$d$ [pc]                            & $12.0641 \pm 0.0024$ & {\it Gaia} EDR3 \\
$\gamma$ [$\mathrm{km\,s^{-1}}$]    & $-24.694 \pm 0.0023$ & Sou18 \\ 
$\dot{\gamma}$ [$\mathrm{m\,s^{-1}\,yr^{-1}}$] & $+0.07250 \pm 0.00035$ & This work \\
$U$ [$\mathrm{km\,s^{-1}}$]         & $-29.9631 \pm 0.0041$ & This work \\
$V$ [$\mathrm{km\,s^{-1}}$]         & $-21.5366 \pm 0.0023$ & This work \\
$W$ [$\mathrm{km\,s^{-1}}$]         & $-10.2250 \pm 0.0045$ & This work \\
Galactic population                 & Young disc & This work \\
\noalign{\smallskip}
\multicolumn{3}{c}{\em Fundamental parameters}\\
\noalign{\smallskip}
$L_\star$ [$10^{-6}\,L_\odot$]      & $25985 \pm 98$    & This work \\
$T_{\mathrm{eff}}$ [K]              & $3600 \pm 16$     & Mar21 \\ 
$\log{g_{\rm spec}}$                & $4.98 \pm 0.12$ & Mar21 \\
{[Fe/H]}                            & $-0.28 \pm 0.07$ & Mar21 \\
$R_\star$ [$R_{\odot}$]             & $0.4144\pm0.0038$ & This work \\
$M_\star$ [$M_{\odot}$]             & $0.413\pm0.011$ & This work \\
\noalign{\smallskip}
\multicolumn{3}{c}{\em Activity and age}\\
\noalign{\smallskip}
$v \sin i_\star$ [$\mathrm{km\,s^{-1}}$] & $<2.0$ & Rei18, Mar21 \\
$P_{\rm rot,phot}$ [d]              & 34.6--48.1 & This work$^{d}$ \\ 
pEW(He~{\sc i} D$_3$) [\AA]         & $-0.019 \pm 0.010$ & Fuh20 \\
pEW(H$\alpha$) [\AA]                & $+0.311 \pm 0.012$ & Fuh20 \\
pEW(Ca~{\sc ii} IRT$_1$) [\AA]      & $+0.742 \pm 0.009$ & Fuh20 \\
pEW(He~{\sc i} IR) [\AA]            & $+0.132 \pm 0.009$ & Fuh20 \\
$\log R'_{\rm HK}$                  & $-4.923^{+0.052}_{-0.059}$ & This work$^{c}$ \\
$\langle B \rangle$ [G]             & $170 \pm 60$ & Rei22 \\ 
$\log{L_{\rm Ca} / L_{\rm bol}}$    & --4.92 & Rei22 \\ 
Age [Gyr]                           & 1--8 & This work$^{d}$ \\
\noalign{\smallskip}
\hline
\end{tabular}
\tablebib{
AF15: \citet{AlonsoFloriano2015};
Arg1903: \citet{Argelander1903}; 
Cab16a: \citet{Caballero2016};
ExoFOP-{\em TESS}: \url{https://exofop.ipac.caltech.edu/tess/};
Fuh20: \citet{Fuhrmeister2020};
{\em Gaia} EDR3: \citet{GaiaEDR3};
Gli69: \citet{Gliese1969};
Lin21: \citet{Lindegren2021};
Mar21: \citet{Marfil2021};
PMSU: \citet{Reid1995};
Rei18: \citet{Reiners2018};
Rei22: \citet{Reiners2022};
Sou18: \citet{Soubiran2018}.
}
\tablefoot{
\tablefoottext{a}{See Table~\ref{tab:phot} for multiband photometry different from {\em TESS} $T$.}
\tablefoottext{b}{See Sect.~\ref{sec:activity} for the $P_{\rm rot}$ determination from ground photometry.}
\tablefoottext{c}{From data compiled by \citet{Perdelwitz2021}.} 
\tablefoottext{d}{\citet{Passegger2019} assumed a mean age of 0.6\,Gyr.}
}
\end{table}

The star GJ~806 was first tabulated in the Bonner Durchmusterung astro-photometric star catalog \citep{Argelander1903} with the designation BD+44~3567.
However, following rules of the International Astronomical Union, in this manuscript we use the designation in the first Catalogue of Nearby Stars by \citet{Gliese1969}. To avoid confusion with the character string `GI', here we use `GJ' \citet{Gliese1988} instead of `Gl' (Gliese) for the acronym.

Because of its relatively brightness ($V \sim$ 10.7\,mag) and closeness ($d \sim$ 12\,pc), GJ~806 has been frequently investigated on stellar radial velocities \citep{Wilson1953, Nidever2002}, parallaxes \citep{Strand1951, Wagman1967}, magnitudes and colors \citep{Leggett1992}, multiplicity \citep{Fischer1992, Kervella2019}, spectral typing \citep{Kirkpatrick1991, Rayner2009}, 
activity \citep{Hawley1996, Wright2004},
kinematics \citep{Reid1995, Montes2001},
or characterization in general \citep{Lepine2005, Mann2015, Schweitzer2019}, 
just to cite a few examples. 

GJ~806 was one of the targets of the primary CARMENES guaranteed time observations sample \citep{Reiners2018}, and as such it has been well characterized in the past \citep[e.g.,][]{2019A&A...627A.161P, Cifuentes2020}. 
Table~\ref{tab:stellar_parameters} summarizes the stellar parameters of GJ~806.
We compiled them from {\em Gaia} results and preparatory works \citep{GaiaEDR3, Soubiran2018} and from CARMENES publications \citep{Fuhrmeister2020, Marfil2021, Reiners2022}, or computed by us.
For simplicity, we refer to Caballero et~al. (subm.), who exhaustively described each parameter.
In the case of $\log R'_{\rm HK}$, we averaged 81 $R'_{\rm HK}$ measurements with Keck/HIRES compiled by \citet{Perdelwitz2021}.
The age and X-ray emission is discussed in Section~\ref{sec:atmos}, while the rotation period range was derived by us from photometry as described below. Additionally, the $\log R'_{\rm HK}$ value of -4.92 and the $\log{L_{\rm Ca} / L_{\rm bol}}$ value of  -4.92 \citep{Reiners2022} are consistent with GJ~806 being a very magnetically inactive star.


\subsection{Stellar rotation from seeing-limited photometry}
\label{sec:activity}

GJ 806 has a published rotation period of 19.9 days \citep{DiezAlonso2019}, however, there are clear indications from the TESS light curve, and from the magnetic field value \citep{Reiners2022} that the rotation period might actually be substantially longer. A quasi-periodic GP on TESS's sector 41 data, masking the transit periods returns a periodicity of ~$53^{+21}_{-14}$ days, significantly longer than the 27 day duration of the TESS sector. 

In order to determine GJ 806's rotational period, we started a photometric follow-up campaign using ground-based telescopes to  measure periodic flux variations related to the stellar rotation. We gathered seeing-limited data from the 1.5 m Ritchey-Chr\'{e}tien telescope at Sierra Nevada Observatory ($V$ and $R$ Johnson filters), the e-EyE telescopes ($B$ and $R$ Johnson filters), the 0.4 m telescopes at Las Cumbres Observatory ($V$ and $B$ Johnson filter), and the 0.8\,m Telescopi Joan Or\'{o} (TJO) at Observatori Astron\`{o}mic del Montsec (OAdM) ($R$ Johnson filter), see Sect.~\ref{sec:groud_obs} for details.

Each photometric data set was fitted using a linear function to model the mean value and slope of the measurements, and Gaussian Processes (GP) to model the periodic flux variations of the star. The linear function was used to describe long term variations present in the data while the GPs were used to model periodic variations. We used the package for Gaussian Processes \texttt{celerite} (\citealp{celerite}) and chose the kernel
\begin{equation}
    k_{ij\; \mathrm{Phot}} = \frac{B}{2+C} e^{-|t_i-t_j|/L} \left[  \cos \left( \frac{2 \pi |t_i-t_j|}{P_{rot}} \right) + (1 + C)  \right]
\label{Eq:StarRot_GPKernel}
,\end{equation}
where $|t_i-t_j|$ is the difference between two epochs or observations, $B$, $C$, $L$ are positive constants, and $P_{rot}$ is the stellar rotational period (see \citealp{celerite} for details). We performed a joint fit of all the photometry available, where each data set had as free parameters the zero point and slope value of the linear function and the GP kernel constants $B$, $C$, $L$. We set as common parameter for all the data sets the rotational period $P_{rot}$.

We started the fit with a global optimization of a log posterior function using \texttt{PyDE}\footnote{\url{https://github.com/hpparvi/PyDE}}. Then we used the results of the optimization to sample the posterior distribution of the parameters using an MCMC procedure with \texttt{emcee} (\citealp{emcee}). The MCMC consisted of 150 chains and ran for 2000 iterations as a burn-in and the main MCMC ran for 10000 iterations. The final parameter values and their respective uncertainties were computed using the percentiles of the posterior distributions: the median values were taken from the 50th percentile and the lower and upper uncertainties were computed using the 16th and 84th percentiles respectively.

\begin{figure*}[ht!]
\includegraphics[width=1\linewidth]{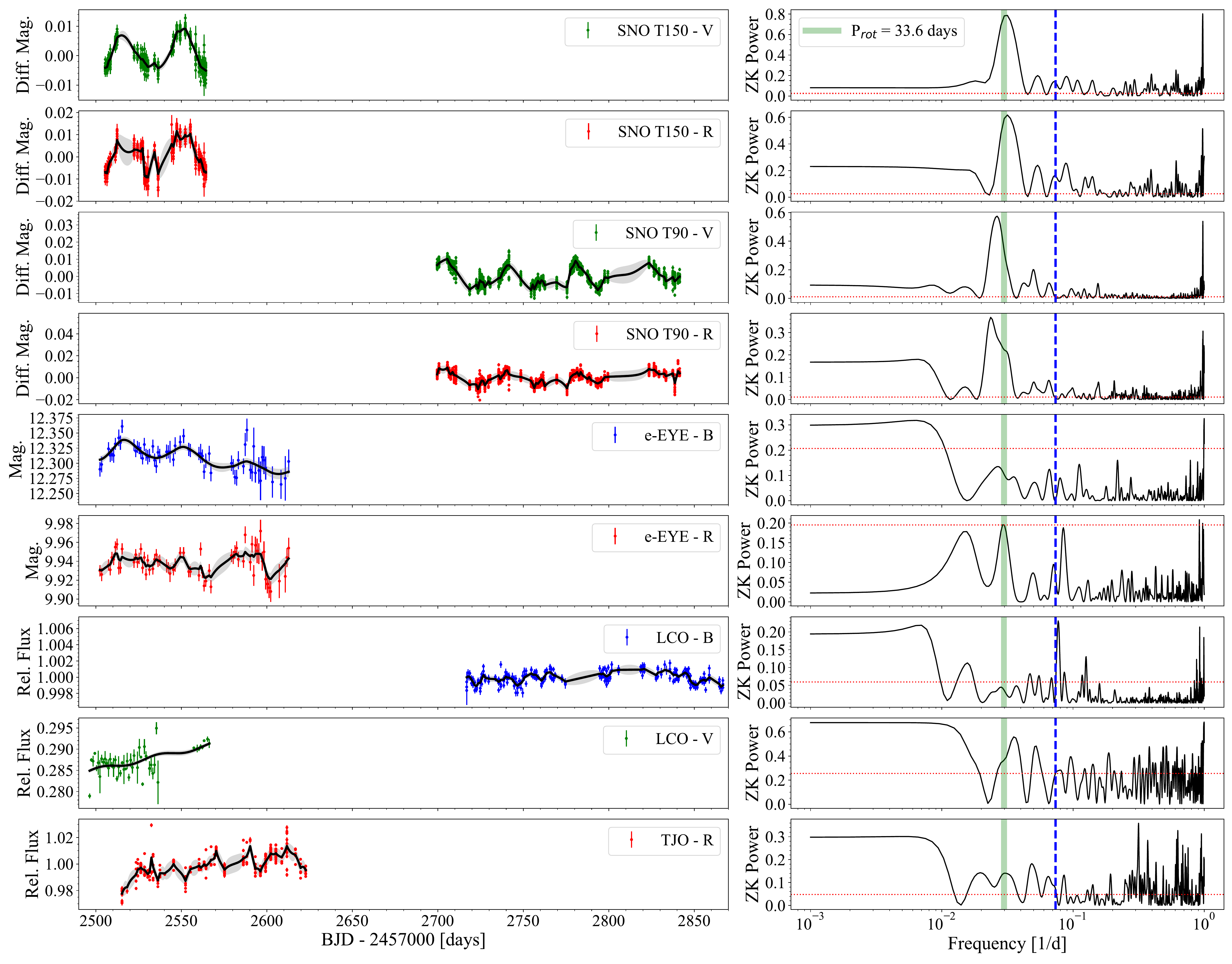}
\caption{Ground-based photometric observations (left) and Generalized Lomb-Scargle (GLS, \citealp{Zechmeister2009}) periodograms (right) for each data set of GJ~806. Fitting a linear function and a kernel with a periodic term to all the data sets using Gaussian Processes we find a stellar rotation period of $33.6^{+1.5}_{-1.0}$ days. The broken vertical blue line marks the 13.6 day period discussed later sections. The horizontal dotted red line 10\% false alarm probability (FAP) levels.}
\label{fig:rotphotometry}
\end{figure*}

Figure \ref{fig:rotphotometry} shows all the photometric data sets analysed here and the best fitted model for each time series. Using the procedure described before we find a stellar rotation period of $P_{rot} = 33.6^{+1.5}_{-1.0}$ days. This is in clear disagreement with the value from \citet{DiezAlonso2019}.  It is noteworthy, however, that a significant peak near 13.6 days appears in the LCO B band photometric time series, and in the e-EYE R band with sightly shorter periodicity. The period will be discusses in the following sections.

\subsection{Activity indicators}
\label{sec:activity}



\begin{figure}[ht!]
\includegraphics[width=1\linewidth]{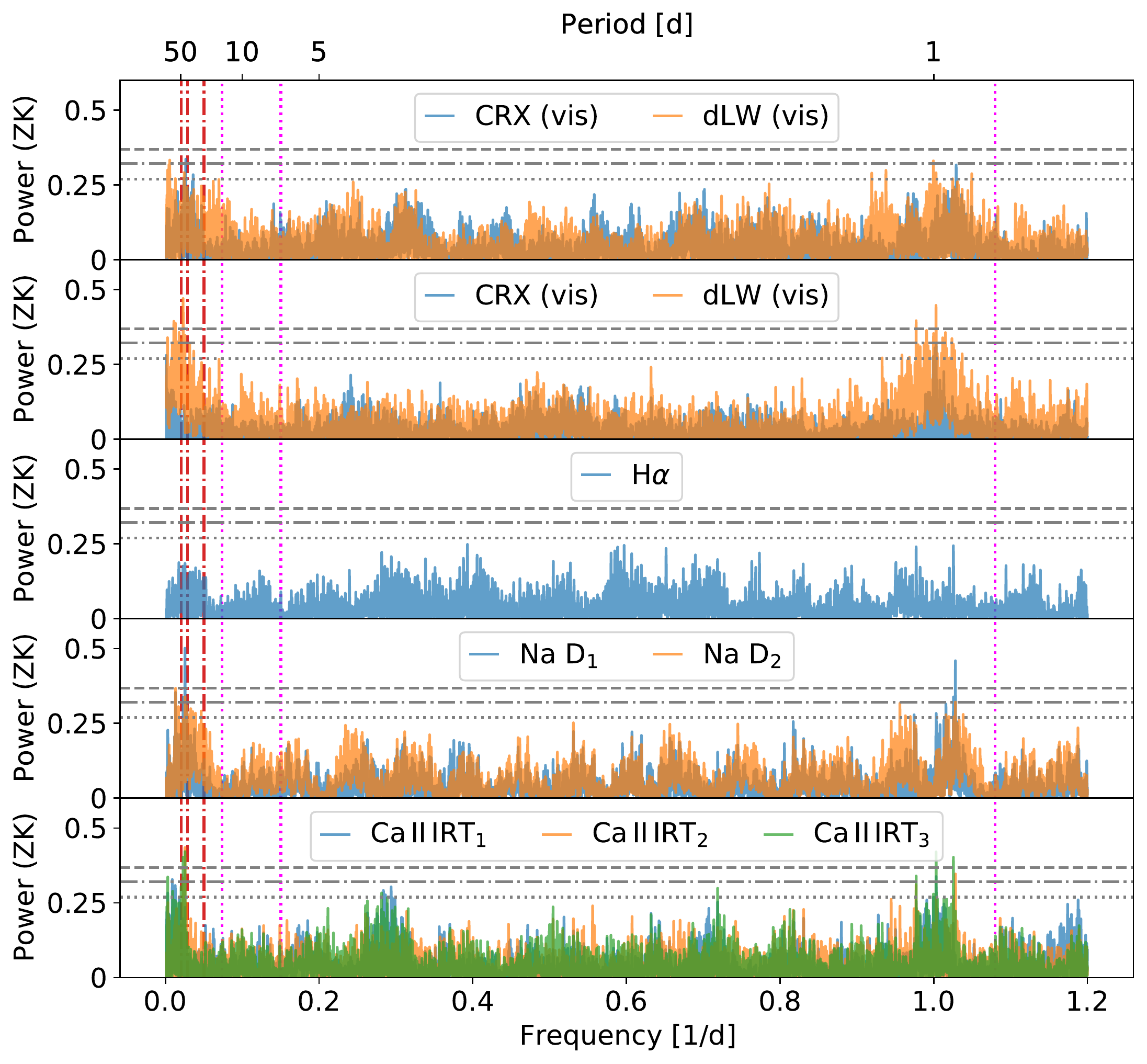}
\caption{Generalized Lomb-Scargle (GLS) periodograms of the activity indices derived from \texttt{SERVAL} for the CARMENES data. In all panels, the broken magenta lines indicate the planetary periodicities of 0.96, 6.6 and 13.6 days, and the broken red lines mark the 19.9, 34.6 and 48.1 day possible rotation periods discussed here. In the top two panels, the GLS for the CRX and dLW indices, are given independently for the visible and infrared channel spectra. The 10\%, 1\% and 0.1\% FAP levels are indicated by gray dotted, dash dotted, and dashed horizontal lines, respectively.}
\label{fig:actindex}
\end{figure}

As mentioned in Section~\ref{sec:Obs} the \texttt{SERVAL} analysis of the CARMENES data provides a set of activity indices (see \citet{Jeffers2022} for a full description and discussion of the different indices). Generalized Lomb-Scargle (GLS) periodograms for some of these indices, derived individually for the visible and the near-infrared spectra, are plotted in Figure~\ref{fig:actindex}.
Also marked are some of the periodicities later discussed in the data analysis section of this paper that we attribute to planetary signals.  

It is readily appreciable from Figure~\ref{fig:actindex} that none of the indices present significant (FAP < 0,1\%) periodic signals at either the rotation or planetary signals that will be later discussed in this paper. Only the D1 and the Ca II IRT$_3$ indices present a significant periodicity at 39.5 days which does not have a counterpart in the analysis of the radial velocity values.

To explore these periodicities in more detail, we use as chromospheric indicators the pseudo equivalent width (pEW) of H$\alpha$ and the two bluer \ion{Ca}{ii} infrared triplet (IRT) lines, following \citet{Fuhrmeister2019}; see their Table 2 for the used integration bands. Additionally we use a TiO bandhead index at 7050 \AA, defined as the ratio of the integrated flux density in two wavelength bands on both sides of the bandhead. There we follow \citet{Schoefer2019}; see their Table 3 for the used wavelength bands. To each time series of these chromospheric indicators a 3$\sigma$-clipping is applied to omit outliers due to flaring or weather and instrumental issues. Afterwards we detrend each  time series with a polynomial of grade three and then use the GLS periodogram \citep{Zechmeister2009} as implemented in PyAstronomy\footnote{\tt https://github.com/sczesla/PyAstronomy} \citep{Czesla2019} to search for periods in the undetrended and detrended time series. For the 67 usable (of 68 total) spectra of GJ 806 we find  a period of 38.629 to 38.996 days in the undetrended data of the four indicators with FAP lower than 0.0036 and a period of  38.690 to  38.934 days in the detrended data with FAP lower than 0.0007. The mean of all eight computed periods (for the four indicators in the undetrended and detrended case) is 38.8 $\pm$ 0.2 days. See Figure~\ref{fig:rothalpha}.

\begin{figure}[ht!]
\includegraphics[width=1\linewidth]{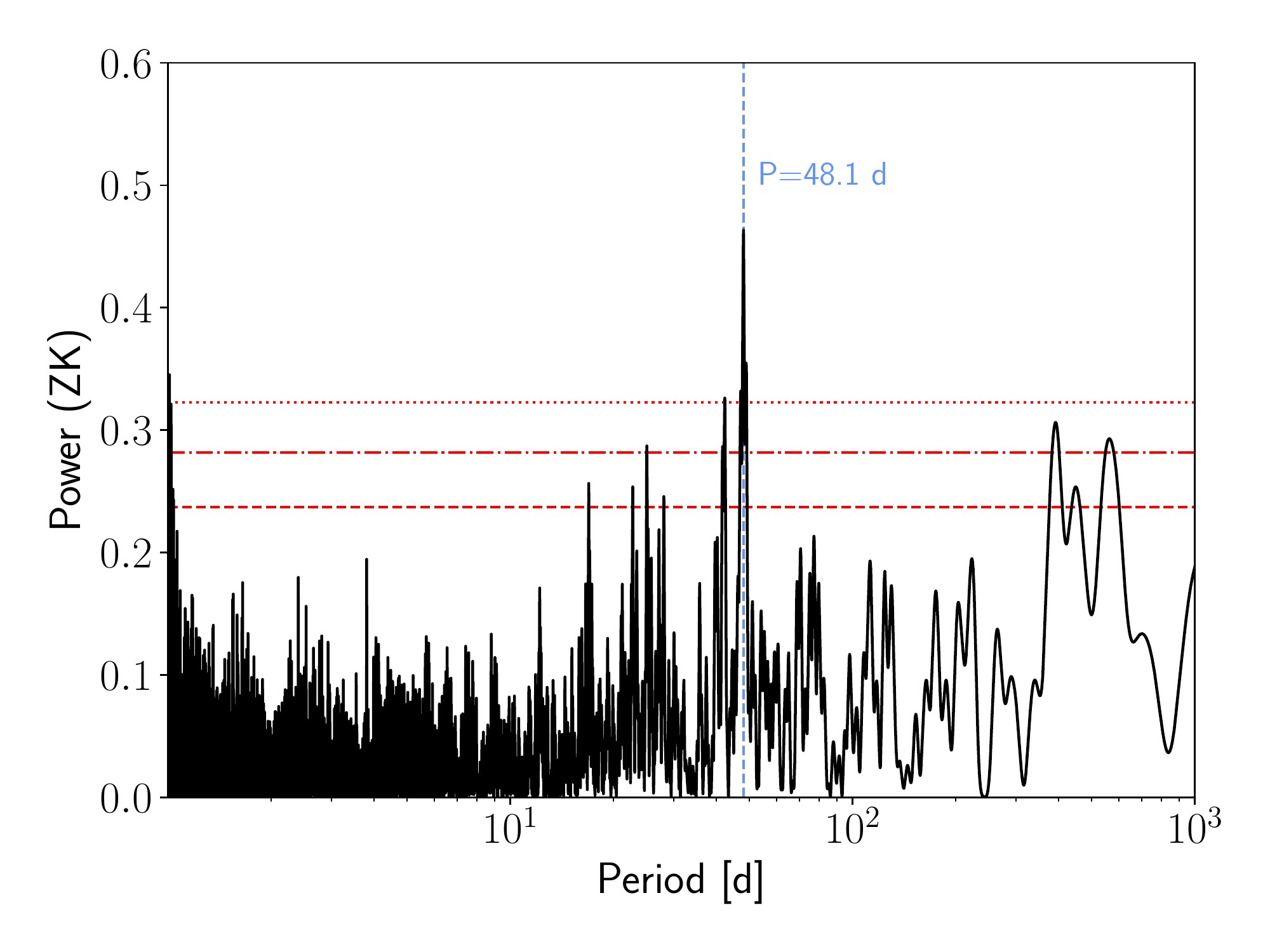}
\caption{A GLS periodogram of the $R_{HK}^\prime$ values from the HIRES spectra. The dashed lines in the GLS panel represent the The 10\%, 1\% and 0.1\% significance levels.}
\label{fig:rothk}
\end{figure}

Finally, to determine the stellar rotation period in a third independent way, we also used the $R_{HK}^\prime$ measurements published by \cite{Perdelwitz2021}, which are based on spectra acquired with HIRES \citep{Vogt1992}. The data reduction is also described in \cite{Perdelwitz2021}. The 81 values with sufficient signal-to-noise (SNR>5 at the Ca~{\sc ii}~H\&K lines) were analyzed with a GLS approach \citep{Zechmeister2009} with a period range of 1-1000~d and an oversampling of 1000. The GLS periodogram (see Figure~\ref{fig:rothk}) yields a clear detection at a period of 48.1~d, with a false-alarm probability below $10^{-6}$. 

\begin{figure*}
    \centering
    \includegraphics[width=\textwidth]{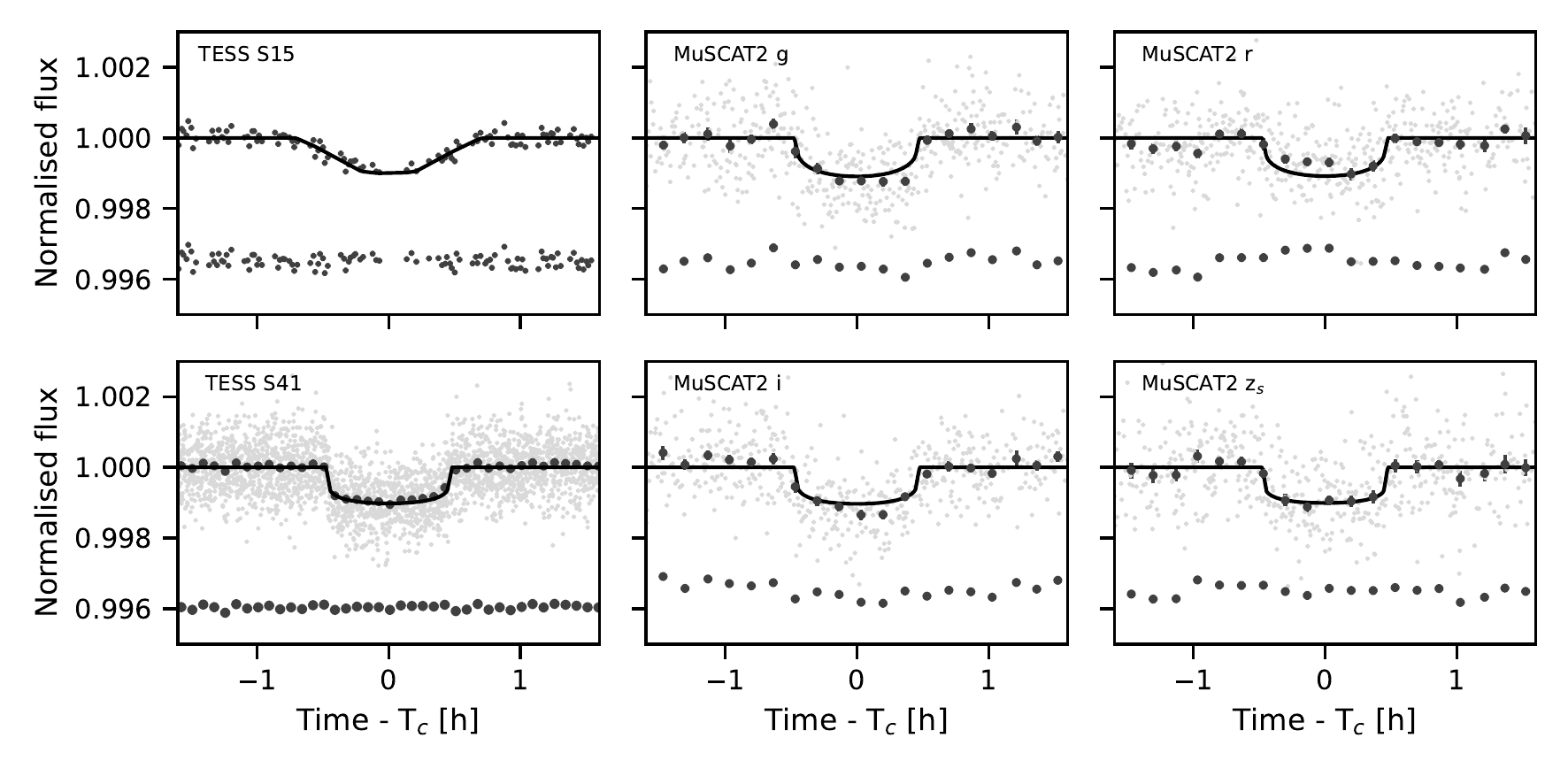}
    \caption{The phase-folded \tess and MuSCAT2 photometry with the median posterior model. The small dots show the original data and the larger dots show data 
    binned to 5~min (\tess) and 10~min (M2) resolution for visualisation. The black line shows the posterior median transit model for each passband and dataset. 
    The model for 30~min cadence TESS QLP light curve is supersampled with 10 samples per exposure.}
    \label{fig:photometry_joint_fit}
\end{figure*}

In summary, using different approaches, we find evidences for GJ 806's rotation period to be at $\sim$34.6, $\sim$38.8 and $\sim$48.1 days. Unfortunately, even considering the large errorbars in these period determinations, the results are not compatible. We have to conclude that while the rotation period of GJ 806 lies most likely between 30-50 days, its true value remains undetermined.

\section{Analysis}
\label{sec:Analysis}

\subsection{Transit photometry} 
\label{sec:photan}

We modelled the transits for the innermost planet (GJ 806 b) using the \tess sector~15 light curve observed in 30~min cadence, the \tess sector~41 light  curve observed in 2~min cadence, and the MuSCAT2 four-colour light curves from three nights jointly using \pytransit \citep{Parviainen2015,Parviainen2020a,Parviainen2020b}, 
and show the photometry with the fitted transits models in Fig.~\ref{fig:photometry_joint_fit}. The model was parameterised using the mid-transit time at epoch zero, the orbital period, the stellar
density, the impact parameter, and the planet-star area ratio (independent of passband or light curve), two quadratic limb darkening coefficients for each passband,
an average white noise estimate for each light curve, and a set of linear model covariate coefficients for each light curve. The limb darkening coefficients were constrained using priors calculated with \ldtk \citep{Parviainen2015b}, the zero epoch and orbital period had wide normal priors centred around the TESS' TOI announcement values \footnote{https://tev.mit.edu/data/},
and the rest of the parameters had uninformative priors that were after the posterior estimation checked not to constrain the parameter posteriors. The parameter posteriors agree well with the joint analysis combining photometry and RV information in Sect.~\ref{sec:jfit}.


\subsection{Radial velocities} 
\label{sec:rvs}


We searched for planetary signals in the different RV datasets using a GLS and computing the theoretical false alarm probability (FAP) as described in \citet{GLS_paper}.
Furthermore, we used \texttt{juliet}\footnote{\url{https://juliet.readthedocs.io/en/latest/index.html}} (\citealp{juliet}) to model the detected signals. This \texttt{python} code is based on other public packages for transit light curves (\texttt{batman}; \citealp{batman}) and RV (\texttt{radvel}; \citealp{radvel}) modeling and allows the inclusion of Gaussian Processes (GPs; \texttt{george}, \citealp{george}; \texttt{celerite}, \citealp{celerite}) to model the presence of systematic effects in the data. Instead of using Markov chain Monte Carlo (MCMC) techniques, \texttt{juliet} uses a nested sampling algorithm to explore all the parameter space and also compute the Bayesian model log-evidence ($\ln Z$). This is performed using the MultiNest algorithm (\citealp{MultiNest}) via its \texttt{python} implementation \texttt{PyMultinest} \citep{PyMultiNest}.
We considered sinusoidal signals with normal priors for the period ($P$) and uniform priors for the central time of transit ($t_0$) and the semi-amplitude ($K_p$) to fit the periodicities found in the RVs. We included an instrumental jitter and systemic velocity terms for each of the individual RV datasets.

\subsubsection{HIRES RVs} 

\begin{figure*}[ht!]
\includegraphics[width=0.49\linewidth]{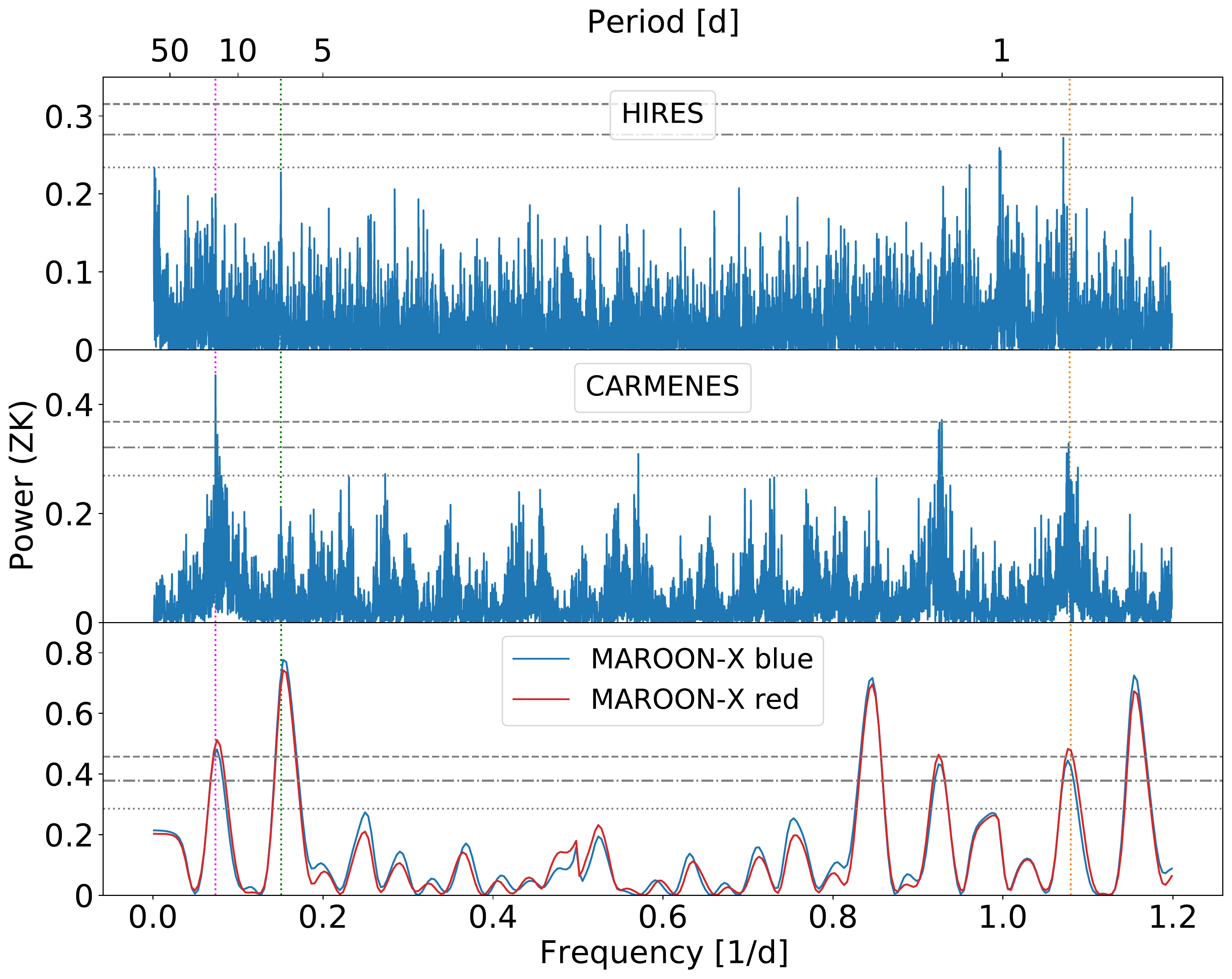}
\includegraphics[width=0.49\linewidth]{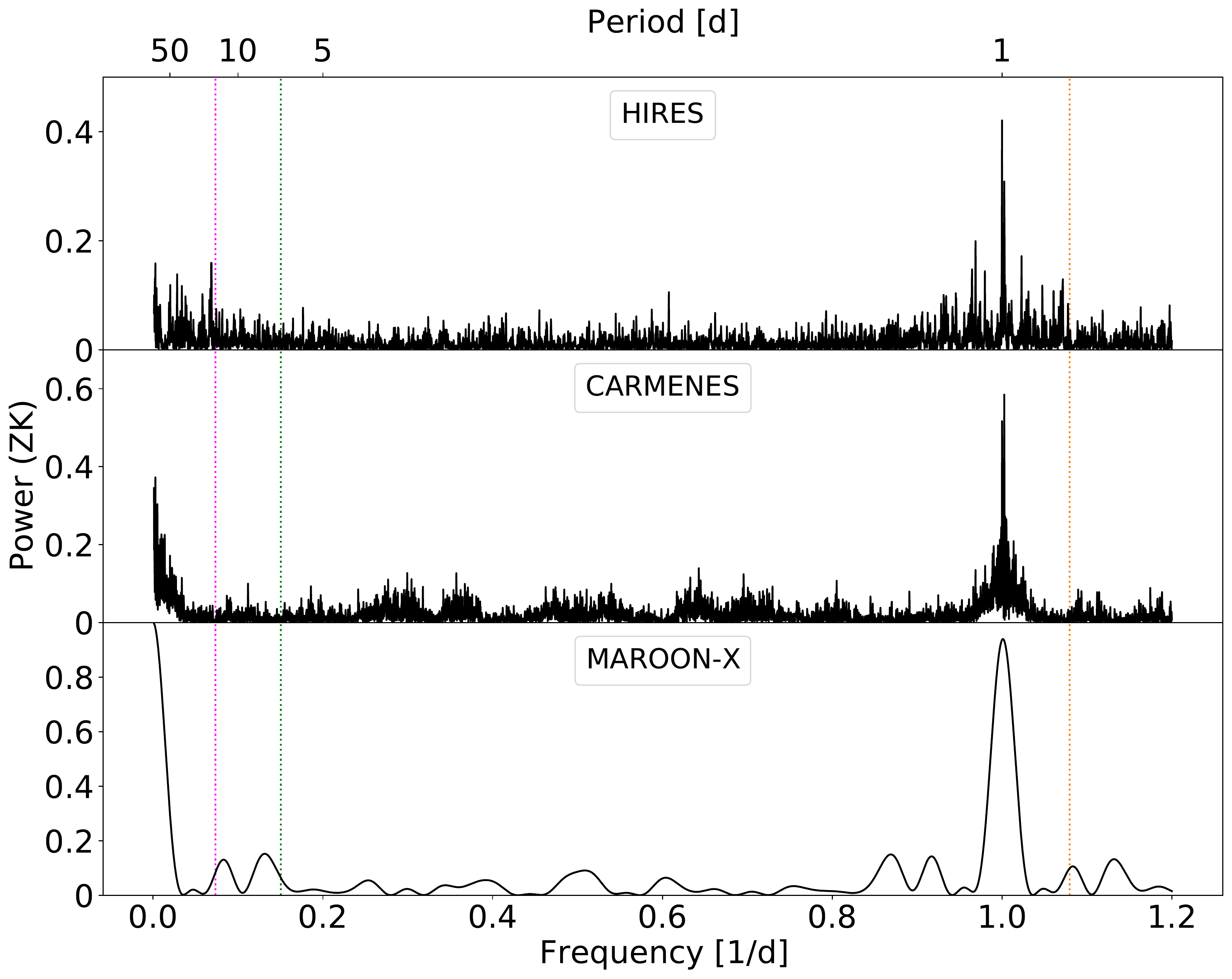}
\caption{Left: Generalized Lomb-Scargle (GLS) periodograms of the HIRES (top), CARMENES (middle) and the two MAROON-X channels (bottom) radial velocity time series. Vertical dotted lines mark the highest peaks at 13.6\,d (magenta), 6.6\,d (green) and 0.92\,d (orange). The 10\%, 1\% and 0.1\% FAP levels are indicated by gray dotted, dash dotted, and dashed lines, respectively. Right: Window functions of the HIRES, CARMENES and MAROON-X channels time series. Vertical dotted lines mark the 13.6\,d (magenta), 6.6\,d (green) and 0.92\,d (orange) periods.  }
\label{fig:singleRvs}
\end{figure*}

Despite the relatively large number of HIRES measurements, and the long baseline of the dataset, the GLS does not show significant peaks (FAP\,$\leq$10\%) at periods greater than 1\,d. There is a significant peak (FAP\,$\sim$1\%) near 0.92\,d but it is too far away in period to be associated to the signal from the transiting planet GJ 806 b (Figure\,\ref{fig:singleRvs}). We forced a fit to the transiting planet using the parameters from the photometry but the retrieved model does not recover any significant planet signal. Thus, the HIRES RVs alone do not allow us to characterize the transiting planet, nor do they contain any indication of additional signals.

Although the HIRES data alone can not find significant planetary signals, two signals at 6.6 and 13.6 days are found when using the other datasets discussed later in this work. If we perform a specific fit for these two signals, with two Keplerians centered around these periods, we find significant detections with semi amplitudes of 2.7$^{+0.6}_{-1.1}$\,m\,s$^{-1}$  and 3.20\,$\pm$\,0.60\,m\,s$^{-1}$, respectively.
Using one Keplerian only for either of the periods or 3 Keplerians including the transiting planet does not change the results, and the inner transiting planet is not detected (see Table\ref{tab:Semiampl}).

\subsubsection{CARMENES RVs} 

\begin{figure*}
    \centering
    \includegraphics[width=0.49\hsize]{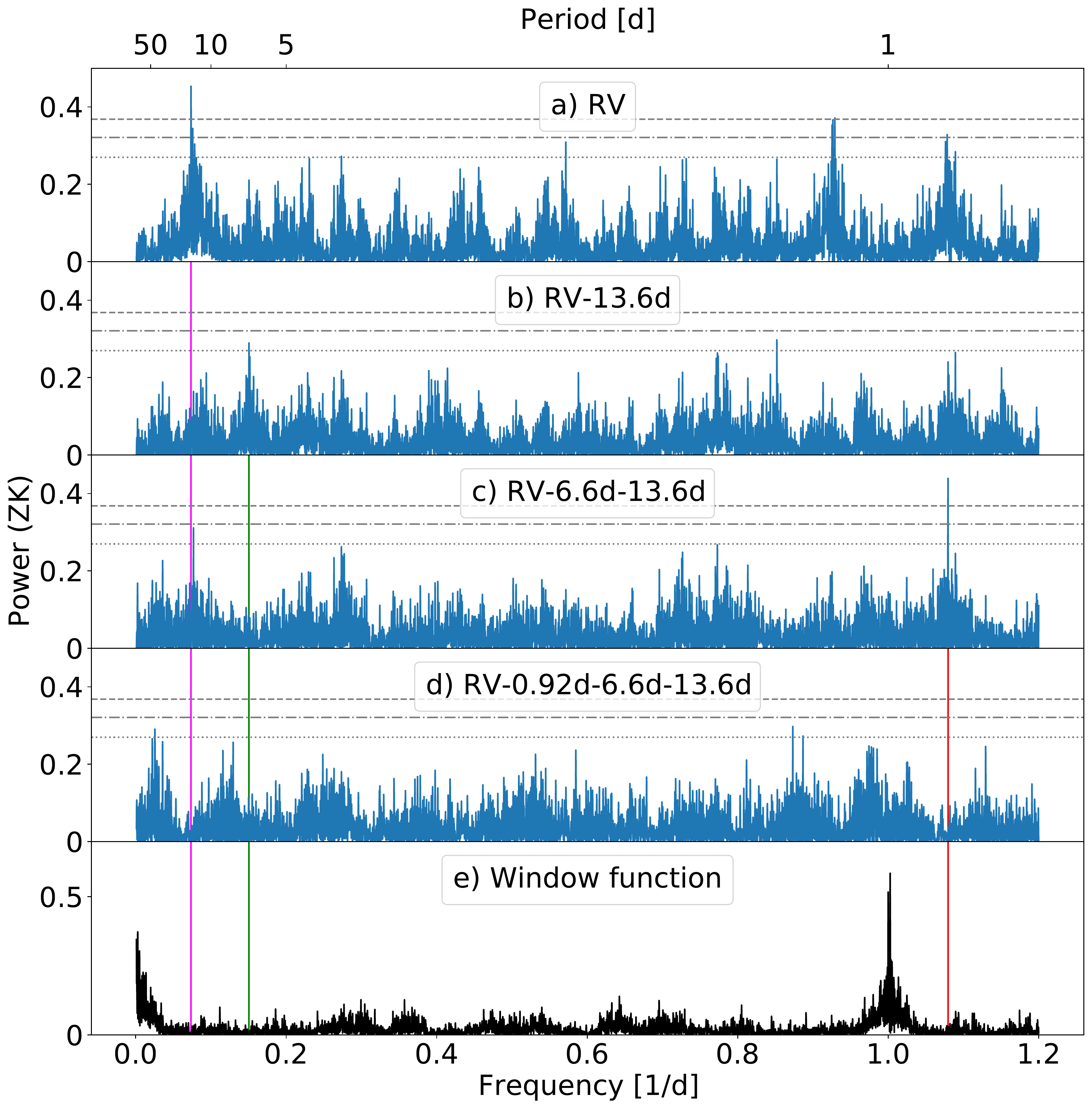}
    \includegraphics[width=0.49\hsize]{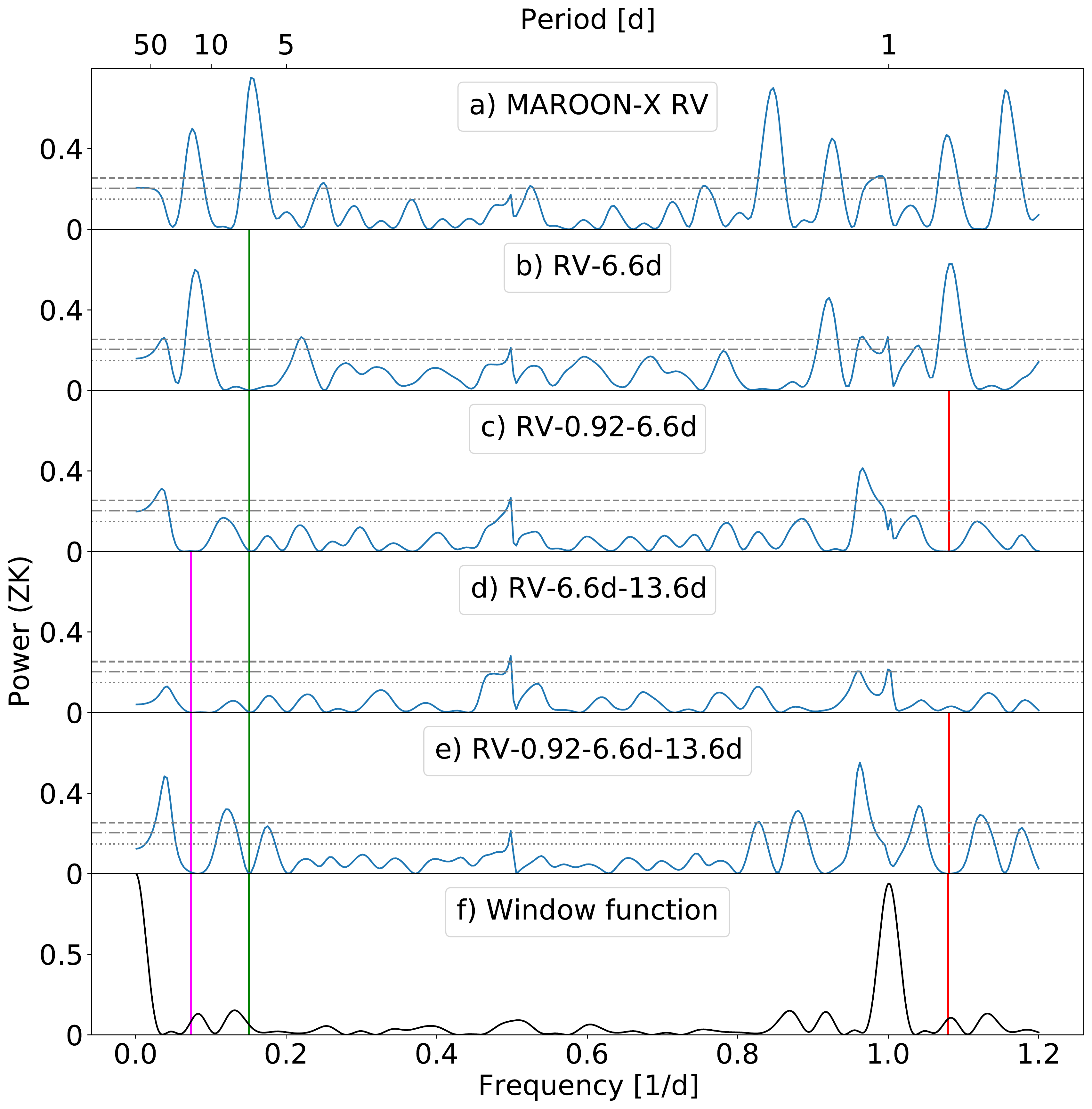}
    \caption{ Generalized Lomb-Scargle (GLS) periodograms of CARMENES (left) and MAROON-X (right) RV measurements and the residual RVs after subtraction of different models. In all panels, the 10\%, 1\% and 0.1\% FAP levels are indicated by gray dotted, dash dotted, dashed lines, respectively. Left Panels) $a)$ GLS of RV dataset. $b)$ GLS of the RV residuals after fitting the 13.6\,d signal (vertical magenta line). $c)$ GLS of the RV residuals after simultaneously fitting the 6.6\,d (vertical green line) and 13.6\,d signals. $d)$ GLS of the RV residuals after simultaneously fitting the transitting planet ($P$\,=\,0.926\,d, vertical red line), 6.6\,d and 13.6\,d signals. Right Panels) $a)$ GLS of RV dataset. $b)$ GLS of the RV residuals after fitting the 6.6\,d signal (vertical green line). $c)$ GLS of the RV residuals after simultaneously fitting the transiting planet ($P$\,=\,0.926\,d, vertical red line) and 6.6\,d signals. $d)$ GLS of the RV residuals after simultaneously fitting the 6.6\,d and 13.6\,d (vertical magenta line) signals. $e)$ GLS of the RV residuals after simultaneously fitting the transiting planet, 6.6\,d and 13.6\,d signals. $f)$ Window function.
    }
    \label{Fig:CARMandMAROON_GLS}
\end{figure*}

The GLS of CARMENES RVs presents a very significant peak (FAP\,$\ll$0.1\%) at $\sim$13.6\,d and a significant peak (FAP\,$\sim$1\%) near 0.92\,d (Figure\,\ref{fig:singleRvs}). Because the periodogram region near the transiting planet may be affected by alias of other signals we studied the signals in order of significance. We fitted the signal at 13.6\,d with a period normal prior ($\mathcal{N}(13.608,0.05)$ [d]). Due to the dispersion in RV, we always used a semi-amplitude uniform prior between 0 and 20\,m/s ($\mathcal{U}(0,20)$ [m/s]) for the fitted signals. For the 6.6\,d signal, we used also a period normal prior ($\mathcal{N}(6.64,0.1)$ [d]). The priors to fit the transiting planet signal were a period normal prior ($\mathcal{N}(0.92632,0.001)$ [d]).
After fitting the signal at 13.6\,d, the most significant signal of the residuals is at $\sim$6.6\,d and the 0.92\,d decreases its significance. Thus, we simultaneously fitted the signals at 6.6\,d and 13.6\,d. The residuals still present a peak near $\sim$13\,d, but the signal of the transiting planet is clearly detected (FAP\,$\ll$0.1\%). After simultaneously fitting the transiting planet and the 6.6\,d and 13.6\,d signals, the GLS of the residuals is mainly flat with only a non-significant peak near 39\,d (FAP\,$\sim$10\%). This fitting process is illustrated in Figure\,\ref{Fig:CARMandMAROON_GLS} (and in Figure\,\ref{Fig: CARMENES GLS freq} with frequency in the x-axis). Compared with the other models, the 3 planets model is preferred for the CARMENES data in terms of Bayesian log-evidence (see Table\,\ref{table - LOG EVIDENCE}) and minimises the squared sum of the residuals and the jitter term contribution.

Because the 6.6\,d and 13.6\,d periods are close to 1:2 ratio, we explored the hypothesis that one signal is an harmonic of the other one. However, in the CARMENES RV analysis, and in the rest of datasets we analyze in the next sections, we found that in general when fitting either of the signals, the other becomes stronger, indicating that we are in fact dealing with either two outer planets in near-resonance or a planet signal and a stellar signal.

\subsubsection{MAROON-X RVs} 


The GLS of the MAROON-X RVs from the red arm, the blue arm and their combination are similar. They clearly present the transiting planet signal (FAP\,$\sim$0.1\%) and well defined peaks at $\sim$6.6\,d (FAP\,$\ll$0.1\%) and $\sim$13.6\,d (FAP\,$\sim$0.1\%) (Figure\,\ref{fig:singleRvs}). In this section, we used the following period priors to fit the 0.9\,d, 6.6\,d, and 13.6\,d signals, respectively: $\mathcal{N}(0.9263,0.001)$ [d] , $\mathcal{N}(6.6,0.1)$ [d] , and $\mathcal{N}(13.6,0.1)$ [d].

In that case, we first fitted the $\sim$6.6\,d as it is the strongest signal. In the GLS of the residuals, the transiting planet then became the biggest peak, followed by the $\sim$13.6\,d signal, both with FAP\,$\ll$0.1\%. Then, we simultaneously fitted the 0.92\,d and 6.6\,d signal.
However, after fitting those signals, the 13.6\,d signal completely disappeared in the GLS of the residuals. There is only a long trend that peaks at near $\sim$30\,d.
On the contrary, if we fit simultaneously the 6.6\,d and 13.6\,d periods, the transiting planet signal disappears in the GLS of the residuals.
When forcing the models to fit simultaneously the 0.92\,d, 6.6\,d and 13.6\,d signals, the 13.6\,d signal is not significantly recovered. While we have no clear explanation for these results, it is possible that the 1-day-alias of 13.6-day signal (at P$\simeq$0.931\,d) affects the MAROON-X planet signal detections when the transiting planet at 0.92d is fitted. Again our fitting process is illustrated in Figure\,\ref{Fig:CARMandMAROON_GLS}.

\subsubsection{CARMENES + HIRES RVs} 

\begin{figure}
    \centering
    \includegraphics[width=\hsize]{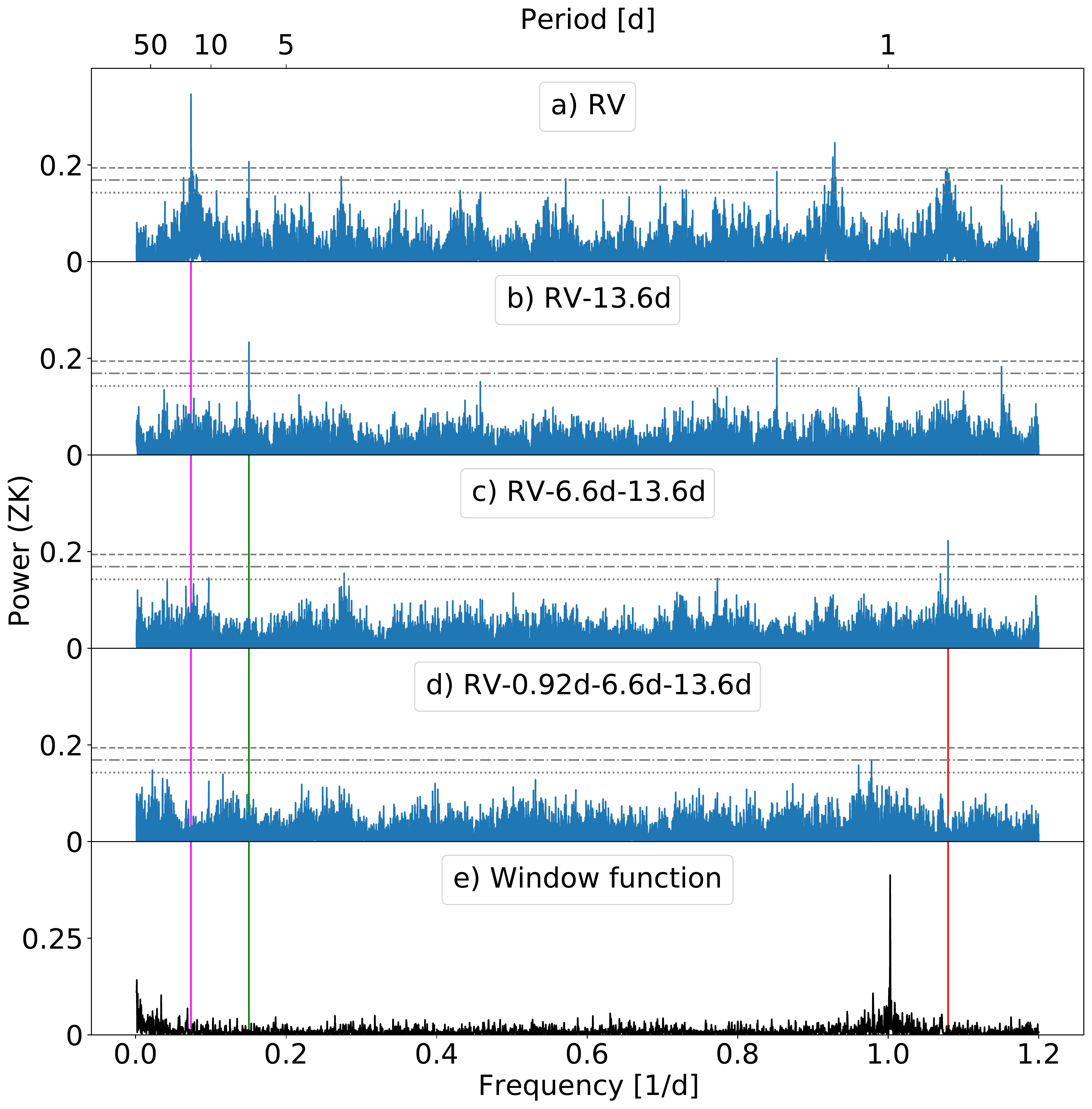}
    \caption{Generalized Lomb-Scargle (GLS) periodograms of the combined CARMENES and HIRES RVs measurements and the residual RVs after subtraction of different models. $a)$ GLS of RV dataset. $b)$ GLS of the RV residuals after fitting the 13.6\,d signal (vertical magenta line). $c)$ GLS of the RV residuals after simultaneously fitting the 6.6\,d (vertical green line) and 13.6\,d signals. $d)$ GLS of the RV residuals after simultaneously fitting the transitting planet ($P$\,=\,0.926\,d, vertical red line), 6.6\,d and 13.6\,d signals. $e)$ Window function.  The 10\%, 1\% and 0.1\% FAP levels are indicated by gray dotted, dash dotted, dashed lines, respectively.
    }
    \label{Fig: CARMENES+HIRES GLS period}
\end{figure}

\begin{table}
\caption[width=\hsize]{
\label{table - LOG EVIDENCE}
Comparative between Bayesian log-evidence ($\Delta\,\ln{Z}$) for the 13.6\,d (1\,pl), 6.6\,d+13.6\,d (2\,pl) and 0.92\,d+6.6\,d+13.6\,d (3\,pl) models using CARMENES+HIRES, and CARMENES+HIRES+MAROON-X RVs datasets. We used the simplest model, the 1 planet model, as a reference.
}
\centering
\resizebox{\columnwidth}{!}{%
\begin{tabular}{lccc}
\hline \hline \vspace{-0.25cm} \\
RV dataset & 13.6\,d & 6.6\,d+13.6\,d & 0.92\,d+6.6\,d+13.6\,d \vspace{0.15cm}\\
\hline \\
CARMENES & 0.0 & 6.8 & 19.8 \vspace{0.05cm}\\
CARMENES+HIRES & 0.0 & 9.6 & 23.6 \vspace{0.05cm}\\
CARMENES+HIRES+MAROON-X & 0 & 41 & 102 \vspace{0.15cm}\\
\hline
\end{tabular}
}
\end{table}

We tried to improve our results by combining the CARMENES and HIRES measurements.
The GLS combining both RV datasets presents peaks near the transiting planet period with FAP lower than 1\%. However, the most significant peak is at 13.6\,d and also displays a significant peak at 6.6\,d (FAP\,$\sim$0.1\%). The GLS is shown in Figure\,\ref{Fig: CARMENES+HIRES GLS period}.

Following the same steps as in the CARMENES RV alone analysis, we fitted the 13.6\,d periodicity with a period normal prior ($\mathcal{N}(13.6,0.1)$ [d]) and, after subtracting this signal, the signal at 6.6\,d increase its significance. Thus, we simultaneously fitted that, with a period normal prior ($\mathcal{N}(6.64,0.1)$ [d]), and 13.6\,d signals refining the 6.6\,d signal properties and constraining the results for 13.6\,d periodicity. When the three periods are simultaneously fitted, using a period normal prior ($\mathcal{N}(0.9263,0.000032)$ [d]) for the transitting planet, the RV residuals are mainly flat without significant peaks.
Compared with the other models, the 3 keplerian signal model is preferred in terms of Bayesian log-evidence (Table\,\ref{table - LOG EVIDENCE}) and minimises the squared sum of the residuals and the jitter term contribution.

We also tested the significance of the 0.92\,d, 6.6\,d and 13.6\,d periods fitting sequentially the three signals but changing the fitting order. We obtained consistent results in all the cases, enhancing the planetary origin of the signals. Clearly the combined analysis of CARMENES + HIRES data is dominated by the CARMENES signals, and is not significantly different from using CARMENES data alone.

\subsubsection{CARMENES + HIRES + MAROON-X RVs} 
\label{sec:threervs}

\begin{figure}
    \centering
    \includegraphics[width=\hsize]{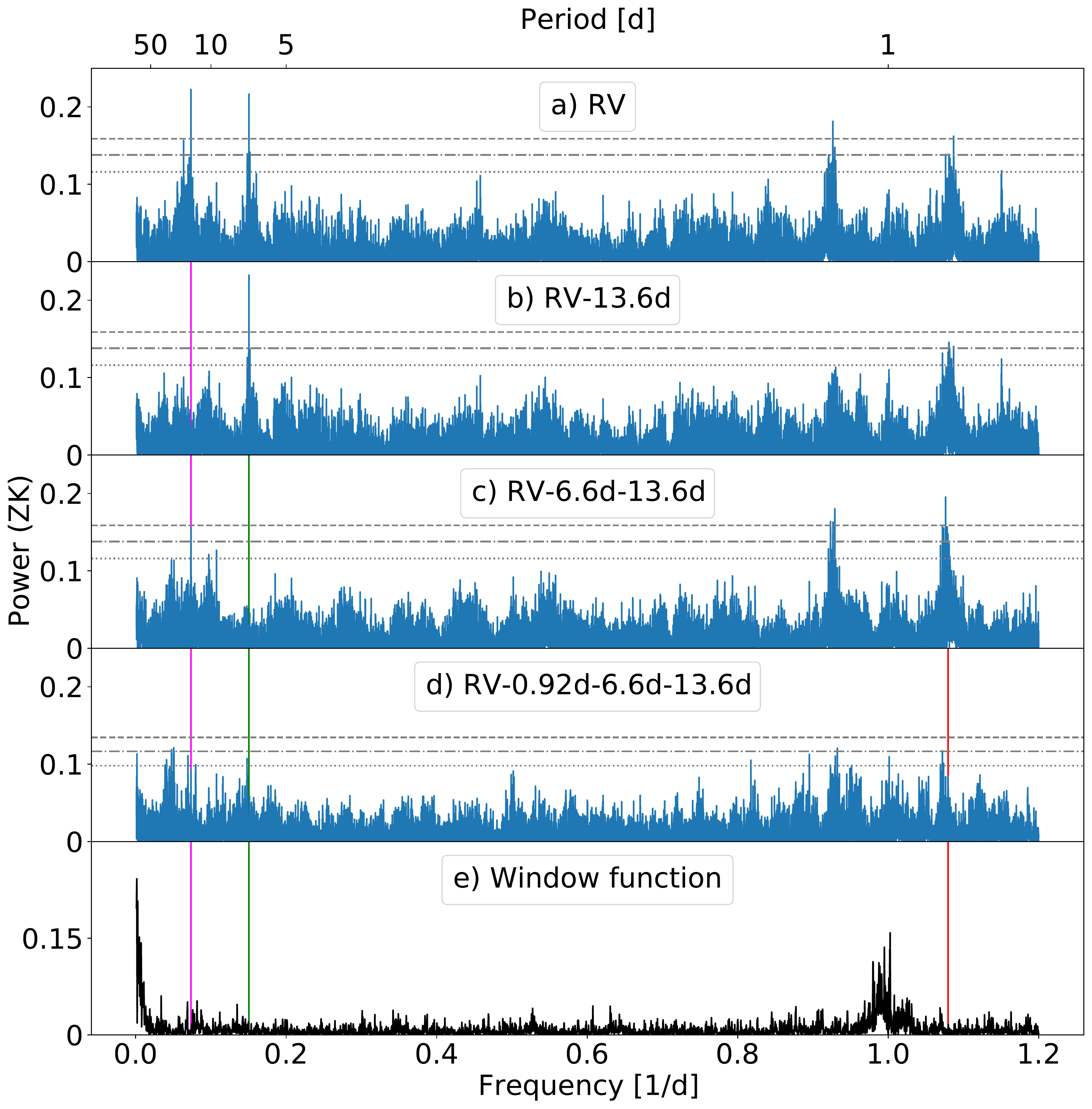} 
    \caption{ Generalized Lomb-Scargle (GLS) periodograms of CARMENES, HIRES and MAROON-X red and blue arm RVs measurements and the residual RVs after subtraction of different models. $a)$ GLS of RV datasets. $b)$ GLS of the RV residuals after fitting the 13.6\,d signal (vertical magenta line). $c)$ GLS of the RV residuals after simultaneously fitting the 6.6\,d (vertical green line) and 13.6\,d signals. $d)$ GLS of the RV residuals after simultaneously fitting the transiting planet ($P$\,=\,0.926\,d, vertical red line), 6.6\,d and 13.6\,d signals. $e)$ Window function.
    The 10\%, 1\% and 0.1\% FAP levels are indicated by gray dotted, dash dotted, dashed lines, respectively.
    }
    \label{Fig: ALL RV GLS period}
\end{figure}

\begin{table*}
\caption[width=\hsize]{The retrieved semi-amplitudes of the three radial velocity signals obtained with each analysis of the different datasets. For the different CARMENES+HIRES+MAROON-X models, the $\Delta\ln{Z}$ values are also given. The priors used in these models are described in Table\,\ref{tab:Priors_TABLE3}
\label{tab:Semiampl}
}
\centering
\begin{tabular}{lcccc}
\hline 
\hline 
\noalign{\smallskip}
Instrument & $\Delta\ln{Z}$ & $K_{0.92\,d}$ [m\,s$^{-1}$] & $K_{6.6\,d}$ [m\,s$^{-1}$] & $K_{13.6\,d}$ [m\,s$^{-1}$] \\
\hline 
\noalign{\smallskip}
CARMENES & &3.30 $\pm$ 0.45 & 3.30 $\pm$ 0.45 & 3.65 $\pm$ 0.42 \\
MAROON-X $^{(a)}$ & & 2.48 $\pm$ 0.15 & 3.45 $\pm$ 0.15 & -- \\
MAROON-X red $^{(a)}$ & & 2.47 $\pm$ 0.20 & 3.53 $\pm$ 0.18 & -- \\
MAROON-X blue $^{(a)}$ & & 2.38 $\pm$ 0.25 & 3.57 $\pm$ 0.21 & -- \\
HIRES $^{(b)}$ & & -- & 2.7$^{+0.6}_{-1.1}$ & 3.20 $\pm$ 0.60 \\
CARMENES+HIRES & &2.65 $\pm$ 0.35 & 3.00 $\pm$ 0.35 & 3.77 $\pm$ 0.32 \\
\noalign{\smallskip}
\hline
\noalign{\smallskip}
\multicolumn{5}{c}{\em CARMENES+HIRES+MAROON-X} \\
2pl (0.92d + 6.6d) & -20.4 &  2.20 $\pm$ 0.21 & 3.60 $\pm$ 0.19 & -- \\
3pl  & 0.0  &  3.36 $\pm$ 0.25 & 3.17 $\pm$ 0.18 & 1.77 $\pm$ 0.30 \\
\textbf{2pl ((CARM+HIRES)-13.6d) + MAROON-X} $^{(c)}$ &  23.5 &  2.38 $\pm$ 0.18  & 3.45 $\pm$ 0.16  & 4.10 $\pm$ 0.20 \\
M2: 2pl+GP$_{\mathrm{QP}}$(CARM) & -8.2  &  2.10 $\pm$ 0.20 & 3.63 $\pm$ 0.18 &  \textit{ 6.26$^{+2.2}_{-1.6}$ } \\
M4: 2pl+GP$_{\mathrm{QP}}$(CARM,HIRES) &  3.3 &  2.20 $\pm$ 0.21 & 3.63 $\pm$ 0.18 &  \textit{ 6.26$^{+2.2}_{-1.6}$ , 8.7$\pm$2.8 }  \\
M6: 2pl+GP$_{\mathrm{QP}}$(CARM+HIRES) & 7.2  &  2.18 $\pm$ 0.21 &  3.61 $\pm$ 0.18  &  \textit{ 7.6$^{+1.8}_{-1.4}$ } \\
M3: 2pl+GP$_{\mathrm{ExpSinSq}}$(CARM) & -3.4  &  2.22 $\pm$ 0.20  & 3.54 $\pm$ 0.17 & \textit{ 4.8$^{+8}_{-1.8}$ } \\
M5: 2pl+GP$_{\mathrm{ExpSinSq}}$(CARM,HIRES) &  10.5  &  2.30 $\pm$ 0.20 & 3.55 $\pm$ 0.18 & \textit{ 4.7$^{+7.5}_{-1.8}$ , 4.2 $\pm$ 0.5 } \\
M7: 2pl+GP$_{\mathrm{ExpSinSq}}$(CARM+HIRES) &  8.1  &  2.26 $\pm$ 0.20 & 3.56 $\pm$ 0.18 & \textit{ 3.6 $\pm$ 0.8 } \\
\hline
\end{tabular}
\tablefoot{The numbers in italics are not determined by a keplerian fit, but are the semi-amplitudes derived from the GP fitting.
$^{(a)}$ The 13.6\,d signal is not detected in the MAROON-X data.
$^{(b)}$ HIRES GLS do not show significant signals at 0.92, 6.6 or 13.6\,d.
$^{(c)}$ Reference Model: 2pl ((CARM+HIRES)-13.6d) + MAROON-X. }
\end{table*}

Finally, we jointly analysed the RV measurements from CARMENES, HIRES and MAROON-X red and blue arm. For a quick inspection of the three datasets combined, we used the following period priors to fit the 0.9\,d, 6.6\,d, and 13.6\,d signals, respectively: $\mathcal{N}(0.9263,0.0001)$ [d], $\mathcal{N}(6.64,0.05)$ [d], and $\mathcal{N}(13.60,0.05)$ [d]. The GLS of the full combined dataset clearly shows the tree signals under study: the transiting planet at 0.93\,d, and the 6.6\,d and 13.6\,d planet candidates. Figure\,\ref{Fig: ALL RV GLS period} displays the GLS.

After fitting the 3 signals under study in the same way as in previous section, the GLS peaks in the residuals remain under FAP$\leq10\%$. The 3 keplerian signal model is still the preferred choice in terms of Bayesian log-evidence (Table\,\ref{table - LOG EVIDENCE}).

Thus it would seem that we obtain a consistent picture of three planetary candidates orbiting GJ~806. But a closer inspection of our analysis shows some inconsistencies in the results. The top section in Table\,\ref{tab:Semiampl} summarizes the fitted semi-amplitude (ultimately mass) to each of the three planet candidates. It is noticeable from the table that the semi-amplitude of the 6.6 day signal is statistically constant no matter what dataset or analysis methods are used. However, the semi-amplitude for the transiting inner planet varies substantially ($>2\sigma$) depending on how many datasets are used, and the third signal at 13.6 days varies its amplitude widely. One would expect that a priori MAROON-X is the best instrument to capture precisely the semi-amplitude of GJ~806b, but the value of the fit to all datasets deviates significantly from the MAROON-X fit alone as a consequence of a possible overestimation of this amplitude in the CARMENES and HIRES data. Moreover, it is possible that the 13.6-day signal, which is detected in the CARMENES and HIRES data, can be of stellar origin, and related to (time-varying?) stellar activity. Having been unable to unambiguously determine the rotation period of the host star as shown in previous sections, this possibility can not be disregarded.

\begin{figure*}[ht!]
\includegraphics[width=1\linewidth]{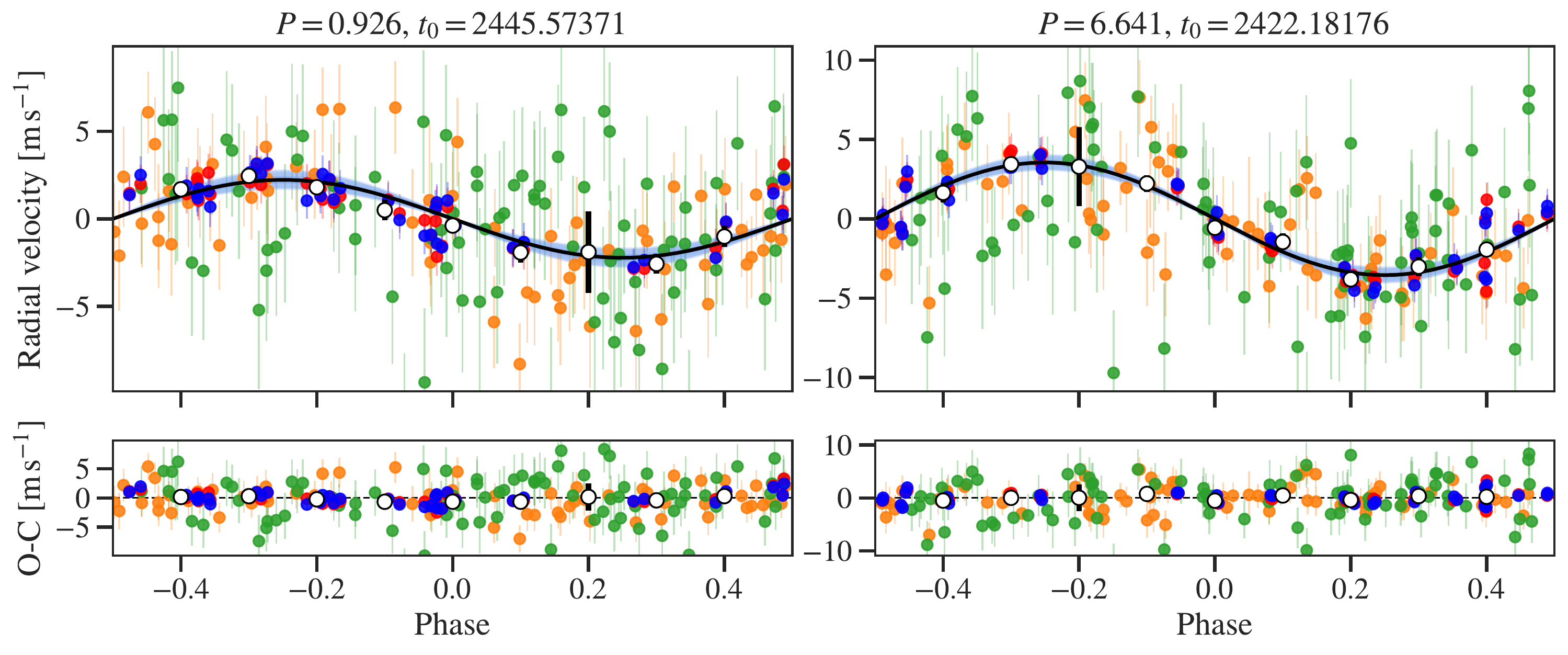}
\caption{RVs phase-folded to the period and central time of transit (shown above each panel, period units are days and central time of transit $t_0$ units are BJD\,$-$\,2\,457\,000) for b (\textit{left}), c (\textit{right}) planets along with the best-fit model (black line) and $3\sigma$ confidence intervals (shaded light blue areas). RVs from CARMENES (orange), HIRES (green) and MAROON-X red channel (red) and blue channel (blue) and binned RV (white dots with black error bars). The instrument error bars include the extra jitter term added in quadrature. 
\label{fig:phasefold}}
\end{figure*}

\begin{figure}[ht!]
\includegraphics[width=1.0\linewidth, angle=0]{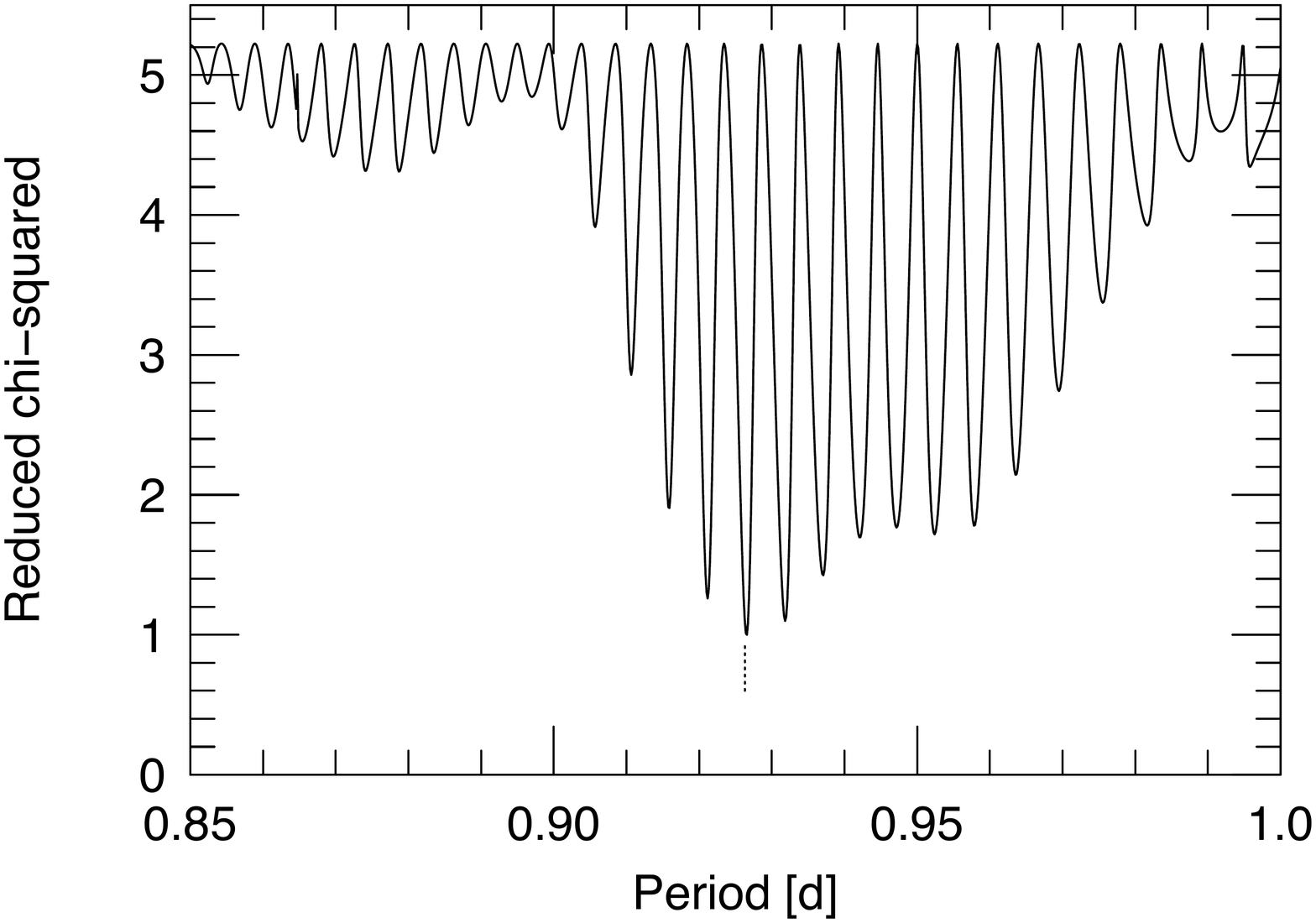}
\caption{Periodogram of the FCO data set showing the reduced $\chi^2$ as a function of input period. The vertical dashed line marks the period of TOI 4481b.}
\label{fig:fco_periodogram}
\end{figure}

To account for this possible flaw in our analysis we undertook a series of additional modelling. For a better comparation between datasets, models, and to avoid influences from the used priors, we used the same priors to fit the same signal in all the models showed in Table\,3. The priors used in each signal are shown in Table\,\ref{tab:Priors_TABLE3}. Furthermore, the priors for the GP kernels are also shown in Table\,\ref{tab:Priors_TABLE3}. Those GP prior distributions are the same when the GP is applied to the three different datasets: CARMENES, CARMENES and HIRES individually, and the joint CARMENES+HIRES dataset. In Table\,\ref{tab:Semiampl} we also report the planetary  semi-amplitudes reported by a series of models, all including the CARMENES+MAROON-X+HIRES data. The first two models contain a simple 2 and 3 keplerian signal fits following the same procedures as described in previous sections. In the rest of the models we remove the 13.6-day signal from the CARMENES and/or the HIRES data in different ways, namely: In the reference model, which we name this way as it will be our final adopted model, we removed the 13.6-day signal by fitting a Keplerian signal to the CARMENES+HIRES data and combined the residuals with the MAROON-X data to fit for the 0.9 and 6.6d signals. In the rest of the models, we fit the 13.6-day signal using Gaussian processes (GPs). The fit is performed using two different kernels (Quasi-periodic and Exponential Sinus Squared), and applying them to the CARMENES data alone, the CARMENES and HIRES data individually sharing the GP Prot hyperparameter, and to the joint CARMENES+HIRES dataset. The different combinations give rise to models 2 to 7. The $\Delta\ln{Z}$  values of each model is also given in Table\,\ref{tab:Semiampl}.

\subsubsection{Floating chunk offset analysis}
\label{sec:float}

As an independent method of deriving the $K$-amplitude of the transiting USP planet we applied the so-called floating chunk offset (FCO) method \citep{Hatzes2014}. This method is relatively insensitive to the presence of other long-period signals and it serves as a check on the $K$-amplitude found by our previous analyses.

Basically, FCO treats measurements taken on different nights as independent data sets with different zero-point offsets. These nightly data sets are then fit using a fixed period and phase  of the transiting planet, but allowing the
$K$-amplitude and nightly zero-point to vary until the $\chi^2$ is minimized. FCO acts as a high pass filter that removes the underlying, long-period signals. It also assumes that stellar activity is "frozen" over a single night. In order to apply FCO, two criteria must be met: 1) The periods of other signals must be much longer than that of the short period transiting planet, and 2) One must have several nights of data with at least 2 RV measurements with good time separation.

There were seven nights of CARMENES data with 3-4 measurements and  only two nights of MAROON-X measurements with good time separation. For the FCO method the number of measurements is relatively sparse, so it is wise to check
that the signal of the USP can be detected in the data. To do this we applied an FCO ``periodogram''. The data were fit using a range of trial input periods and allowing the $K$-amplitude and phase to vary. Figure~\ref{fig:fco_periodogram} shows the resulting $\chi^2$ fit as a function of trial input period. Although there are a large number of ``alias'' periods  present, the best fit is found at the period of the transiting planet.

Applying the FCO method results in an RV amplitude for the USP of $K_b$ = 3.05 $\pm$ 0.32 m\,s$^{-1}$. However, the presence of additional periods (6.6 d, 13.6 d) found in our previous analyses are short enough that they may introduce a systematic error in the FCO $K$-amplitude. To check this possibility, we generated a synthetic data set consisting of the orbit of the USP planet with $K_b$ = 3 m\,s$^{-1}$ and including the periodic signals (6.6 d and 13. 6 d) found by the other analyses. The 13.6-day signal was only added to the synthetic ``CARMENES'' data set. For all input signals the data were sampled using the same time stamps as the observations. No noise was added in  this simulation in order to assess a possible offset in the ``perfect'' case. The simulation  resulted in  $K_b$ = 3.5 m\,s$^{-1}$ implying a systematic offset of $+$0.5 m\,s$^{-1}$ in the FCO value. Applying this correction would result in a final FCO amplitude of $K_b$ = 2.55 $\pm$ 0.32  m\,s$^{-1}$. However, given that this offset is comparable to the error in the FCO K-amplitude all that can be said with certainty is that the FCO amplitude recovers the true amplitude within the errors and that the value is entirely consistent with the result derived in Section 4.2.5.

\subsubsection{Best model adoption}
\label{sec:adopt}

\begin{figure}[ht!]
\includegraphics[width=1\linewidth]{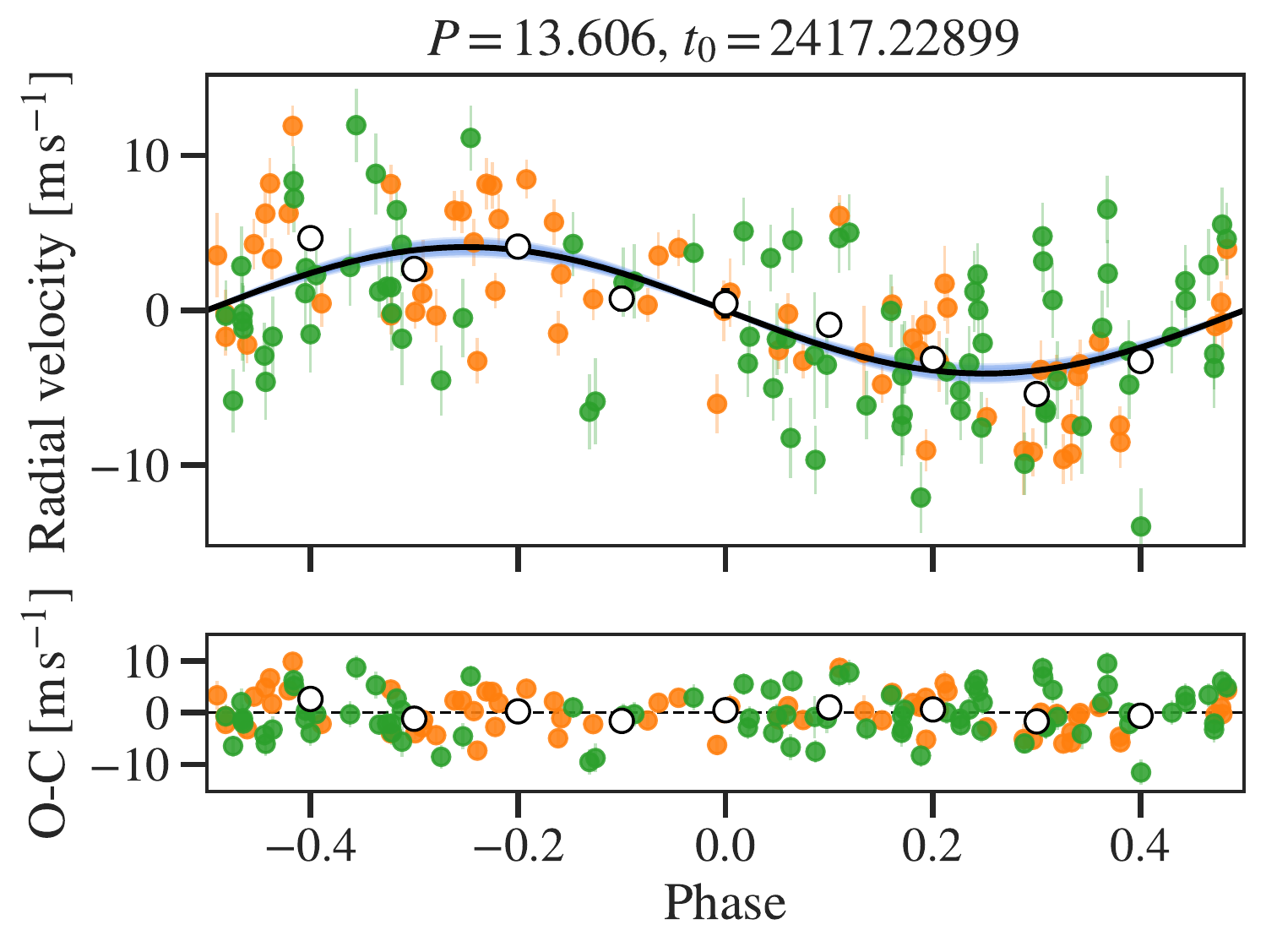}
\caption{RVs phase-folded to the period and central time of transit (shown above each panel, period units are days and central time of transit $t_0$ units are BJD\,$-$\,2\,457\,000) for the 13.6\,d signal along with the best-fit model (black line) and $3\sigma$ confidence intervals (shaded light blue areas). RVs from CARMENES (orange) and HIRES (green) and binned RV (white dots with black error bars). The instrument error bars include the extra jitter term added in quadrature. 
\label{fig:phasefold 13d}}
\end{figure}

As we discussed in previous sections, while the CARMENES data alone would suggest a three planet solution, we have reasonable doubts on the nature of the 13.6-day signal. First, MAROON-X data cover nearly two cycles of the 13.6-day signal and the data have enough precision to confidently detect it, but does not retrieve this signal if the transiting planet signal is fitted. A likely explanation is that the 1-day-alias of the 13.6-day signal at $P = 0.931$ d, may affect the MAROON-X planet signal detections as it is extremely close to the transiting planet period. But it is also possible that the signal originates from transient stellar activity in the CARMENES data. Future longer-term RV monitoring observations of the system may solve this issue. 

All this taken in consideration, here we adopted the so-called reference model (2pl ((CARM+HIRES)-13.6d) + MAROON-X) in Table~\ref{tab:Semiampl} as our final solution of the RV analysis based on its $\Delta\ln{Z}$ value and its simplicity over the use of GPs. This model's solution is also consistent with GJ~806b's semi-amplitude solution from MAROON-X data and has a semi-amplitude for the 13.6-day signal which is close to the one from CARMENES data alone.

\subsection{Joint fit}
\label{sec:jfit}

\begin{figure*}
    \centering
    \includegraphics[width=0.49\hsize]{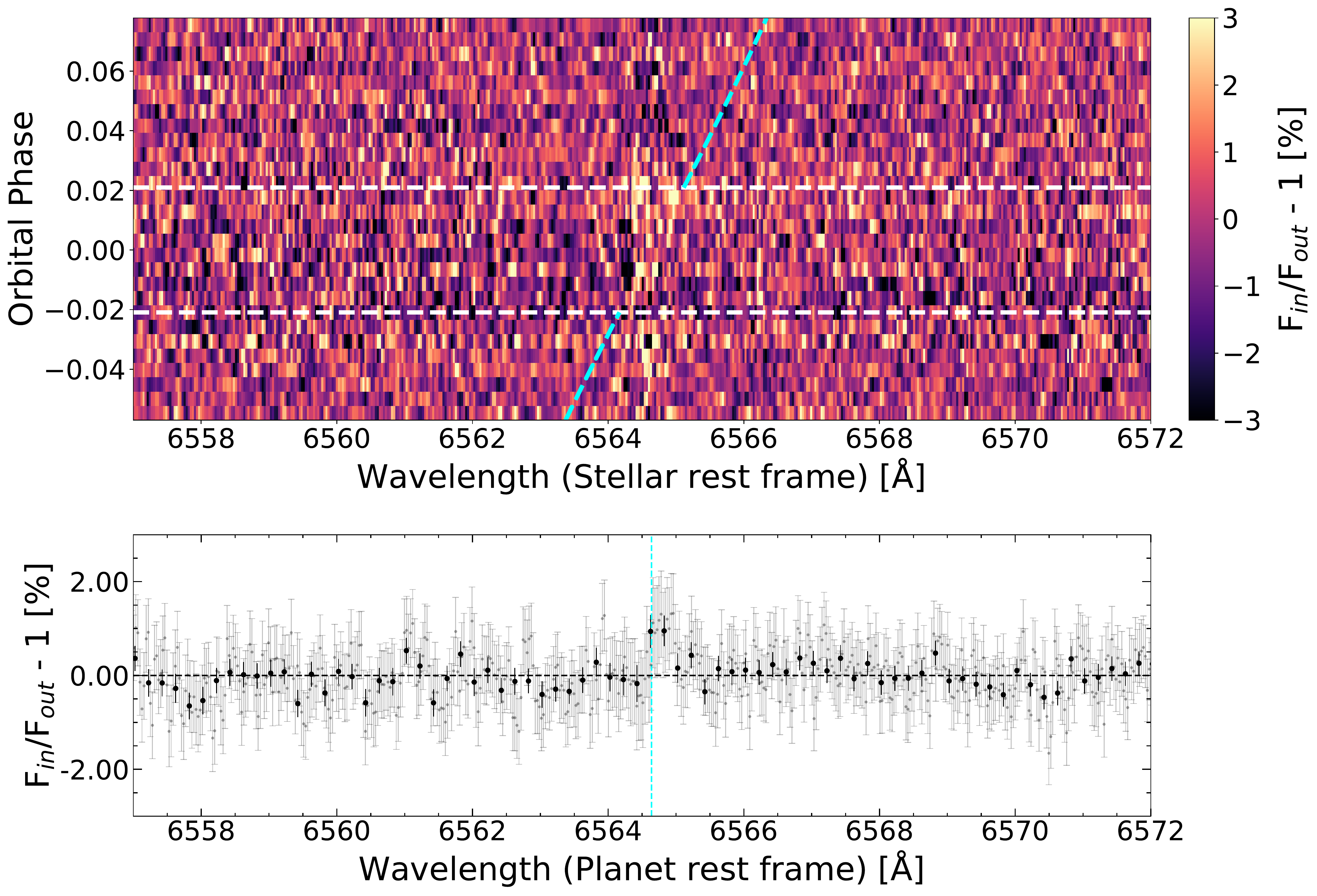}
    \includegraphics[width=0.49\hsize]{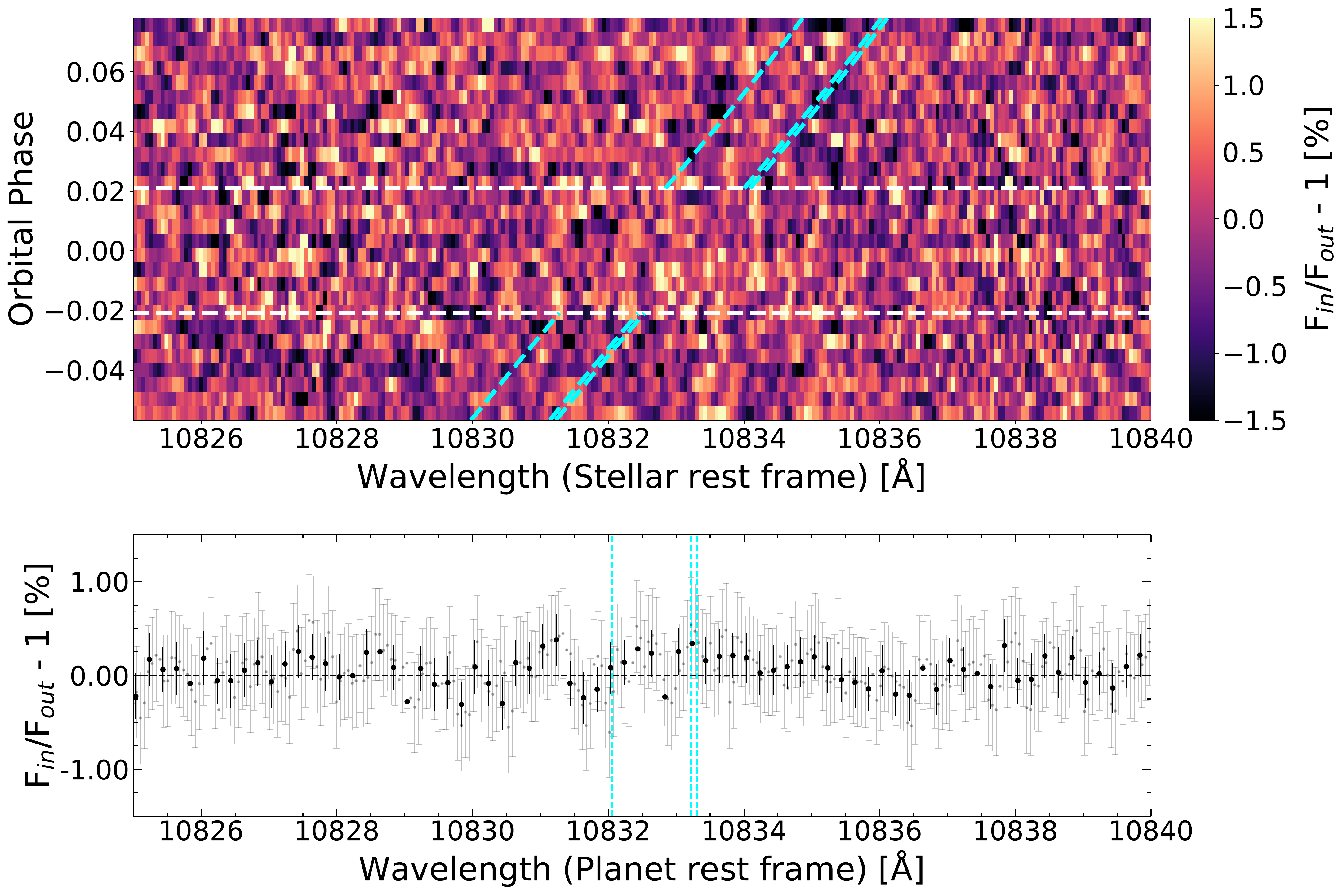}
    \caption{\label{Fig:TRANSMISSION_SPECTRA_Ha_He}
    Residuals maps and transmission spectra around the H$_{\alpha}$ line (left) and He I triplet lines (right). \textit{Top panels}: Residual maps in the stellar rest frame. Planet orbital phase is shown in the vertical axis, wavelength is in the horizontal axis, and relative absorption is colour-coded. Dashed white horizontal lines indicate the first and fourth contacts. Cyan lines show the theoretical trace of the planetary signals. \textit{Bottom panels}: Transmission spectra obtained combining all the spectra between the first and fourth contacts. All the wavelengths in this figure are referenced in the vacuum.
    }
\end{figure*}

We simultaneously modeled the \textit{TESS} and MuSCAT2 photometry and CARMENES, HIRES and MAROON-X RVs using \texttt{juliet} to obtain the most precise parameters of the GJ~806 planetary system.
For the joint fit we adopted the reference model from the RV analysis in Sec.\,\ref{sec:threervs} and only considered transits for the planet b. In the process, we took into account the error propagation from the 13.6\,d signal subtraction into the CARMENES and HIRES data.

To fit the photometric datasets with \texttt{juliet}, we adopted a linear limb darkening law for the MuSCAT2 photometry and a quadratic limb darkening law for the \textit{TESS} light curve.
The limb darkening coefficients were parameterized with a uniform sampling prior $(q_1,q_2)$, introduced by \citet{Kipping2013}. Additionally, rather than fitting directly the impact parameter of the orbit ($b$) 
and the planet-to-star radius ratio ($p$\,$=$\,$R_p/R_{\star}$), we considered the uninformative sample ($r_1$,$r_2$) parametrization introduced in \citet{Espinoza2018}.
The parameters $r_1$ and $r_2$ ensure a full exploration of the physically plausible values of $p$ and $b$, with uniform priors sampling.
We fixed all the photometric dilution factors to 1 and we added a relative flux offset and a jitter term to \textit{TESS} data and for each filter of MuSCAT2.

To save computational time, we narrowed our priors based on the results from the photometry and RV models, but we kept them wide enough to ensure a full exploration of the posterior distribution.
The priors used in the joint fit are listed in Table\,\ref{table - Joint Fit PRIORS}.
The median and 68.3\% credible intervals of the posterior distributions and the derived planetary parameters are reported in Table\,\ref{tab:JointFitPOST}.
Figure\,\ref{fig: JF CORNER PLOT} presents the corner plot of the posterior distributions.
Figure\,\ref{fig:phasefold} displays the phase-folded RVs models for the 2 planets and Figure\,\ref{fig: RV + MODEL JF} displays the RV time series together with the model. The rms of the residuals is 2.9\,m/s and the errorbars median error is 2.7\,m/s. We also explored an eccentric solution for planet c, however this solution is statistically indistinguishable from the circular model ($\Delta \log$Z\,$\sim$\,1), with an eccentricity value not well constrained ($ecc_{c}$\,=\,0.07$\pm$0.05), and with model parameters  consistent within uncertainties with the values reported in Table\,\ref{tab:JointFitPOST}.

The final analysis result in an inner ultra-short period planet with a radius of $1.331\pm0.023 R_{\oplus}$ and a mass of $1.90\pm0.17 M_{\oplus}$, and an outer planet with minimum mass of $5.80\pm 0.30 M_{\oplus}$.




\begin{table}[ht!]
\caption[width=\hsize]{
\label{tab:JointFitPOST}
Parameters and $1\sigma$ uncertainties for the \texttt{juliet} joint fit model for GJ\,806 planetary system. Priors and description for each parameter are presented in Table\,\ref{table - Joint Fit PRIORS}. The adopted stellar properties used to derive the planetary parameters are the ones from Table\,\ref{tab:stellar_parameters}.
}
\centering
\resizebox{\columnwidth}{!}{%
\begin{tabular}{lcc}
\hline \hline
\vspace{-0.25cm} \\
Parameter &  \,b  & \,c  \\
\hline
\vspace{-0.30cm} \\
\multicolumn{3}{c}{\textit{ Stellar parameters }} \vspace{0.15cm} \\
$\rho_{\star}$ [kg\,m$^{-3}$] & \multicolumn{2}{c}{  8600$^{+450}_{-460}$ } \vspace{0.20cm} \\
\multicolumn{3}{c}{\textit{ Planet parameters }} \vspace{0.15cm} \\
$P$ [d] & 0.9263237\,(9)  & 6.64064\,(25)    \vspace{0.05cm}\\
$t_{0}$\,$^{(a)}$ &  2445.57371\,(15)  &  2422.182$\pm$0.050  \vspace{0.05cm}\\
$K$ [$\mathrm{m\,s^{-1}}$] &  2.25$\pm$0.20  &  3.55$\pm$0.17    \vspace{0.05cm}\\
$r_{1}$ & 0.53$^{+0.04}_{-0.05}$  &  --     \vspace{0.05cm}\\
$r_{2}$ &  0.0295$\pm$0.0004  &  --    \vspace{0.20cm}\\
\multicolumn{3}{c}{\textit{ Photometry parameters }} \vspace{0.15cm} \\
$q_{1,\textit{TESS}}$ & \multicolumn{2}{c}{ 0.31$^{+ 0.20}_{-0.13 }$ } \vspace{0.05cm} \\
$q_{2,\textit{TESS}}$ & \multicolumn{2}{c}{ 0.30$^{+0.30}_{-0.20}$ } \vspace{0.05cm} \\
$M_{\textit{TESS}}$ (ppm) & \multicolumn{2}{c}{ $-4\pm$4 } \vspace{0.05cm} \\
$\sigma_{\textit{TESS}}$ (ppm) & \multicolumn{2}{c}{ 207$\pm$7 } \vspace{0.05cm} \\
$q_{1,\textit{MuSCAT2-g}}$ & \multicolumn{2}{c}{0.83$^{+0.11}_{-0.20}$ } \vspace{0.05cm} \\
$M_{\textit{MuSCAT2-g}}$ (ppm) & \multicolumn{2}{c}{ 16$\pm$40 } \vspace{0.05cm} \\
$\sigma_{\textit{MuSCAT2-g}}$ (ppm) & \multicolumn{2}{c}{ 4$^{+40}_{-3}$ } \vspace{0.05cm} \\
$q_{1,\textit{MuSCAT2-r}}$ & \multicolumn{2}{c}{0.80$^{+ 0.15}_{- 0.25}$ } \vspace{0.05cm} \\
$M_{\textit{MuSCAT2-r}}$ (ppm) & \multicolumn{2}{c}{ $-2\pm$60 } \vspace{0.05cm} \\
$\sigma_{\textit{MuSCAT2-r}}$ (ppm) & \multicolumn{2}{c}{ 5$^{+50}_{-4}$ } \vspace{0.05cm} \\
$q_{1,\textit{MuSCAT2-i}}$ & \multicolumn{2}{c}{ 0.88$^{+0.08}_{-0.13 }$ } \vspace{0.05cm} \\
$M_{\textit{MuSCAT2-i}}$ (ppm) & \multicolumn{2}{c}{ 70$\pm$30 } \vspace{0.05cm} \\
$\sigma_{\textit{MuSCAT2-i}}$ (ppm) & \multicolumn{2}{c}{ 7$^{+70}_{-6}$ } \vspace{0.05cm} \\
$q_{1,\textit{MuSCAT2-z}}$ & \multicolumn{2}{c}{ 0.38$\pm$0.22 } \vspace{0.05cm} \\
$M_{\textit{MuSCAT2-z}}$ (ppm) & \multicolumn{2}{c}{ 6$\pm$35 } \vspace{0.05cm} \\
$\sigma_{\textit{MuSCAT2-z}}$ (ppm) & \multicolumn{2}{c}{ 12$^{+170}_{- 11 }$ } \vspace{0.20cm} \\
\noalign{\smallskip} 
\multicolumn{3}{c}{\textit{ RV parameters }}\\
\noalign{\smallskip} 
$\gamma_{CARMENES}$ [$\mathrm{m\,s^{-1}}$] & \multicolumn{2}{c}{ 0.35$\pm$0.35 } \vspace{0.05cm} \\
$\sigma_{CARMENES}$ [$\mathrm{m\,s^{-1}}$] & \multicolumn{2}{c}{ 0.40$^{+0.55}_{-0.21}$ } \vspace{0.05cm} \\
$\gamma_{HIRES}$ [$\mathrm{m\,s^{-1}}$] & \multicolumn{2}{c}{ 0.45$\pm$0.45 } \vspace{0.05cm} \\
$\sigma_{HIRES}$ [$\mathrm{m\,s^{-1}}$] & \multicolumn{2}{c}{ 3.05$\pm$0.40 } \vspace{0.05cm} \\
$\gamma_{MAROON-X_{red}}$ [$\mathrm{m\,s^{-1}}$] & \multicolumn{2}{c}{ $-0.30\pm$0.20 } \vspace{0.05cm} \\
$\sigma_{MAROON-X_{red}}$ [$\mathrm{m\,s^{-1}}$] & \multicolumn{2}{c}{ 1.05$^{+0.15}_{-0.12 }$ } \vspace{0.05cm} \\
$\gamma_{MAROON-X_{blue}}$ [$\mathrm{m\,s^{-1}}$] & \multicolumn{2}{c}{ $-0.30\pm$0.18  } \vspace{0.05cm} \\
$\sigma_{MAROON-X_{blue}}$ [$\mathrm{m\,s^{-1}}$] & \multicolumn{2}{c}{ 0.98$^{+0.14}_{-0.12}$ } \vspace{0.15cm} \\
\hline
\noalign{\smallskip} 
\multicolumn{3}{c}{\textit{ Derived parameters }} \\ 
\noalign{\smallskip} 
$p = {R}_{\mathrm{p}}/{R}_{\star}$  &  0.0294$\pm$0.0004  &  --   \vspace{0.05cm}\\
$b = (a/{R}_{\star}) \cos{ i_{\mathrm{p}} }$ &  0.300$^{+0.060}_{-0.070}$  &  -- \vspace{0.05cm}\\
$a/{R}_{\star}$  & 7.30$\pm$0.13  & 27.13$^{+0.45}_{-0.50}$  \vspace{0.05cm} \\  
$i_{\mathrm{p}}$ (deg)  &  87.7$^{+0.6}_{-0.5}$  &  --  \vspace{0.05cm}\\
$t_T$ [h]  & 0.930$\pm$0.010 & -- \vspace{0.05cm} \\  
$R_{\mathrm{p}}$ [${R}_\oplus$] & 1.331$\pm$0.023  &  -- \vspace{0.05cm}\\
$M_{\mathrm{p}}$ [${M}_\oplus$]\,$^{(b)}$ &  1.90$\pm$0.17  & >5.80$\pm$0.30   \vspace{0.05cm}\\
$\rho_{\mathrm{p}}$ [$\mathrm{g\,cm^{-3}}$]  &  4.40$\pm$0.45  & --  \vspace{0.05cm}\\
$g_{\mathrm{p}}$ [$\mathrm{m\,s^{-2}}$]  &  10.4$\pm$1.0  &  --  \vspace{0.05cm}\\
$a_{\mathrm{p}}$ [AU]  & 0.01406$\pm$0.00030  & 0.0523$\pm$0.0011  \vspace{0.05cm}\\
$T_{\mathrm{eq}}$ [K]\,$^{(c)}$  & 940$\pm$10  &  490$\pm$5  \vspace{0.05cm}\\
${S}$ [$\mathrm{S}_\oplus$]  & 130$\pm$6  &  9.5$\pm$0.4   \vspace{0.05cm}\\
\hline
\end{tabular}
}
\end{table}




\begin{table}[ht!]
\caption[width=\hsize]{
\label{tab:13d_values}
Planetary parameters and $1\sigma$ uncertainties for the 13.6\,d signal. The adopted system properties used to derive the planetary parameters are the ones from Table\,\ref{tab:stellar_parameters} and \ref{tab:JointFitPOST}.
}
\centering
\begin{tabular}{lc}
\hline \hline
\noalign{\smallskip} 
Signal parameter &  Value  \\
\hline
\noalign{\smallskip} 

$P$ [d] & 13.60588\,(65)   \vspace{0.05cm}\\
$t_{0}$\,$^{(a)}$ &  2417.23$\pm$0.12  \vspace{0.05cm}\\
$K$ [$\mathrm{m\,s^{-1}}$] &  4.10$\pm$0.20   \vspace{0.05cm}\\
\hline
\noalign{\smallskip} 
$a/{R}_{\star}$  & 48.8$\pm$0.8  \vspace{0.05cm} \\  
$M_{\mathrm{p}}$ [${M}_\oplus$]\,$^{(b)}$ &   >8.50$\pm$0.45   \vspace{0.05cm}\\
$a_{\mathrm{p}}$ [AU]  & 0.0844$\pm$0.0017  \vspace{0.05cm}\\
$T_{\mathrm{eq}}$ [K]\,$^{(c)}$  & 385$\pm$4  \vspace{0.05cm}\\
${S}$ [$\mathrm{S}_\oplus$]  & 3.6$\pm$0.15  \vspace{0.05cm}\\
\hline
\end{tabular}
\tablefoot{ $^{(a)}$ Central time of transit ($t_0$) units are BJD\,$-$\,2457000. $^{(b)}$ The mass is a lower limit ($M_p \sin{i_p}$) since it is only detected in the RV data. $^{(c)}$ Equilibrium temperatures were calculated assuming zero Bond albedo. }
\end{table}

\subsubsection{The 13.6-day signal as a planet}

In our joint fit we determine the planetary nature of GJ~806b and GJ~806c, but we cannot validate the 13.6-day signal as a planet. However, if in the future this signal can be validated, the derived parameters from our reference model in Sec.\,\ref{sec:threervs} are given in Table\,\ref{tab:13d_values}. And the phase-folded CARMENES and HIRES RVs are shown in Figure\,\ref{fig:phasefold 13d}. The planet would likely be a temperate sub-Neptune, with a minimum mass of $8.50\pm0.45 M_{\oplus}$

\subsection{A search for an extended atmosphere}
\label{sec:atmos}

GJ~806b is so far the second ultra-short period  planet ($P<1d$) with lower than Earth's mean density discovered around an M dwarf. It joins TOI-1685 b in this special category \citep{bluhm21} although GJ~806 is about 2 mag brighter than TOI-1685.  As will be discussed in the next section, this implies that GJ~806b might posses some type of volatile envelope, either possible remnants of a primordial atmosphere
\citep{Howe2020} or a secondary atmospheres formed through outgassing \citep{Swain2021}. 
We use a conservative rotational period of the star of 48.1 days to estimate the expected high energy emission and subsequently the mass loss rate expected in the planet (a shorter rotation period would lead to a larger XUV flux estimate). The stellar relation between rotation and X-ray activity by \citet{wri11} is used to calculate an X-ray (5--100~\AA) luminosity of $L_{\rm X}=2.1\times 10^{27}$~erg\,s$^{-1}$. The relations in \citet{san11} yield a GJ~806 extreme ultraviolet (EUV, 100-920~\AA) luminosity of $L_{\rm EUV}=2.0\times 10^{28}$~erg\,s$^{-1}$, and a expected mass loss rate in GJ~806~b of $1.5\times 10^{11}$~g\,s$^{-1}$. This X-ray emission is consistent with an age of $\sim$4~Gyr.
Thus, with the aim of detecting a possible extended atmosphere, as described in Section\,\ref{sec:carmrvs}, we took transit observations of GJ~806b with CARMENES on 30 November 2021 to measure the H${\alpha}$ and He~{\sc i} planetary absorption.



The transmission spectroscopy data analysis was performed following the same methodology previously employed for CARMENES M dwarf planets in \citet{Palle2020} and \citet{Orell2022}. The resulting transmission spectrum centered in the  spectral regions of the H$\alpha$ and He I triplet is shown in Figure\,\ref{Fig:TRANSMISSION_SPECTRA_Ha_He}. The He I triplet transmission spectrum shows a flat spectrum, while the H$\alpha$ transmission spectrum has an emission-like features that is a result of stellar variability during the transit. Overall, we found no significant absorption in either of the two line tracers, and we could only place a 3$\sigma$ upper limit to the excess absorption of 1.5\,\% and 0.7\,\% for H$\alpha$ and He I, respectively.

\section{Discussion}
\label{sec:disc}

\subsection{GJ~806b planet properties}
\label{sec:plpr}

\begin{figure*}[ht!]
\includegraphics[width=0.45\linewidth]{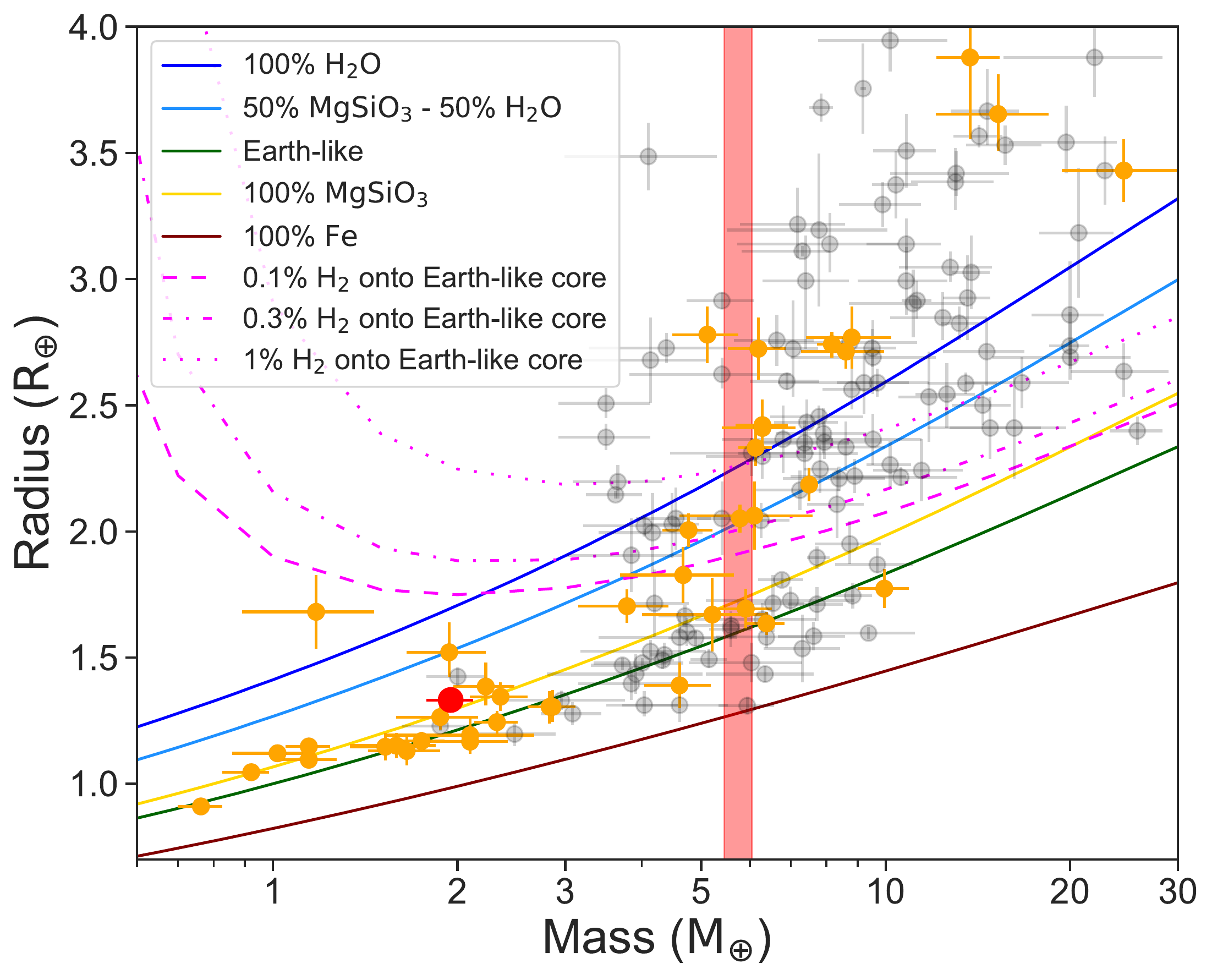}
\includegraphics[width=0.45\linewidth]{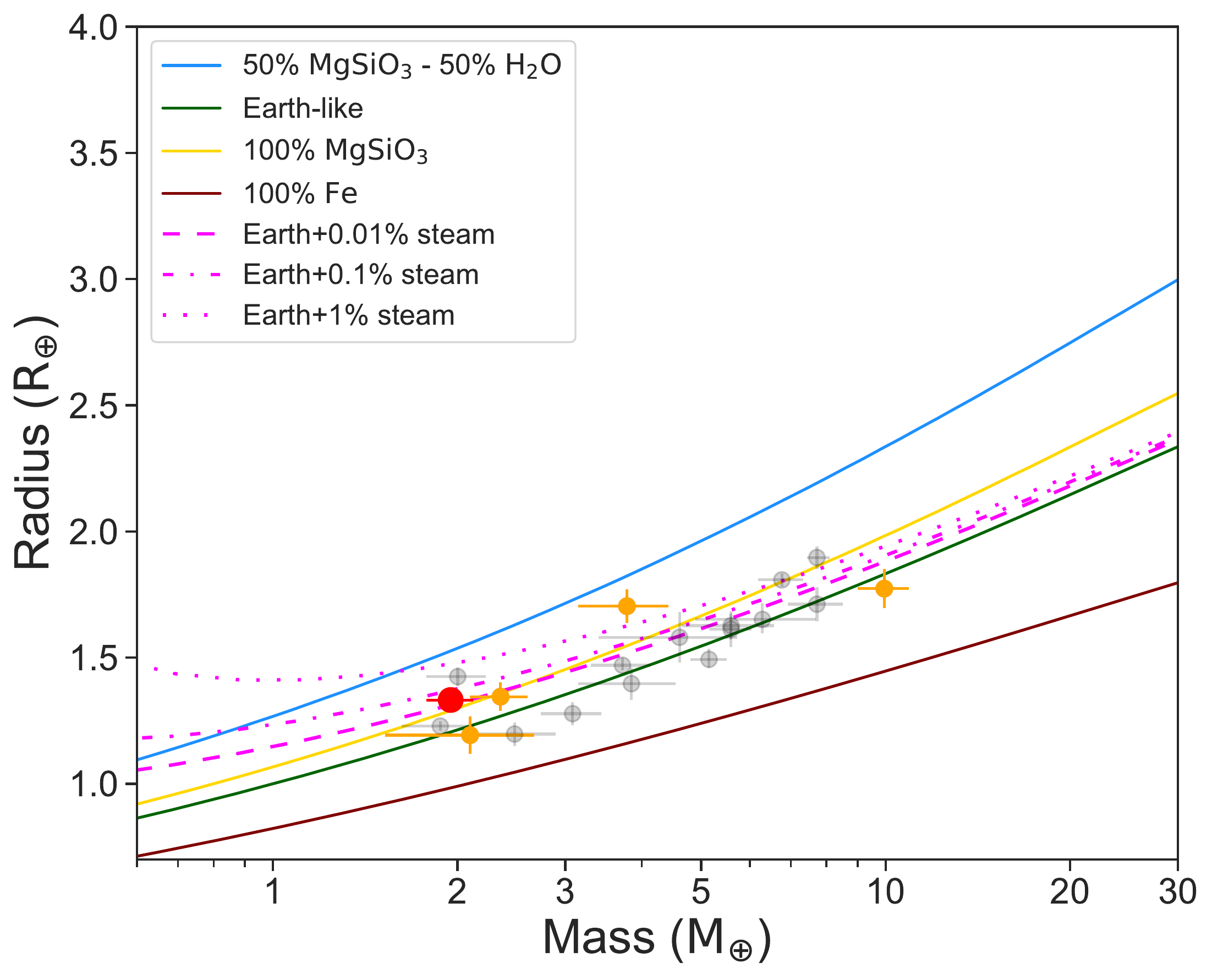}
\caption{Mass radius diagrams of all known planets (left) and ultra-short period planets only (right). Only planets with mass determination better than 30\% and radius determination better than 10\%, according to the TEPCat database \citep{Southworth2011} are shown. The orange dots represent the known planet orbiting around M dwarfs  ($T_{\rm eff} < 4000\,\mathrm{K}$) and the grey dots represent those orbiting around other stellar types. 
Overplotted are theoretical models for the planet's internal composition from \citet{Zeng2019} (left) and \citet{Turbet2020} (right). GJ~806b is marked with a red dot, while the radius band that GJ~806c occupies is shaded in red.
}
\label{fig:mrdiagram}
\end{figure*}

As established in previous sections, the GJ806 system is composed of an inner ultra-short period planet with a radius of $1.331\pm0.023 R_{\oplus}$ and a mass of $1.90\pm0.17 M_{\oplus}$, and at least one outer planet with a 6.6 day period and with minimum mass of $5.80\pm 0.30 M_{\oplus}$. No signs of transits for this outer planet were detected in the TESS light curves (not shown). It is the transiting USP inner planet, however, that makes this system especially interesting. 

Figure \ref{fig:mrdiagram} shows a mass radius diagram for all know planets with precise mass and radius determinations, where planets around M dwarf stellar types ($T_{\rm eff} < 4000\,\mathrm{K}$) are distinguished with a different color. With a mean density of $4.40 \pm 0.45$ g\,cm$^{-3}$ GJ806b lies in the pure MgSiO$_3$ model. Thus, GJ806b belongs to the growing population of rocky planets in the mass regime of 1-3 $M_{\oplus}$, the majority of which have been discovered orbiting around M dwarf stars. GJ806b is nearly the same size as two other benchmark targets discovered by CARMENES: GJ~357~b \citep{Luque2019} and GJ~486~b \citep{trifonov2021}, although with a much lower density and higher equilibrium temperature.

Focusing on the USP population, as previously mentioned GJ~806b is, so far, the second ultra-short period  planet ($P<1d$) with lower than Earth's mean density discovered around an M dwarf (third considering the full population of USPs), and the one with the lowest mass (second considering all USPs). Although for the lowest density M dwarf planet, TOI-1685 b, there are some conflicting reports on its final density value \citep{bluhm21,Hirano2021}.

The comparison with planetary models including a light envelope of H/He (\citet{Zeng2019}; Figure \ref{fig:mrdiagram}, left) indicate that such an envelope is very unlikely, well below the 0.1\% mass fraction. This is in agreement with our estimation of a very large mass loss rate and the non-detection of an extended atmosphere in Section  \ref{sec:atmos}. It seems that if GJ~806b ever had a primordial H/He atmosphere it was lost long time ago.

\begin{figure}
   \centering
   \includegraphics[width=\hsize]{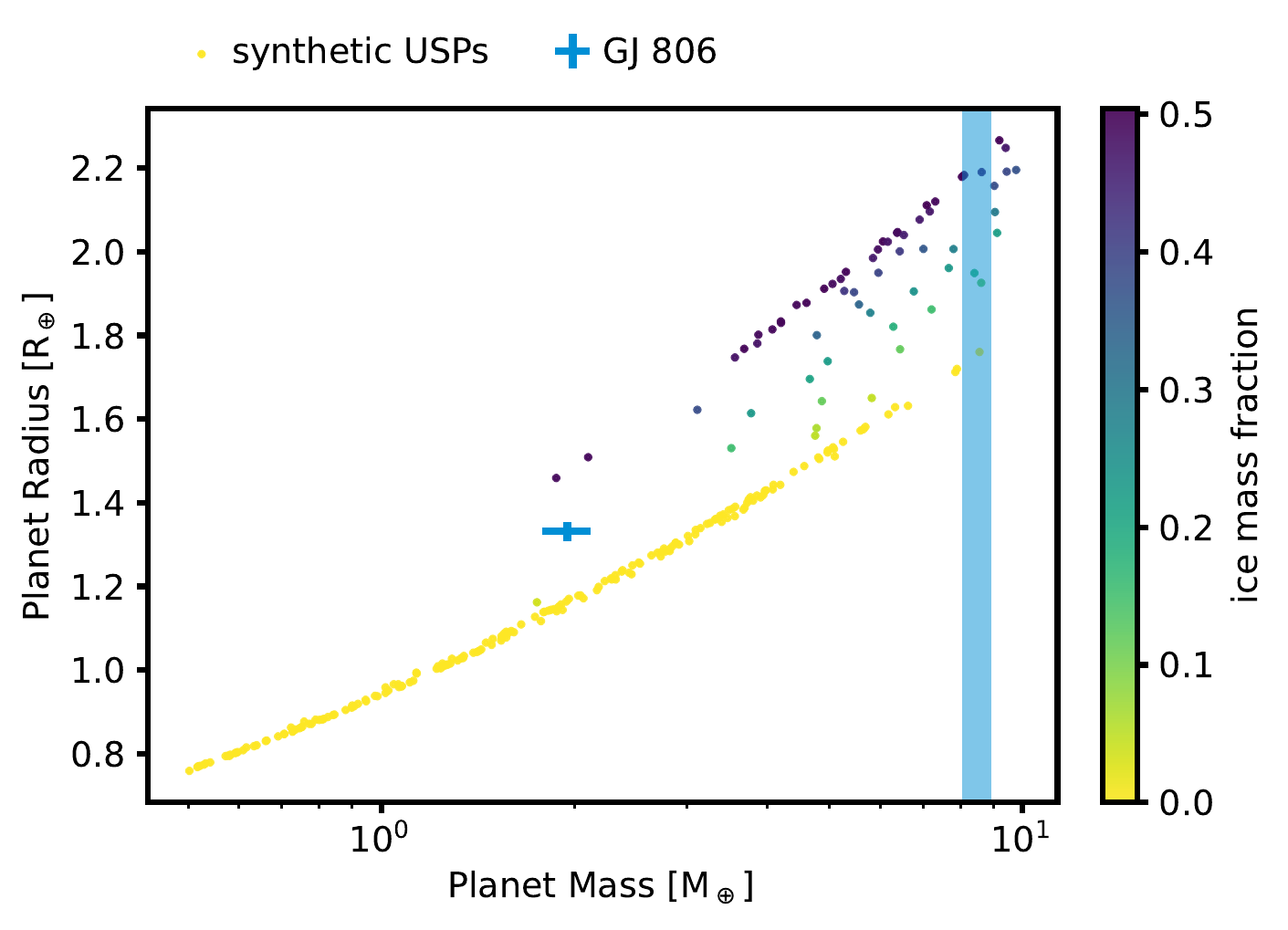}
   \caption{Mass-radius diagram of simulated ultra-short-period planets with periods $<\SI{2}{\day}$ (from \citet{Burn2021}) and GJ~806. Most planets simulated in a core accretion framework (markers color-coded by water ice fraction in their cores), are either purely rocky or contain about $50\,\%$ ice. None of them has an extended atmosphere. GJ~806~b (blue marker with error bars) is located in the unpopulated space between the two synthetic groups. GJ~806~c's radius is not constrained (blue stripe).}
   
    \label{fig:synthetic_M-R}
\end{figure}

Consistent with this hypothesis is a comparison to synthetic planets computed with the Generation III Bern model of planet formation~\citep{Emsenhuber2021a,Schlecker2021,Schlecker2021b,Mishra2021b}. The population of M~dwarf planets with \SI{0.5}{M_\odot} host stars presented in \citet{Burn2021} contains ultrashort-period planets ($P \lesssim \SI{2}{\day}$) that are typically of either Earth-like, purely rocky composition or contain $\sim 50\,\%$ water ice (see Fig.~\ref{fig:synthetic_M-R}). A fact supported by the observational population studies of small planets around M dwarfs \citep{Luque2022}. 
Due to atmospheric photoevaporation by their host star, these planets never retain their atmospheres.

GJ~806b's location in the mass-radius diagram strongly suggests that it is a planet devoid of any extended atmosphere. It further occupies a region that is completely unpopulated by synthetic USPs, which may imply that it is a bare core with an intermediate volatile content.
If it has accreted all its solids in the form of planetesimals, such an outcome is most likely to occur when the growing planetary core has accreted both inside and outside the water ice line~\citep{Burn2021}.
This suggests that the planet has migrated inwards significantly during the disk phase, although a giant impact event offers a valid alternative explanation~\citep{Emsenhuber2021b}.

Simulations of multi-planet systems with N-body interactions show that a GJ~806-like orbit configuration commonly originates from resonant migration during the disk phase: Once planets lock into a mean-motion resonance, the innermost planet is pushed inside the disk cavity by inward migration of the external planet \citep{Ataiee2021, Schlecker2022}. GJ~806~b adds to a small sample of USP planets around mid- and early M dwarfs, contributing to constraints on the scaling of migration traps with stellar host mass.

\cite{Turbet2020} discussed the possibility that planets with a substantial water envelope can develop a supercritical steam atmosphere, which would translate into slightly larger planetary radius, and a slightly lower bulk density than Earth-like planets. Figure \ref{fig:mrdiagram} (right) indicates that in such a case, GJ806b would contain a water mass fraction between 0.1 and 0.01\%. 


On the other hand, \cite{Dorn2019} discuss how differences in the observed bulk density of small USP planets may occur as a function of radial location and time of planet formation, leading to a class of super-Earths that would have no core and be rich in Ca and Al. This class of planets would have densities 10-20\% lower than Earth's and have very different interior dynamics, outgassing histories, and magnetic fields compared to the majority of super-Earths. This scenario is also compatible with the observed properties of GJ~806b. 

With an equilibrium temperature of 940 K, well above the $T= 880$\,K boundary in which rocks start to melt \citep{Mansfield2019}, GJ806b is probably a lava world, at least in parts of its surface. \citet{Dorn2021} demonstrated how the storage capacity of volatiles in magma oceans has significant implications for the bulk composition. They found that models with and without rock melting and water partitioning lead to deviations in planet radius of up to 16\% for fixed bulk compositions and planet mass. While GJ~806b is not a water world, accurate modeling of the mantle melting and volatile redistribution will be needed in order to accurately estimate the bulk water content.



\begin{figure*}
   \centering
   \includegraphics[width=\hsize]{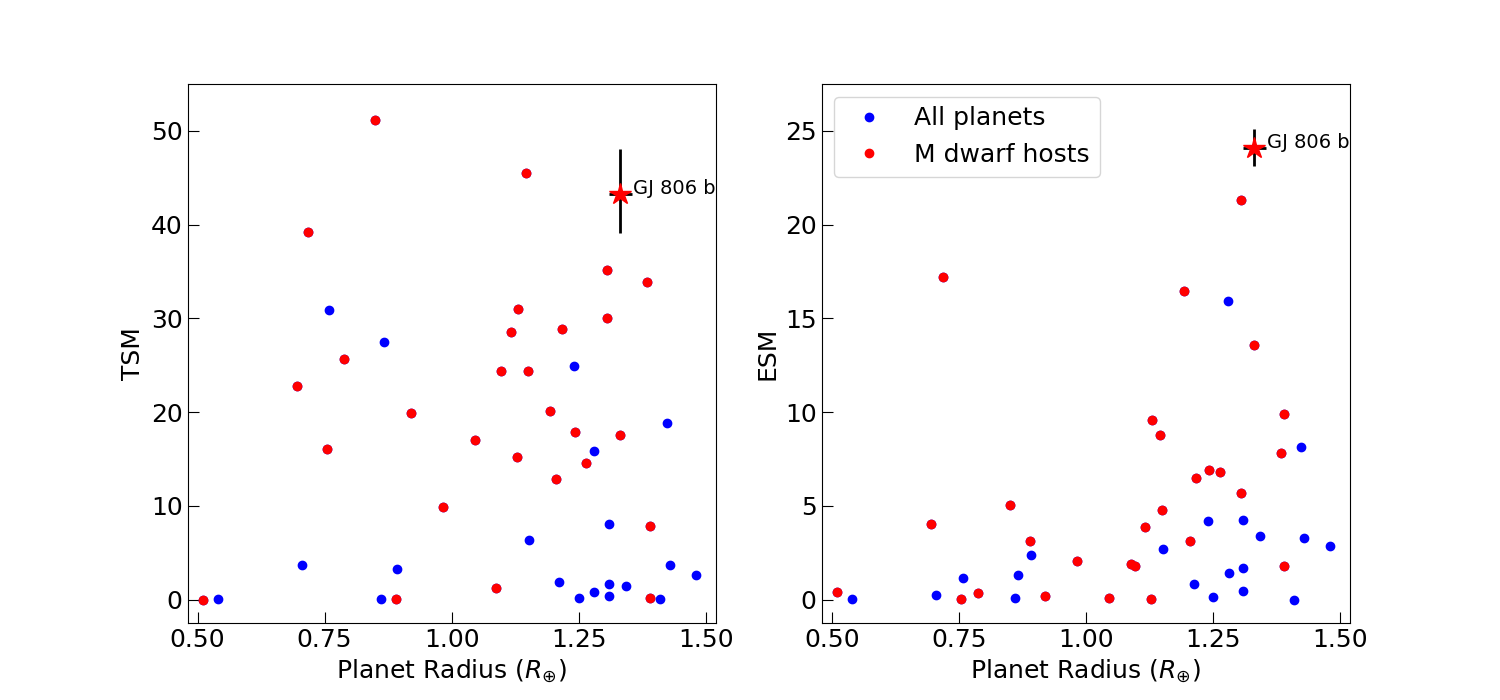}
   \caption{A plot of the TSM and ESM values versus planetary radius for terrestrial-sized planets ($R< 1.5 R_{\odot}$. Red dots correspond to planets orbiting around M dwarfs ($T_{eff}< 4000$\,K), while blue dots correspond to all other spectral type hosts. Planet parameters were taken from the NASA Exoplanet Archive as of 30 March 2022. GJ806b is marked with a red star.}
    \label{Fig:TSM_ESM}
\end{figure*}

\subsection{Atmospheric characterization prospects}
\label{sec:futatmos}

We computed the transmission spectroscopy metric (TSM) and emission spectroscopy metric (ESM), as defined by \cite{Kempton2018PASP..130k4401K}, to evaluate the prospects for atmospheric characterization of the innermost planet GJ~806 b.
Using the stellar and planetary parameters reported in Tables~\ref{tab:stellar_parameters} and \ref{tab:JointFitPOST}, we obtained $\mathrm{TSM} = 44.3_{-4.3}^{+5.1}$ and $\mathrm{ESM} = 24.1 \pm 1.0$. Both metric values are well above the respective thresholds of 10 and 7.5 suggested by \cite{Kempton2018PASP..130k4401K} for terrestrial planets, hence classifying GJ~806 b as high priority target for transit and eclipse spectroscopic observations. Compared to all confirmed planets with radius $R_{\mathrm{p}} < 1.5 \, R_{\oplus}$ and given mass measurements from the NASA Exoplanet Archive, GJ~806 b has the third highest TSM and the highest ESM values (see Figure~\ref{Fig:TSM_ESM}). 

Competing targets for transmission spectroscopy value are L98-59 b ($\mathrm{TSM}=51.2$, \citealp{kostov2019,demangeon2021}), LTT 1445 c ($\mathrm{TSM}=45.6$, \citealp{winters2022}), GJ~367 b ($\mathrm{TSM}=39.2$, \citealp{lam2021}), GJ~486 b ($\mathrm{TSM}=35.2$, \citealp{trifonov2021}), and L98-59 c ($\mathrm{TSM}=33.9$, \citealp{demangeon2021}).
For emission spectroscopy they are GJ~486 b ($\mathrm{ESM}=21.3$), GJ~367 b ($\mathrm{ESM}=17.2$), GJ~1252 b ($\mathrm{ESM}=16.5$, \citealp{shporer2020}), and TOI-431 b ($\mathrm{ESM}=15.9$, \citealp{osborn2021}).

\begin{figure*}
   \centering
   \includegraphics[width=\hsize]{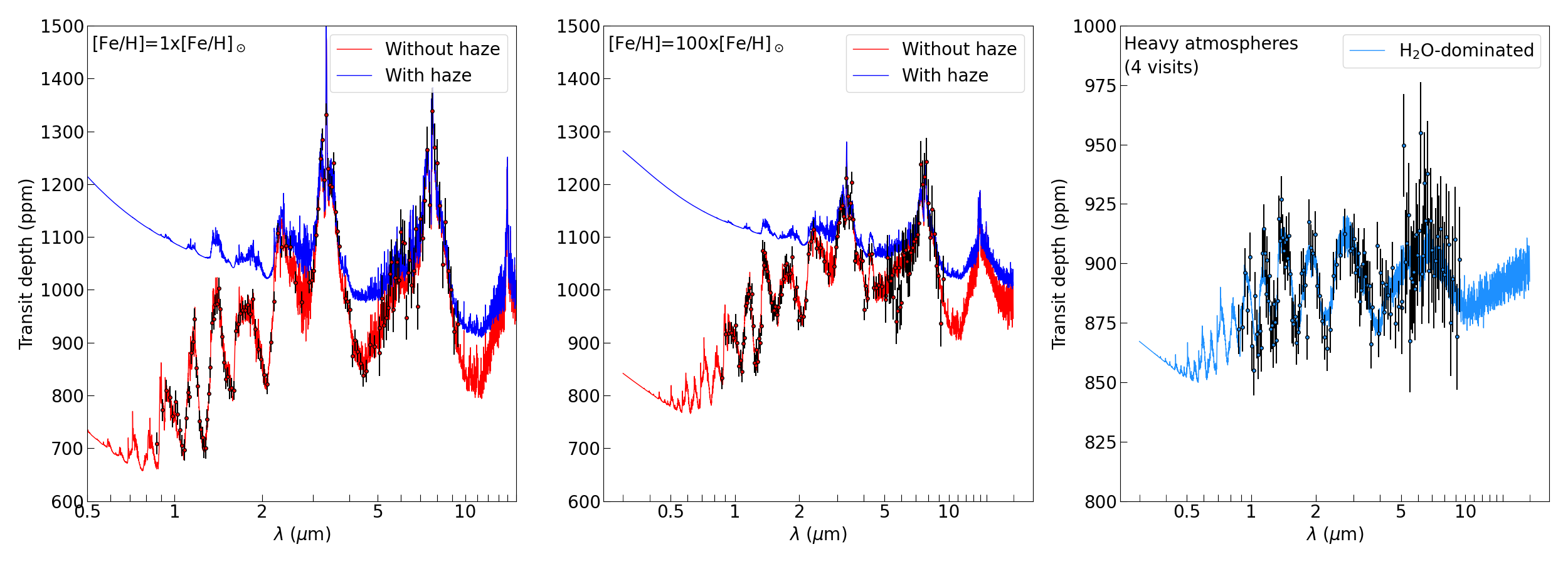}
   \caption{Synthetic \textit{JWST} transmission atmospheric spectra of GJ~806 b. Left panel: Fiducial models with solar abundances and clear atmosphere (solid red line) or haze (solid blue line), and simulated spectral measurements after one transit observation with JWST NIRISS-SOSS, NIRSpec-G395M, and MIRI-LRS configurations. Middle panel: Analogous models with enhanced metallicity by a factor of 100. Right panel: Fiducial model for a secondary atmosphere made of water, and simulated spectral measurements after combining 4 transits for each observing mode.}
    \label{Fig:JWSTspectra}
\end{figure*}

We made use of the online Exoplanet Characterization Toolkit (ExoCTK, \citealp{bourque2021_exoctk})\footnote{\url{https://exoctk.stsci.edu}} and of the \textit{JWST} Exposure Time Calculator (ETC)\footnote{\url{https://jwst.etc.stsci.edu}} to assess the observability of GJ~806 with various spectroscopic modes. The largest spectral coverage can be achieved by combining NIRISS-SOSS (0.8--2.8\,$\mu$m), NIRSpec-G395H (2.87--5.27\,$\mu$m) and MIRI-LRS (5--12\,$\mu$m) instrumental modes. However, the 1.1-1.3\,$\mu$m region is likely to saturate the NIRISS detector at the end of the first group, according to the latest version of the ETC (v2.0). We generated synthetic \textit{JWST} spectra for a range of atmospheric scenarios using the photo-chemical model ChemKM \citep{molaverdikhani19a, molaverdikhani19b, molaverdikhani20}, the radiative transfer code petitRADTRANS \citep{molliere19}, and \texttt{ExoTETHyS}\footnote{\url{https://github.com/ucl-exoplanets/ExoTETHyS}} \citep{morello21} to incorporate the instrumental response, including realistic noise and error bars (but assuming no saturation). We considered four models with H/He gaseous envelope, 1$\times$ or 100$\times$ solar abundances, without or with haze, and a fifth model with H$_2$O-dominated atmosphere. The models are displayed in Figure~\ref{Fig:JWSTspectra}.
The spectroscopic modulations are of several hundreds of parts per million (ppm) for the cases of H/He-dominated atmospheres, mostly attributable to H$_2$O and CH$_4$ absorption. The spectral features are dampened by a factor $\sim$2 in the cases with 100$\times$ solar metallicity. The presence of haze significantly dampens the spectral features at wavelengths shorter than 2 $\mu$m, but some features remain detectable with just one transit observation even in case of enhanced metallicity and haze. Similar trends with enhanced metallicity or haze were also observed in simulations made for other planets (e.g., \citealp{espinoza22}). Interestingly, the spectrum for the H$_2$O-dominated atmosphere presents absorption features of $\sim$60 ppm, that might be detected with high significance by combining a few visits. For reference, we report here our one-visit estimated error bars of 18-22 ppm for the \textit{JWST} NIRISS-SOSS and NIRSpec-G395H modes with median spectral resolution of R$\sim$50, and 42-45 ppm for the MIRI-LRS with wavelength bin sizes of 0.1--0.2\,$\mu$m.

Owing to the brightness of the host star GJ~806, the near-IR atmospheric features could also be explored with the \textit{Hubble Space Telescope} (\textit{HST}) Wide Field Camera 3 (WFC3). In particular, we report one-visit estimated error bars of 21-25 ppm for the \textit{HST} WFC3-G141 scanning mode (1.075--1.7\,$\mu$m) using 18 bins, and 27-30 ppm for the WFC3-G102 scanning mode (0.8--1.15\,$\mu$m) using 12 bins.



\subsection{Radio emission detection perspectives}
\label{sec:radio}



The ground-based detection of direct radio emission from Earth-sized exoplanets is not possible, as the associated frequency falls below the $\simeq$10 MHz Earth's ionosphere cutoff. 
However, in the case of star-planet interaction, the radio emission
arises  from the magnetosphere of the host star, induced by the exoplanet crossing the star magnetosphere, and the  relevant magnetic field is that of the star,  $B_\star$, not the exoplanet magnetic field.  Since M-dwarf stars have magnetic fields ranging from about 100 G and up to above 2--3\,kG, their auroral emission falls in the range from a few hundred MHz up to a few GHz. This interaction is expected to  yield detectable auroral radio emission via the cyclotron emission mechanism (e.g., \citealt{Turnpenney2018, Vedantham2020, PerezTorres2021}). 

\begin{figure}
\centering
\vspace{-10pt}
\includegraphics[width=0.49\textwidth]{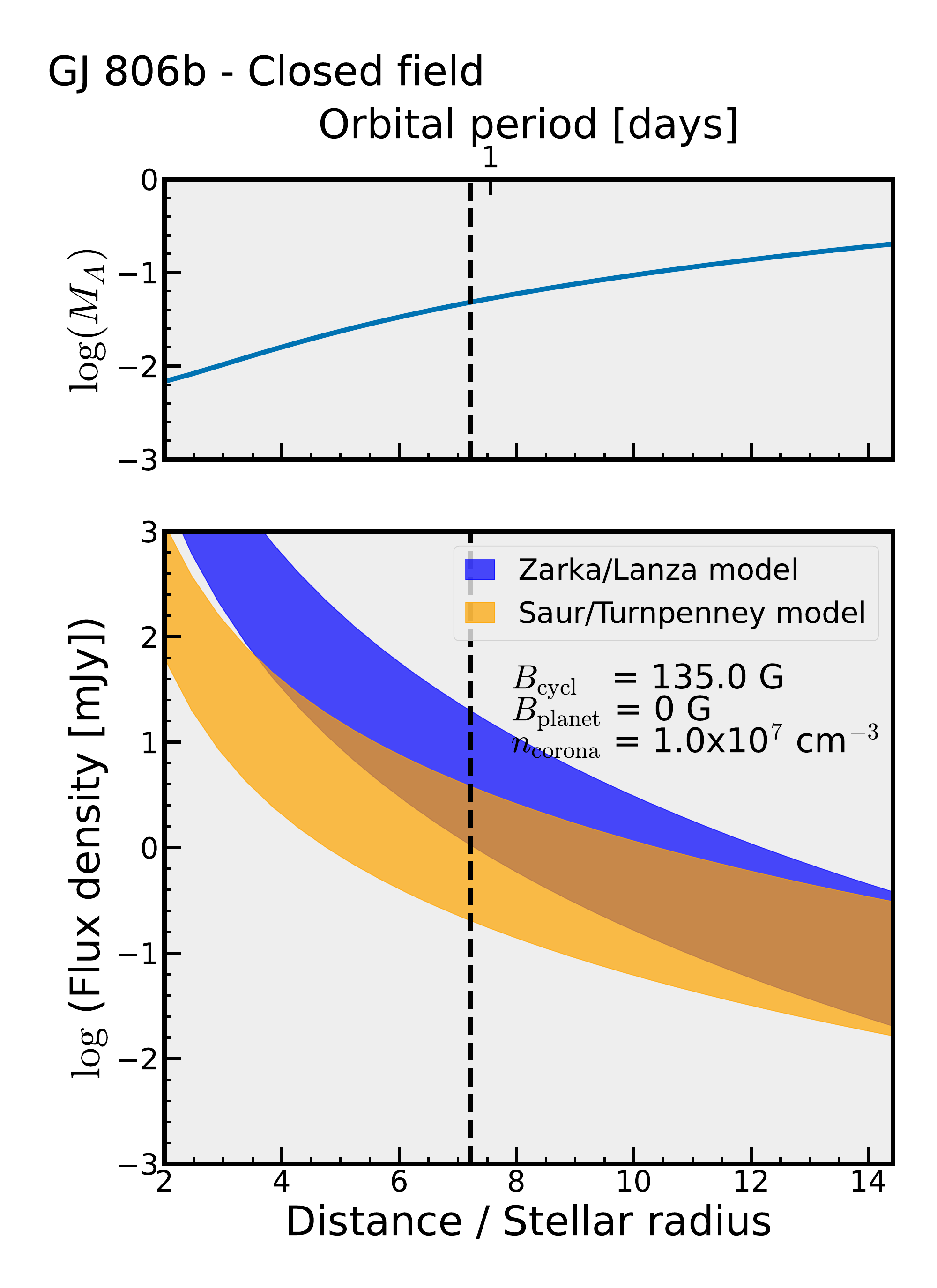}
\vspace{-0pt}
\caption{Expected flux density for auroral radio emission arising from star-planet interaction in the system GJ~806, as a function of orbital distance. The interaction is expected to be in the sub-Alfv\'enic regime (i.e. $M_A = v_{\rm rel}/v_{\rm Alfv} \leq 1$; top panel) at the location of the planet GJ~806b (vertical dashed line). }
\label{fig:wolf1069-spi-radio} 
\end{figure}

We followed the prescriptions in Appendix B of \citet{PerezTorres2021} to estimate the flux density expected to arise from the interaction between the planet GJ~806b and its host star, at a frequency of $\sim$380\,MHz, which corresponds to the cyclotron frequency of the star magnetic field of 135 G, taken from \citet{Reiners2022}. 
We show here the radio emission expected to arise from star-planet interaction for a closed dipolar magnetic field geometry, where the interaction between the planet and its host star happens in the sub-Alfv\'enic regime.
Figure~\ref{fig:wolf1069-spi-radio} shows the predicted flux density as a function of the planet orbital distance, assuming it is not magnetized.  The yellow shaded area encompasses the range of values from 0.005 up to 0.05 for the efficiency factor, $\epsilon$, in converting Poyinting flux into  electron cyclotron maser (ECM) radio emission. The flux density arising from star planet interaction is expected to be from $\sim160\, \mu$Jy up to $\sim$25 mJy for the assumed parameters. Thus, radio observations from this system look very promising to probe Sub-Alfv\'enic interaction and, eventually, independently detect radio emission from it. Given the peak frequency, observations at 400 MHz, or even at smaller frequencies, would be ideal to probe those scenarios.

\section{Conclusions}
\label{sec:Concl}

In this work, we presented the discovery of a multi planetary system around the bright and nearby M1.5 V star GJ~806, using ground-based photometric observations and radial velocity measurements from CARMENES, MAROON-X and HIRES spectrographs. The star host at least two planets: an ultra-short period (0.93 d) rocky super-Earth, GJ~806b, with a radius of $1.331\pm0.023 R_{\oplus}$, a mass of $1.90\pm 0.17 M_{\oplus}$, a mean density of $4.40 \pm0.45$\,g\,cm$^{-3}$, and an equilibrium temperature of $940 \pm 10$\,K, and a second non-transiting super-Earth, GJ~806c, has an orbital period of 6.6 d, a mass of $5.80\pm 0.30 M_{\oplus}$, and an equilibrium temperature of $490\pm 5$\,K. The radial velocity data, CARMENES data in particular, shows evidences for what would be a third super-Earth mass (M = $8.50\pm 0.45 M_{\oplus}$) planet with a period of 13.6 days, but we are unable to unambiguously discard that this signal is not induced by stellar activity, and are thus unable to confirm its planetary nature at this time.

We also estimated the extreme ultraviolet luminosity of GJ~806, and the inferred expected mass loss rate of GJ~806b. We report the results of a primary transit observation, taken with CARMENES, in search for a possible extended atmosphere focusing on the $H_{\alpha}$ and $He I$ absorptions lines. We found no significant absorption in either of the two
line tracers, but we could set a $3\sigma$ upper limit to the excess absorption of 1.5\,\% and 0.6\,\% for H${\alpha}$ and He~{\sc i}, respectively.

GJ~806b's relatively low bulk density makes it likely that the planet hosts some type of volatile atmosphere or relatively large mass water fraction, and makes it a very suitable target for atmospheric exploration with JWST and the upcoming ELTs. With a TSM of 43 and an ESM of 24, GJ~806b is the third-ranked terrestrial planet ($R < 1.5 R_{\oplus}$) around an M dwarf suitable for transmission spectroscopy studies using the JWST, and the most promising terrestrial planet for emission spectroscopy studies. We provide simulations of the characterization prospects with the JWST and the HST space telescopes. Additionally, GJ~806b is an excellent target for the detection of radio emission via star-planet interactions.

\begin{acknowledgements}

CARMENES is an instrument for the Centro Astron\'{o}mico Hispano-Alem\'{a}n de Calar Alto (CAHA, Almer\'{\i}a, Spain). CARMENES is funded by the German Max-Planck-Gesellschaft (MPG), the Spanish Consejo Superior de Investigaciones Cient\'{\i}ficas (CSIC), the European Union through FEDER/ERF FICTS-2011-02 funds, and the members of the CARMENES Consortium (Max-Planck-Institut f\"{u}r Astronomie, Instituto de Astrof\'{\i}sica de Andaluc\'{\i}a, Landessternwarte K\"{o}nigstuhl, Institut de Ci\`{e}ncies de l'Espai, Institut f\"{u}r Astrophysik G\"{o}ttingen, Universidad Complutense de Madrid, Th\"{u}ringer Landessternwarte Tautenburg, Instituto de Astrof\'{\i}sica de Canarias, Hamburger Sternwarte, Centro de Astrobiolog\'{\i}a and Centro Astron\'{o}mico Hispano-Alem\'{a}n), with additional contributions by the Spanish Ministry of Economy, the German Science Foundation through the Major Research Instrumentation Programme and DFG Research Unit FOR2544 ``Blue Planets around Red Stars'', the Klaus Tschira Stiftung, the states of Baden-W\"{u}rttemberg and Niedersachsen, and by the Junta de Andaluc\'{\i}a.

This paper includes data collected by the \tess{} mission. Funding for the \tess{} mission is provided by the NASA Explorer Program. We acknowledge the use of public TOI Release data from pipelines at the \tess{} Science Office and at the \tess{} Science Processing Operations Center. Resources supporting this work were provided by the NASA High-End Computing (HEC) Program through the NASA Advanced Supercomputing (NAS) Division at Ames Research Center for the production of the SPOC data products. This research has made use of the Exoplanet Follow-up Observation Program website, which is operated by the California Institute of Technology, under contract with the National Aeronautics and Space Administration under the Exoplanet Exploration Program.


The development of the MAROON-X spectrograph was funded by the David and Lucile Packard Foundation, the Heising-Simons Foundation, the Gemini Observatory, and the University of Chicago. The MAROON-X team acknowledges support for this work from the NSF (award number 2108465) and NASA (through the \textit{TESS} Cycle 4 GI program, grant number 80NSSC22K0117). This work was enabled by observations made from the Gemini North telescope, located within the Maunakea Science Reserve and adjacent to the summit of Maunakea. We are grateful for the privilege of observing the Universe from a place that is unique in both its astronomical quality and its cultural significance.

This article is partly based on observations made with the MuSCAT2 instrument, developed by ABC, at Telescopio Carlos S\'{a}nchez operated on the island of Tenerife by the IAC in the Spanish Observatorio del Teide.

This work makes use of observations from the LCOGT network. Part of the LCOGT telescope time was granted by NOIRLab through the Mid-Scale Innovations Program (MSIP). MSIP is funded by NSF.

PPP, BC, DV and MRM would like to acknowledge the following iSHELL observers: Claire Geneser, Ahmad Sohani, John Berberian, Patrick Nercessian, Jennah Fayaz, Kevin I Collins and Ian Helm. 

Based on observations collected at the Observatorio de Sierra Nevada, operated by the Instituto de AstrofÌsica de AndalucÌa (IAA-CSIC).
This work makes use of observations from the Las Cumbres Observatory global telescope network.
The Joan Oró Telescope (TJO) of the Montsec Observatory (OAdM) is owned by the Generalitat de Catalunya and operated by the Institute for Space Studies of Catalonia (IEEC).


This material is based upon work supported by the National Science Foundation Graduate Research Fellowship under Grant No. DGE 1746045.

We acknowledge financial support from the Agencia Estatal de Investigaci\'on of the Ministerio de Ciencia e Innovaci\'on and the ERDF ``A way of making Europe'' through projects 
  PID2019-109522GB-C5[1:4]	
  PGC2018-098153-B-C3[1,3]	
  ID2019-109522GB-C5[1:4] 
and the Centre of Excellence ``Severo Ochoa'' and ``Mar\'ia de Maeztu'' awards to the Instituto de Astrof\'isica de Canarias (CEX2019-000920-S), Instituto de Astrof\'isica de Andaluc\'ia (SEV-2017-0709), and Centro de Astrobiolog\'ia (MDM-2017-0737);
the Generalitat de Catalunya/CERCA programme;
NASA (Exoplanet Research Program Award \#80NSSC20K0251, TESS Cycle~3 Guest Investigator Program Award \#80NSSC21K0349, JPL Research and Technology Development, and Keck Observatory Data Analysis); 
the USA National Science Foundation (Astronomy and Astrophysics Grants \#1716202 and 2006517); 
the Mt. Cuba Astronomical Foundation; 
and MANY MORE, in anonymous style...

R.L. acknowledges funding from University of La Laguna through the Margarita Salas Fellowship from the Spanish Ministry of Universities ref. UNI/551/2021-May 26, and under the EU Next Generation funds.

R.L. acknowledges financial support from the Spanish Ministerio de Ciencia e Innovación, through project PID2019-109522GB-C52, and the Centre of Excellence "Severo Ochoa" award to the Instituto de Astrofísica de Andalucía (SEV-2017-0709).

This research was supported by the Excellence Cluster ORIGINS which is funded by the Deutsche Forschungsgemeinschaft (DFG, German Research Foundation) under Germany's Excellence Strategy - EXC-2094 - 390783311.

J.L-B. acknowledges financial support received from "la Caixa" Foundation (ID 100010434) and from the European Unions Horizon 2020 research and innovation programme under the Marie Slodowska-Curie grant agreement No 847648, with fellowship code LCF/BQ/PI20/11760023. This research has also been partly funded by the Spanish State Research Agency (AEI) Projects No.PID2019-107061GB-C61 and No. MDM-2017-0737 Unidad de Excelencia "Mar\'ia de Maeztu"- Centro de Astrobiolog\'ia (INTA-CSIC).

This work is partly supported by JSPS KAKENHI Grant Numbers
JP17H04574, JP18H05439, Grant-in-Aid for JSPS Fellows, Grant Number
JP20J21872, JST CREST Grant Number JPMJCR1761, and the Astrobiology
Center of National Institutes of Natural Sciences (NINS) (Grant Number
AB031010). This article is based on observations made with the MuSCAT2
instrument, developed by ABC, at Telescopio Carlos Sánchez operated on
the island of Tenerife by the IAC in the Spanish Observatorio del
Teide.

MPT acknowledges financial support from the State Agency for Research of the Spanish MCIU through the
"Center of Excellence Severo Ochoa" award to the Instituto de Astrofísica de Andalucía (SEV-2017-0709)
and through the grant PID2020-117404GB-C21 (MCI/AEI/FEDER, UE).

Funding for the TESS mission is provided by NASA's Science Mission directorate.

This paper includes data collected with the TESS mission, obtained from the MAST data archive at the Space Telescope Science Institute (STScI). Funding for the TESS mission is provided by the NASA Explorer Program. STScI is operated by the Association of Universities for Research in Astronomy, Inc., under NASA contract NAS 5–26555.

SVJ acknowledges the support of the DFG priority programme SPP 1992 Exploring the Diversity of Extrasolar Planets (JE 701/5-1)

The results reported herein benefited from collaborations and/or information exchange within the program “Alien Earths” (supported by the National Aeronautics and Space Administration under agreement No. 80NSSC21K0593) for NASA’s Nexus for Exoplanet System Science (NExSS) research coordination network sponsored by NASA’s Science Mission Directorate.

\end{acknowledgements}

\bibliographystyle{aa.bst} 
\bibliography{biblio.bib}

\begin{thebibliography}{143}
\expandafter\ifx\csname natexlab\endcsname\relax\def\natexlab#1{#1}\fi

\bibitem[{{Aller} {et~al.}(2020){Aller}, {Lillo-Box}, {Jones}, {Miranda}, \&
  {Barcel{\'o} Forteza}}]{2020A&A...635A.128A}
{Aller}, A., {Lillo-Box}, J., {Jones}, D., {Miranda}, L.~F., \& {Barcel{\'o}
  Forteza}, S. 2020, \aap, 635, A128

\bibitem[{{Alonso-Floriano} {et~al.}(2015){Alonso-Floriano}, {Morales},
  {Caballero}, {Montes}, {Klutsch}, {Mundt}, {Cort{\'e}s-Contreras}, {Ribas},
  {Reiners}, {Amado}, {Quirrenbach}, \& {Jeffers}}]{AlonsoFloriano2015}
{Alonso-Floriano}, F.~J., {Morales}, J.~C., {Caballero}, J.~A., {et~al.} 2015,
  \aap, 577, A128

\bibitem[{{Ambikasaran} {et~al.}(2014){Ambikasaran}, {Foreman-Mackey},
  {Greengard}, {Hogg}, \& {O'Neil}}]{george}
{Ambikasaran}, S., {Foreman-Mackey}, D., {Greengard}, L., {Hogg}, D.~W., \&
  {O'Neil}, M. 2014

\bibitem[{{Argelander}(1903)}]{Argelander1903}
{Argelander}, F. W.~A. 1903, Eds Marcus and Weber's Verlag, 0

\bibitem[{{Ataiee} \& {Kley}(2021)}]{Ataiee2021}
{Ataiee}, S. \& {Kley}, W. 2021, \aap, 648, A69

\bibitem[{{Bluhm} {et~al.}(2021){Bluhm}, {Pall{\'e}}, {Molaverdikhani},
  {Kemmer}, {Hatzes}, {Kossakowski}, {Stock}, {Caballero}, {Lillo-Box},
  {B{\'e}jar}, {Soto}, {Amado}, {Brown}, {Cadieux}, {Cloutier}, {Collins},
  {Collins}, {Cort{\'e}s-Contreras}, {Doyon}, {Dreizler}, {Espinoza}, {Fukui},
  {Gonz{\'a}lez-{\'A}lvarez}, {Henning}, {Horne}, {Jeffers}, {Jenkins},
  {Jensen}, {Kaminski}, {Kielkopf}, {Kusakabe}, {K{\"u}rster},
  {Lafreni{\`e}re}, {Luque}, {Murgas}, {Montes}, {Morales}, {Narita},
  {Passegger}, {Quirrenbach}, {Sch{\"o}fer}, {Reffert}, {Reiners}, {Ribas},
  {Ricker}, {Seager}, {Schweitzer}, {Schwarz}, {Tamura}, {Trifonov},
  {Vanderspek}, {Winn}, {Zechmeister}, \& {Zapatero Osorio}}]{bluhm21}
{Bluhm}, P., {Pall{\'e}}, E., {Molaverdikhani}, K., {et~al.} 2021, \aap, 650,
  A78

\bibitem[{{Bourque} {et~al.}(2021){Bourque}, {Espinoza}, {Filippazzo}, {Fix},
  {King}, {Martlin}, {Medina}, {Batalha}, {Fox}, {Fowler}, {Fraine}, {Hill},
  {Lewis}, {Stevenson}, {Valenti}, \& {Wakeford}}]{bourque2021_exoctk}
{Bourque}, M., {Espinoza}, N., {Filippazzo}, J., {et~al.} 2021, {The Exoplanet
  Characterization Toolkit (ExoCTK)}

\bibitem[{{Brown} {et~al.}(2013){Brown}, {Baliber}, {Bianco}, {Bowman},
  {Burleson}, {Conway}, {Crellin}, {Depagne}, {De Vera}, {Dilday}, {Dragomir},
  {Dubberley}, {Eastman}, {Elphick}, {Falarski}, {Foale}, {Ford}, {Fulton},
  {Garza}, {Gomez}, {Graham}, {Greene}, {Haldeman}, {Hawkins}, {Haworth},
  {Haynes}, {Hidas}, {Hjelstrom}, {Howell}, {Hygelund}, {Lister}, {Lobdill},
  {Martinez}, {Mullins}, {Norbury}, {Parrent}, {Paulson}, {Petry}, {Pickles},
  {Posner}, {Rosing}, {Ross}, {Sand}, {Saunders}, {Shobbrook}, {Shporer},
  {Street}, {Thomas}, {Tsapras}, {Tufts}, {Valenti}, {Vander Horst}, {Walker},
  {White}, \& {Willis}}]{brown13}
{Brown}, T.~M., {Baliber}, N., {Bianco}, F.~B., {et~al.} 2013, \pasp, 125, 1031

\bibitem[{{Buchner} {et~al.}(2014){Buchner}, {Georgakakis}, {Nandra}, {Hsu},
  {Rangel}, {Brightman}, {Merloni}, {Salvato}, {Donley}, \&
  {Kocevski}}]{PyMultiNest}
{Buchner}, J., {Georgakakis}, A., {Nandra}, K., {et~al.} 2014, \aap, 564, A125

\bibitem[{Burn {et~al.}(2021)Burn, Schlecker, Mordasini, Emsenhuber, Alibert,
  Henning, Klahr, \& Benz}]{Burn2021}
Burn, R., Schlecker, M., Mordasini, C., {et~al.} 2021, A\&A, 656, A72

\bibitem[{{Butler} {et~al.}(2017){Butler}, {Vogt}, {Laughlin}, {Burt},
  {Rivera}, {Tuomi}, {Teske}, {Arriagada}, {Diaz}, {Holden}, \&
  {Keiser}}]{Butler2017}
{Butler}, R.~P., {Vogt}, S.~S., {Laughlin}, G., {et~al.} 2017, \aj, 153, 208

\bibitem[{{Caballero} {et~al.}(2016){Caballero}, {Gu{\`a}rdia}, {L{\'o}pez del
  Fresno}, {Zechmeister}, {de Juan}, {Alonso-Floriano}, {Amado}, {Colom{\'e}},
  {Cort{\'e}s-Contreras}, {Garc{\'\i}a-Piquer}, {Gesa}, {de Guindos}, {Hagen},
  {Helmling}, {Hern{\'a}ndez Casta{\~n}o}, {K{\"u}rster}, {L{\'o}pez-Santiago},
  {Montes}, {Morales Mu{\~n}oz}, {Pavlov}, {Quirrenbach}, {Reiners}, {Ribas},
  {Seifert}, \& {Solano}}]{Caballero2016}
{Caballero}, J.~A., {Gu{\`a}rdia}, J., {L{\'o}pez del Fresno}, M., {et~al.}
  2016, in Society of Photo-Optical Instrumentation Engineers (SPIE) Conference
  Series, Vol. 9910, Observatory Operations: Strategies, Processes, and Systems
  VI, ed. A.~B. {Peck}, R.~L. {Seaman}, \& C.~R. {Benn}, 99100E

\bibitem[{{Caldwell} {et~al.}(2020){Caldwell}, {Tenenbaum}, {Twicken},
  {Jenkins}, {Ting}, {Smith}, {Hedges}, {Fausnaugh}, {Rose}, \&
  {Burke}}]{Caldwell2020}
{Caldwell}, D.~A., {Tenenbaum}, P., {Twicken}, J.~D., {et~al.} 2020, Research
  Notes of the American Astronomical Society, 4, 201

\bibitem[{{Cifuentes} {et~al.}(2020){Cifuentes}, {Caballero},
  {Cort{\'e}s-Contreras}, {Montes}, {Abell{\'a}n}, {Dorda}, {Holgado},
  {Zapatero Osorio}, {Morales}, {Amado}, {Passegger}, {Quirrenbach}, {Reiners},
  {Ribas}, {Sanz-Forcada}, {Schweitzer}, {Seifert}, \&
  {Solano}}]{Cifuentes2020}
{Cifuentes}, C., {Caballero}, J.~A., {Cort{\'e}s-Contreras}, M., {et~al.} 2020,
  \aap, 642, A115

\bibitem[{{Collins} {et~al.}(2017){Collins}, {Kielkopf}, {Stassun}, \&
  {Hessman}}]{collins17}
{Collins}, K.~A., {Kielkopf}, J.~F., {Stassun}, K.~G., \& {Hessman}, F.~V.
  2017, \aj, 153, 77

\bibitem[{{Colome} \& {Ribas}(2006)}]{colome2006}
{Colome}, J. \& {Ribas}, I. 2006, IAU Special Session, 6, 11

\bibitem[{{Cutri} {et~al.}(2021){Cutri}, {Wright}, {Conrow}, {Fowler},
  {Eisenhardt}, {Grillmair}, {Kirkpatrick}, {Masci}, {McCallon}, {Wheelock},
  {Fajardo-Acosta}, {Yan}, {Benford}, {Harbut}, {Jarrett}, {Lake}, {Leisawitz},
  {Ressler}, {Stanford}, {Tsai}, {Liu}, {Helou}, {Mainzer}, {Gettngs},
  {Gonzalez}, {Hoffman}, {Marsh}, {Padgett}, {Skrutskie}, {Beck}, {Papin}, \&
  {Wittman}}]{Cutri2014}
{Cutri}, R.~M., {Wright}, E.~L., {Conrow}, T., {et~al.} 2021, VizieR Online
  Data Catalog, II/328

\bibitem[{{Czesla} {et~al.}(2019){Czesla}, {Schr{\"o}ter}, {Schneider},
  {Huber}, {Pfeifer}, {Andreasen}, \& {Zechmeister}}]{Czesla2019}
{Czesla}, S., {Schr{\"o}ter}, S., {Schneider}, C.~P., {et~al.} 2019, {PyA:
  Python astronomy-related packages}

\bibitem[{{Demangeon} {et~al.}(2021){Demangeon}, {Zapatero Osorio}, {Alibert},
  {Barros}, {Adibekyan}, {Tabernero}, {Antoniadis-Karnavas}, {Camacho},
  {Su{\'a}rez Mascare{\~n}o}, {Oshagh}, {Micela}, {Sousa}, {Lovis}, {Pepe},
  {Rebolo}, {Cristiani}, {Santos}, {Allart}, {Allende Prieto}, {Bossini},
  {Bouchy}, {Cabral}, {Damasso}, {Di Marcantonio}, {D'Odorico}, {Ehrenreich},
  {Faria}, {Figueira}, {G{\'e}nova Santos}, {Haldemann}, {Hara}, {Gonz{\'a}lez
  Hern{\'a}ndez}, {Lavie}, {Lillo-Box}, {Lo Curto}, {Martins}, {M{\'e}gevand},
  {Mehner}, {Molaro}, {Nunes}, {Pall{\'e}}, {Pasquini}, {Poretti}, {Sozzetti},
  \& {Udry}}]{demangeon2021}
{Demangeon}, O.~D.~S., {Zapatero Osorio}, M.~R., {Alibert}, Y., {et~al.} 2021,
  \aap, 653, A41

\bibitem[{{D{\'\i}ez Alonso} {et~al.}(2019){D{\'\i}ez Alonso}, {Caballero},
  {Montes}, {de Cos Juez}, {Dreizler}, {Dubois}, {Jeffers}, {Lalitha}, {Naves},
  {Reiners}, {Ribas}, {Vanaverbeke}, {Amado}, {B{\'e}jar},
  {Cort{\'e}s-Contreras}, {Herrero}, {Hidalgo}, {K{\"u}rster}, {Logie},
  {Quirrenbach}, {Rau}, {Seifert}, {Sch{\"o}fer}, \& {Tal-Or}}]{DiezAlonso2019}
{D{\'\i}ez Alonso}, E., {Caballero}, J.~A., {Montes}, D., {et~al.} 2019, \aap,
  621, A126

\bibitem[{{Dorn} {et~al.}(2019){Dorn}, {Harrison}, {Bonsor}, \&
  {Hands}}]{Dorn2019}
{Dorn}, C., {Harrison}, J.~H.~D., {Bonsor}, A., \& {Hands}, T.~O. 2019, \mnras,
  484, 712

\bibitem[{{Dorn} \& {Lichtenberg}(2021)}]{Dorn2021}
{Dorn}, C. \& {Lichtenberg}, T. 2021, \apjl, 922, L4

\bibitem[{Emsenhuber {et~al.}(2021{\natexlab{a}})Emsenhuber, Mordasini, Burn,
  Alibert, Benz, \& Asphaug}]{Emsenhuber2021a}
Emsenhuber, A., Mordasini, C., Burn, R., {et~al.} 2021{\natexlab{a}}, A\&A,
  656, A69

\bibitem[{Emsenhuber {et~al.}(2021{\natexlab{b}})Emsenhuber, Mordasini, Burn,
  Alibert, Benz, \& Asphaug}]{Emsenhuber2021b}
Emsenhuber, A., Mordasini, C., Burn, R., {et~al.} 2021{\natexlab{b}}, A\&A,
  656, A70

\bibitem[{{Espinoza}(2018)}]{Espinoza2018}
{Espinoza}, N. 2018, Research Notes of the American Astronomical Society, 2,
  209

\bibitem[{{Espinoza} {et~al.}(2019){Espinoza}, {Kossakowski}, \&
  {Brahm}}]{juliet}
{Espinoza}, N., {Kossakowski}, D., \& {Brahm}, R. 2019, \mnras, 490, 2262

\bibitem[{{Espinoza} {et~al.}(2022){Espinoza}, {Pall{\'e}}, {Kemmer}, {Luque},
  {Caballero}, {Cifuentes}, {Herrero}, {S{\'a}nchez B{\'e}jar}, {Stock},
  {Molaverdikhani}, {Morello}, {Kossakowski}, {Schlecker}, {Amado}, {Bluhm},
  {Cort{\'e}s-Contreras}, {Henning}, {Kreidberg}, {K{\"u}rster}, {Lafarga},
  {Lodieu}, {Morales}, {Oshagh}, {Passegger}, {Pavlov}, {Quirrenbach},
  {Reffert}, {Reiners}, {Ribas}, {Rodr{\'\i}guez}, {Rodr{\'\i}guez L{\'o}pez},
  {Schweitzer}, {Trifonov}, {Chaturvedi}, {Dreizler}, {Jeffers}, {Kaminski},
  {Jos{\'e} L{\'o}pez-Gonz{\'a}lez}, {Lillo-Box}, {Montes}, {Nowak}, {Pedraz},
  {Vanaverbeke}, {Zapatero Osorio}, {Zechmeister}, {Collins}, {Girardin},
  {Guerra}, {Naves}, {Crossfield}, {Matthews}, {Howell}, {Ciardi}, {Gonzales},
  {Matson}, {Beichman}, {Schlieder}, {Barclay}, {Vezie}, {Villase{\~n}or},
  {Daylan}, {Mireies}, {Dragomir}, {Twicken}, {Jenkins}, {Winn}, {Latham},
  {Ricker}, \& {Seager}}]{espinoza22}
{Espinoza}, N., {Pall{\'e}}, E., {Kemmer}, J., {et~al.} 2022, arXiv e-prints,
  arXiv:2202.01240

\bibitem[{{Feroz} {et~al.}(2009){Feroz}, {Hobson}, \& {Bridges}}]{MultiNest}
{Feroz}, F., {Hobson}, M.~P., \& {Bridges}, M. 2009, \mnras, 398, 1601

\bibitem[{{Fischer} \& {Marcy}(1992)}]{Fischer1992}
{Fischer}, D.~A. \& {Marcy}, G.~W. 1992, \apj, 396, 178

\bibitem[{Foreman-Mackey(2016)}]{CORNER_PLOT}
Foreman-Mackey, D. 2016, The Journal of Open Source Software, 1, 24

\bibitem[{{Foreman-Mackey} {et~al.}(2017){Foreman-Mackey}, {Agol}, {Angus}, \&
  {Ambikasaran}}]{celerite}
{Foreman-Mackey}, D., {Agol}, E., {Angus}, R., \& {Ambikasaran}, S. 2017, AJ,
  154, 220

\bibitem[{{Foreman-Mackey} {et~al.}(2013){Foreman-Mackey}, {Hogg}, {Lang}, \&
  {Goodman}}]{emcee}
{Foreman-Mackey}, D., {Hogg}, D.~W., {Lang}, D., \& {Goodman}, J. 2013, \pasp,
  125, 306

\bibitem[{{Fuhrmeister} {et~al.}(2020){Fuhrmeister}, {Czesla}, {Hildebrandt},
  {Nagel}, {Schmitt}, {Jeffers}, {Caballero}, {Hintz}, {Johnson},
  {Sch{\"o}fer}, {Zechmeister}, {Reiners}, {Ribas}, {Amado}, {Quirrenbach},
  {Nortmann}, {Bauer}, {B{\'e}jar}, {Cort{\'e}s-Contreras}, {Dreizler},
  {Galad{\'\i}-Enr{\'\i}quez}, {Hatzes}, {Kaminski}, {K{\"u}rster}, {Lafarga},
  \& {Montes}}]{Fuhrmeister2020}
{Fuhrmeister}, B., {Czesla}, S., {Hildebrandt}, L., {et~al.} 2020, \aap, 640,
  A52

\bibitem[{{Fuhrmeister} {et~al.}(2019){Fuhrmeister}, {Czesla}, {Schmitt},
  {Johnson}, {Sch{\"o}fer}, {Jeffers}, {Caballero}, {Zechmeister}, {Reiners},
  {Ribas}, {Amado}, {Quirrenbach}, {Bauer}, {B{\'e}jar},
  {Cort{\'e}s-Contreras}, {D{\'\i}ez Alonso}, {Dreizler},
  {Galad{\'\i}-Enr{\'\i}quez}, {Guenther}, {Kaminski}, {K{\"u}rster},
  {Lafarga}, \& {Montes}}]{Fuhrmeister2019}
{Fuhrmeister}, B., {Czesla}, S., {Schmitt}, J.~H.~M.~M., {et~al.} 2019, \aap,
  623, A24

\bibitem[{{Fulton} {et~al.}(2018){Fulton}, {Petigura}, {Blunt}, \&
  {Sinukoff}}]{radvel}
{Fulton}, B.~J., {Petigura}, E.~A., {Blunt}, S., \& {Sinukoff}, E. 2018, \pasp,
  130, 044504

\bibitem[{{Gaia Collaboration} {et~al.}(2018){Gaia Collaboration}, {Brown},
  {Vallenari}, {Prusti}, {de Bruijne}, {Babusiaux}, {Bailer-Jones}, {Biermann},
  {Evans}, {Eyer}, {Jansen}, {Jordi}, {Klioner}, {Lammers}, {Lindegren},
  {Luri}, {Mignard}, {Panem}, {Pourbaix}, {Randich}, {Sartoretti}, {Siddiqui},
  {Soubiran}, {van Leeuwen}, {Walton}, {Arenou}, {Bastian}, {Cropper},
  {Drimmel}, {Katz}, {Lattanzi}, {Bakker}, {Cacciari}, {Casta{\~n}eda},
  {Chaoul}, {Cheek}, {De Angeli}, {Fabricius}, {Guerra}, {Holl}, {Masana},
  {Messineo}, {Mowlavi}, {Nienartowicz}, {Panuzzo}, {Portell}, {Riello},
  {Seabroke}, {Tanga}, {Th{\'e}venin}, {Gracia-Abril}, {Comoretto},
  {Garcia-Reinaldos}, {Teyssier}, {Altmann}, {Andrae}, {Audard},
  {Bellas-Velidis}, {Benson}, {Berthier}, {Blomme}, {Burgess}, {Busso},
  {Carry}, {Cellino}, {Clementini}, {Clotet}, {Creevey}, {Davidson}, {De
  Ridder}, {Delchambre}, {Dell'Oro}, {Ducourant},
  {Fern{\'a}ndez-Hern{\'a}ndez}, {Fouesneau}, {Fr{\'e}mat}, {Galluccio},
  {Garc{\'\i}a-Torres}, {Gonz{\'a}lez-N{\'u}{\~n}ez}, {Gonz{\'a}lez-Vidal},
  {Gosset}, {Guy}, {Halbwachs}, {Hambly}, {Harrison}, {Hern{\'a}ndez},
  {Hestroffer}, {Hodgkin}, {Hutton}, {Jasniewicz}, {Jean-Antoine-Piccolo},
  {Jordan}, {Korn}, {Krone-Martins}, {Lanzafame}, {Lebzelter}, {L{\"o}ffler},
  {Manteiga}, {Marrese}, {Mart{\'\i}n-Fleitas}, {Moitinho}, {Mora}, {Muinonen},
  {Osinde}, {Pancino}, {Pauwels}, {Petit}, {Recio-Blanco}, {Richards},
  {Rimoldini}, {Robin}, {Sarro}, {Siopis}, {Smith}, {Sozzetti}, {S{\"u}veges},
  {Torra}, {van Reeven}, {Abbas}, {Abreu Aramburu}, {Accart}, {Aerts},
  {Altavilla}, {{\'A}lvarez}, {Alvarez}, {Alves}, {Anderson}, {Andrei},
  {Anglada Varela}, {Antiche}, {Antoja}, {Arcay}, {Astraatmadja}, {Bach},
  {Baker}, {Balaguer-N{\'u}{\~n}ez}, {Balm}, {Barache}, {Barata}, {Barbato},
  {Barblan}, {Barklem}, {Barrado}, {Barros}, {Barstow}, {Bartholom{\'e}
  Mu{\~n}oz}, {Bassilana}, {Becciani}, {Bellazzini}, {Berihuete}, {Bertone},
  {Bianchi}, {Bienaym{\'e}}, {Blanco-Cuaresma}, {Boch}, {Boeche}, {Bombrun},
  {Borrachero}, {Bossini}, {Bouquillon}, {Bourda}, {Bragaglia}, {Bramante},
  {Breddels}, {Bressan}, {Brouillet}, {Br{\"u}semeister}, {Brugaletta},
  {Bucciarelli}, {Burlacu}, {Busonero}, {Butkevich}, {Buzzi}, {Caffau},
  {Cancelliere}, {Cannizzaro}, {Cantat-Gaudin}, {Carballo}, {Carlucci},
  {Carrasco}, {Casamiquela}, {Castellani}, {Castro-Ginard}, {Charlot},
  {Chemin}, {Chiavassa}, {Cocozza}, {Costigan}, {Cowell}, {Crifo}, {Crosta},
  {Crowley}, {Cuypers}, {Dafonte}, {Damerdji}, {Dapergolas}, {David}, {David},
  {de Laverny}, {De Luise}, {De March}, {de Martino}, {de Souza}, {de Torres},
  {Debosscher}, {del Pozo}, {Delbo}, {Delgado}, {Delgado}, {Di Matteo},
  {Diakite}, {Diener}, {Distefano}, {Dolding}, {Drazinos}, {Dur{\'a}n},
  {Edvardsson}, {Enke}, {Eriksson}, {Esquej}, {Eynard Bontemps}, {Fabre},
  {Fabrizio}, {Faigler}, {Falc{\~a}o}, {Farr{\`a}s Casas}, {Federici},
  {Fedorets}, {Fernique}, {Figueras}, {Filippi}, {Findeisen}, {Fonti},
  {Fraile}, {Fraser}, {Fr{\'e}zouls}, {Gai}, {Galleti}, {Garabato},
  {Garc{\'\i}a-Sedano}, {Garofalo}, {Garralda}, {Gavel}, {Gavras}, {Gerssen},
  {Geyer}, {Giacobbe}, {Gilmore}, {Girona}, {Giuffrida}, {Glass}, {Gomes},
  {Granvik}, {Gueguen}, {Guerrier}, {Guiraud}, {Guti{\'e}rrez-S{\'a}nchez},
  {Haigron}, {Hatzidimitriou}, {Hauser}, {Haywood}, {Heiter}, {Helmi}, {Heu},
  {Hilger}, {Hobbs}, {Hofmann}, {Holland}, {Huckle}, {Hypki}, {Icardi},
  {Jan{\ss}en}, {Jevardat de Fombelle}, {Jonker}, {Juh{\'a}sz}, {Julbe},
  {Karampelas}, {Kewley}, {Klar}, {Kochoska}, {Kohley}, {Kolenberg},
  {Kontizas}, {Kontizas}, {Koposov}, {Kordopatis}, {Kostrzewa-Rutkowska},
  {Koubsky}, {Lambert}, {Lanza}, {Lasne}, {Lavigne}, {Le Fustec}, {Le
  Poncin-Lafitte}, {Lebreton}, {Leccia}, {Leclerc}, {Lecoeur-Taibi},
  {Lenhardt}, {Leroux}, {Liao}, {Licata}, {Lindstr{\o}m}, {Lister}, {Livanou},
  {Lobel}, {L{\'o}pez}, {Managau}, {Mann}, {Mantelet}, {Marchal}, {Marchant},
  {Marconi}, {Marinoni}, {Marschalk{\'o}}, {Marshall}, {Martino}, {Marton},
  {Mary}, {Massari}, {Matijevi{\v{c}}}, {Mazeh}, {McMillan}, {Messina},
  {Michalik}, {Millar}, {Molina}, {Molinaro}, {Moln{\'a}r}, {Montegriffo},
  {Mor}, {Morbidelli}, {Morel}, {Morris}, {Mulone}, {Muraveva}, {Musella},
  {Nelemans}, {Nicastro}, {Noval}, {O'Mullane}, {Ord{\'e}novic},
  {Ord{\'o}{\~n}ez-Blanco}, {Osborne}, {Pagani}, {Pagano}, {Pailler},
  {Palacin}, {Palaversa}, {Panahi}, {Pawlak}, {Piersimoni}, {Pineau}, {Plachy},
  {Plum}, {Poggio}, {Poujoulet}, {Pr{\v{s}}a}, {Pulone}, {Racero}, {Ragaini},
  {Rambaux}, {Ramos-Lerate}, {Regibo}, {Reyl{\'e}}, {Riclet}, {Ripepi}, {Riva},
  {Rivard}, {Rixon}, {Roegiers}, {Roelens}, {Romero-G{\'o}mez}, {Rowell},
  {Royer}, {Ruiz-Dern}, {Sadowski}, {Sagrist{\`a} Sell{\'e}s}, {Sahlmann},
  {Salgado}, {Salguero}, {Sanna}, {Santana-Ros}, {Sarasso}, {Savietto},
  {Schultheis}, {Sciacca}, {Segol}, {Segovia}, {S{\'e}gransan}, {Shih},
  {Siltala}, {Silva}, {Smart}, {Smith}, {Solano}, {Solitro}, {Sordo}, {Soria
  Nieto}, {Souchay}, {Spagna}, {Spoto}, {Stampa}, {Steele},
  {Steidelm{\"u}ller}, {Stephenson}, {Stoev}, {Suess}, {Surdej}, {Szabados},
  {Szegedi-Elek}, {Tapiador}, {Taris}, {Tauran}, {Taylor}, {Teixeira},
  {Terrett}, {Teyssand ier}, {Thuillot}, {Titarenko}, {Torra Clotet}, {Turon},
  {Ulla}, {Utrilla}, {Uzzi}, {Vaillant}, {Valentini}, {Valette}, {van Elteren},
  {Van Hemelryck}, {van Leeuwen}, {Vaschetto}, {Vecchiato}, {Veljanoski},
  {Viala}, {Vicente}, {Vogt}, {von Essen}, {Voss}, {Votruba}, {Voutsinas},
  {Walmsley}, {Weiler}, {Wertz}, {Wevers}, {Wyrzykowski}, {Yoldas},
  {{\v{Z}}erjal}, {Ziaeepour}, {Zorec}, {Zschocke}, {Zucker}, {Zurbach}, \&
  {Zwitter}}]{2018A&A...616A...1G}
{Gaia Collaboration}, {Brown}, A.~G.~A., {Vallenari}, A., {et~al.} 2018, \aap,
  616, A1

\bibitem[{{Gaia Collaboration} {et~al.}(2021){Gaia Collaboration}, {Smart},
  {Sarro}, {Rybizki}, {Reyl{\'e}}, {Robin}, {Hambly}, {Abbas}, {Barstow}, {de
  Bruijne}, \& et~al.}]{GaiaEDR3}
{Gaia Collaboration}, {Smart}, R.~L., {Sarro}, L.~M., {et~al.} 2021, \aap, 649,
  A6

\bibitem[{{Gliese}(1969)}]{Gliese1969}
{Gliese}, W. 1969, Veroeffentlichungen des Astronomischen Rechen-Instituts
  Heidelberg, 22, 1

\bibitem[{{Gliese} \& {Jahrei{\ss}}(1988)}]{Gliese1988}
{Gliese}, W. \& {Jahrei{\ss}}, H. 1988, \apss, 142, 49

\bibitem[{{Gonz{\'a}lez-{\'A}lvarez} {et~al.}(2022){Gonz{\'a}lez-{\'A}lvarez},
  {Zapatero Osorio}, {Sanz-Forcada}, {Caballero}, {Reffert}, {B{\'e}jar},
  {Hatzes}, {Herrero}, {Jeffers}, {Kemmer}, {L{\'o}pez-Gonz{\'a}lez}, {Luque},
  {Molaverdikhani}, {Morello}, {Nagel}, {Quirrenbach}, {Rodr{\'\i}guez},
  {Rodr{\'\i}guez-L{\'o}pez}, {Schlecker}, {Schweitzer}, {Stock}, {Passegger},
  {Trifonov}, {Amado}, {Baker}, {Boyd}, {Cadieux}, {Charbonneau}, {Collins},
  {Doyon}, {Dreizler}, {Espinoza}, {F{\H{u}}r{\'e}sz}, {Furlan}, {Hesse},
  {Howell}, {Jenkins}, {Kidwell}, {Latham}, {McLeod}, {Montes}, {Morales},
  {O'Dwyer}, {Pall{\'e}}, {Pedraz}, {Reiners}, {Ribas}, {Quinn}, {Schnaible},
  {Seager}, {Skinner}, {Smith}, {Schwarz}, {Shporer}, {Vanderspek}, \&
  {Winn}}]{Gonzalez2022}
{Gonz{\'a}lez-{\'A}lvarez}, E., {Zapatero Osorio}, M.~R., {Sanz-Forcada}, J.,
  {et~al.} 2022, \aap, 658, A138

\bibitem[{{Hatzes}(2014)}]{Hatzes2014}
{Hatzes}, A.~P. 2014, \aap, 568, A84

\bibitem[{{Hawley} {et~al.}(1996){Hawley}, {Gizis}, \& {Reid}}]{Hawley1996}
{Hawley}, S.~L., {Gizis}, J.~E., \& {Reid}, I.~N. 1996, \aj, 112, 2799

\bibitem[{{Hirano} {et~al.}(2021){Hirano}, {Livingston}, {Fukui}, {Narita},
  {Harakawa}, {Ishikawa}, {Miyakawa}, {Kimura}, {Nakayama}, {Fujita}, {Hori},
  {Stassun}, {Bieryla}, {Cadieux}, {Ciardi}, {Collins}, {Ikoma}, {Vanderburg},
  {Barclay}, {Brasseur}, {de Leon}, {Doty}, {Doyon}, {Esparza-Borges},
  {Esquerdo}, {Furlan}, {Gaidos}, {Gonzales}, {Hodapp}, {Howell}, {Isogai},
  {Jacobson}, {Jenkins}, {Jensen}, {Kawauchi}, {Kotani}, {Kudo}, {Kurita},
  {Kurokawa}, {Kusakabe}, {Kuzuhara}, {Lafreni{\`e}re}, {Latham}, {Massey},
  {Mori}, {Murgas}, {Nishikawa}, {Nishiumi}, {Omiya}, {Paegert}, {Palle},
  {Parviainen}, {Quinn}, {Ricker}, {Schwarz}, {Seager}, {Tamura}, {Tenenbaum},
  {Terada}, {Vanderspek}, {Vievard}, {Watanabe}, \& {Winn}}]{Hirano2021}
{Hirano}, T., {Livingston}, J.~H., {Fukui}, A., {et~al.} 2021, \aj, 162, 161

\bibitem[{{H{\o}g} {et~al.}(2000){H{\o}g}, {Fabricius}, {Makarov}, {Urban},
  {Corbin}, {Wycoff}, {Bastian}, {Schwekendiek}, \& {Wicenec}}]{Hog2000}
{H{\o}g}, E., {Fabricius}, C., {Makarov}, V.~V., {et~al.} 2000, \aap, 355, L27

\bibitem[{{Howe} {et~al.}(2020){Howe}, {Adams}, \& {Meyer}}]{Howe2020}
{Howe}, A.~R., {Adams}, F.~C., \& {Meyer}, M.~R. 2020, \apj, 894, 130

\bibitem[{{Jeffers} {et~al.}(2022){Jeffers}, {Barnes}, {Scheofer},
  {Quirrenbach}, {Zechmeister}, {Amado}, {Caballero}, {Fernandez}, {Rodriguez},
  {Ribas}, {Reiners}, {Cardona Guillen}, {Cifuentes}, {Czesla}, {Hatzes},
  {Kurster}, {Montes}, {Morales}, {Pedraz}, \& {Sadegi}}]{Jeffers2022}
{Jeffers}, S.~V., {Barnes}, J.~R., {Scheofer}, P., {et~al.} 2022, arXiv
  e-prints, arXiv:2203.00415

\bibitem[{{Jenkins} {et~al.}(2016){Jenkins}, {Twicken}, {McCauliff},
  {Campbell}, {Sanderfer}, {Lung}, {Mansouri-Samani}, {Girouard}, {Tenenbaum},
  {Klaus}, {Smith}, {Caldwell}, {Chacon}, {Henze}, {Heiges}, {Latham},
  {Morgan}, {Swade}, {Rinehart}, \& {Vanderspek}}]{Jenkins2016}
{Jenkins}, J.~M., {Twicken}, J.~D., {McCauliff}, S., {et~al.} 2016, in
  \procspie, Vol. 9913, Software and Cyberinfrastructure for Astronomy IV,
  99133E

\bibitem[{{Kemmer} {et~al.}(2020){Kemmer}, {Stock}, {Kossakowski}, {Kaminski},
  {Molaverdikhani}, {Schlecker}, {Caballero}, {Amado}, {Astudillo-Defru},
  {Bonfils}, {Ciardi}, {Collins}, {Espinoza}, {Fukui}, {Hirano}, {Jenkins},
  {Latham}, {Matthews}, {Narita}, {Pall{\'e}}, {Parviainen}, {Quirrenbach},
  {Reiners}, {Ribas}, {Ricker}, {Schlieder}, {Seager}, {Vanderspek}, {Winn},
  {Almenara}, {B{\'e}jar}, {Bluhm}, {Bouchy}, {Boyd}, {Christiansen},
  {Cifuentes}, {Cloutier}, {Collins}, {Cort{\'e}s-Contreras}, {Crossfield},
  {Crouzet}, {de Leon}, {Della-Rose}, {Delfosse}, {Dreizler}, {Esparza-Borges},
  {Essack}, {Forveille}, {Figueira}, {Galad{\'\i}-Enr{\'\i}quez}, {Gan},
  {Glidden}, {Gonzales}, {Guerra}, {Harakawa}, {Hatzes}, {Henning}, {Herrero},
  {Hodapp}, {Hori}, {Howell}, {Ikoma}, {Isogai}, {Jeffers}, {K{\"u}rster},
  {Kawauchi}, {Kimura}, {Klagyivik}, {Kotani}, {Kurokawa}, {Kusakabe},
  {Kuzuhara}, {Lafarga}, {Livingston}, {Luque}, {Matson}, {Morales}, {Mori},
  {Muirhead}, {Murgas}, {Nishikawa}, {Nishiumi}, {Omiya}, {Reffert},
  {Rodr{\'\i}guez L{\'o}pez}, {Santos}, {Sch{\"o}fer}, {Schwarz}, {Shiao},
  {Tamura}, {Terada}, {Twicken}, {Ueda}, {Vievard}, {Watanabe}, \&
  {Zechmeister}}]{GJ3473}
{Kemmer}, J., {Stock}, S., {Kossakowski}, D., {et~al.} 2020, \aap, 642, A236

\bibitem[{{Kempton} {et~al.}(2018){Kempton}, {Bean}, {Louie}, {Deming}, {Koll},
  {Mansfield}, {Christiansen}, {L{\'o}pez-Morales}, {Swain}, {Zellem},
  {Ballard}, {Barclay}, {Barstow}, {Batalha}, {Beatty}, {Berta-Thompson},
  {Birkby}, {Buchhave}, {Charbonneau}, {Cowan}, {Crossfield}, {de Val-Borro},
  {Doyon}, {Dragomir}, {Gaidos}, {Heng}, {Hu}, {Kane}, {Kreidberg}, {Mallonn},
  {Morley}, {Narita}, {Nascimbeni}, {Pall{\'e}}, {Quintana}, {Rauscher},
  {Seager}, {Shkolnik}, {Sing}, {Sozzetti}, {Stassun}, {Valenti}, \& {von
  Essen}}]{Kempton2018PASP..130k4401K}
{Kempton}, E. M.~R., {Bean}, J.~L., {Louie}, D.~R., {et~al.} 2018, \pasp, 130,
  114401

\bibitem[{{Kervella} {et~al.}(2019){Kervella}, {Arenou}, {Mignard}, \&
  {Th{\'e}venin}}]{Kervella2019}
{Kervella}, P., {Arenou}, F., {Mignard}, F., \& {Th{\'e}venin}, F. 2019, \aap,
  623, A72

\bibitem[{{Kipping}(2013)}]{Kipping2013}
{Kipping}, D.~M. 2013, \mnras, 435, 2152

\bibitem[{{Kirkpatrick} {et~al.}(1991){Kirkpatrick}, {Henry}, \&
  {McCarthy}}]{Kirkpatrick1991}
{Kirkpatrick}, J.~D., {Henry}, T.~J., \& {McCarthy}, Donald~W., J. 1991, \apjs,
  77, 417

\bibitem[{{K{\"o}nigl} {et~al.}(2017){K{\"o}nigl}, {Giacalone}, \&
  {Matsakos}}]{konigl2017}
{K{\"o}nigl}, A., {Giacalone}, S., \& {Matsakos}, T. 2017, \apjl, 846, L13

\bibitem[{{Kostov} {et~al.}(2019){Kostov}, {Schlieder}, {Barclay}, {Quintana},
  {Col{\'o}n}, {Brand e}, {Collins}, {Feinstein}, {Hadden}, {Kane},
  {Kreidberg}, {Kruse}, {Lam}, {Matthews}, {Montet}, {Pozuelos}, {Stassun},
  {Winters}, {Ricker}, {Vanderspek}, {Latham}, {Seager}, {Winn}, {Jenkins},
  {Afanasev}, {Armstrong}, {Arney}, {Boyd}, {Barentsen}, {Barkaoui}, {Batalha},
  {Beichman}, {Bayliss}, {Burke}, {Burdanov}, {Cacciapuoti}, {Carson},
  {Charbonneau}, {Christiansen}, {Ciardi}, {Clampin}, {Collins}, {Conti},
  {Coughlin}, {Covone}, {Crossfield}, {Delrez}, {Domagal-Goldman}, {Dressing},
  {Ducrot}, {Essack}, {Everett}, {Fauchez}, {Foreman-Mackey}, {Gan}, {Gilbert},
  {Gillon}, {Gonzales}, {Hamann}, {Hedges}, {Hocutt}, {Hoffman}, {Horch},
  {Horne}, {Howell}, {Hynes}, {Ireland }, {Irwin}, {Isopi}, {Jensen}, {Jehin},
  {Kaltenegger}, {Kielkopf}, {Kopparapu}, {Lewis}, {Lopez}, {Lissauer}, {Mann},
  {Mallia}, {Mandell}, {Matson}, {Mazeh}, {Monsue}, {Moran}, {Moran}, {Morley},
  {Morris}, {Muirhead}, {Mukai}, {Mullally}, {Mullally}, {Murray}, {Narita},
  {Palle}, {Pidhorodetska}, {Quinn}, {Relles}, {Rinehart}, {Ritsko},
  {Rodriguez}, {Rowden}, {Rowe}, {Sebastian}, {Sefako}, {Shahaf}, {Shporer},
  {Ta{\~n}{\'o}n Reyes}, {Tenenbaum}, {Ting}, {Twicken}, {van Belle}, {Vega},
  {Volosin}, {Walkowicz}, \& {Youngblood}}]{kostov2019}
{Kostov}, V.~B., {Schlieder}, J.~E., {Barclay}, T., {et~al.} 2019, \aj, 158, 32

\bibitem[{{Kreidberg}(2015)}]{batman}
{Kreidberg}, L. 2015, Publications of the Astronomical Society of the Pacific,
  127, 1161

\bibitem[{{Lam} {et~al.}(2021){Lam}, {Csizmadia}, {Astudillo-Defru}, {Bonfils},
  {Gandolfi}, {Padovan}, {Esposito}, {Hellier}, {Hirano}, {Livingston},
  {Murgas}, {Smith}, {Collins}, {Mathur}, {Garcia}, {Howell}, {Santos}, {Dai},
  {Ricker}, {Vanderspek}, {Latham}, {Seager}, {Winn}, {Jenkins}, {Albrecht},
  {Almenara}, {Artigau}, {Barrag{\'a}n}, {Bouchy}, {Cabrera}, {Charbonneau},
  {Chaturvedi}, {Chaushev}, {Christiansen}, {Cochran}, {De Meideiros},
  {Delfosse}, {D{\'\i}az}, {Doyon}, {Eigm{\"u}ller}, {Figueira}, {Forveille},
  {Fridlund}, {Gaisn{\'e}}, {Goffo}, {Georgieva}, {Grziwa}, {Guenther},
  {Hatzes}, {Johnson}, {Kab{\'a}th}, {Knudstrup}, {Korth}, {Lewin}, {Lissauer},
  {Lovis}, {Luque}, {Melo}, {Morgan}, {Morris}, {Mayor}, {Narita}, {Osborne},
  {Palle}, {Pepe}, {Persson}, {Quinn}, {Rauer}, {Redfield}, {Schlieder},
  {S{\'e}gransan}, {Serrano}, {Smith}, {{\v{S}}ubjak}, {Twicken}, {Udry}, {Van
  Eylen}, \& {Vezie}}]{lam2021}
{Lam}, K. W.~F., {Csizmadia}, S., {Astudillo-Defru}, N., {et~al.} 2021,
  Science, 374, 1271

\bibitem[{{Leggett}(1992)}]{Leggett1992}
{Leggett}, S.~K. 1992, \apjs, 82, 351

\bibitem[{{L{\'e}pine} \& {Shara}(2005)}]{Lepine2005}
{L{\'e}pine}, S. \& {Shara}, M.~M. 2005, \aj, 129, 1483

\bibitem[{{Lindegren} {et~al.}(2021){Lindegren}, {Klioner}, {Hern{\'a}ndez},
  {Bombrun}, {Ramos-Lerate}, {Steidelm{\"u}ller}, {Bastian}, {Biermann}, {de
  Torres}, {Gerlach}, {Geyer}, {Hilger}, {Hobbs}, {Lammers}, {McMillan},
  {Stephenson}, {Casta{\~n}eda}, {Davidson}, {Fabricius}, {Gracia-Abril},
  {Portell}, {Rowell}, {Teyssier}, {Torra}, {Bartolom{\'e}}, {Clotet},
  {Garralda}, {Gonz{\'a}lez-Vidal}, {Torra}, {Abbas}, {Altmann}, {Anglada
  Varela}, {Balaguer-N{\'u}{\~n}ez}, {Balog}, {Barache}, {Becciani}, {Bernet},
  {Bertone}, {Bianchi}, {Bouquillon}, {Brown}, {Bucciarelli}, {Busonero},
  {Butkevich}, {Buzzi}, {Cancelliere}, {Carlucci}, {Charlot}, {Cioni},
  {Crosta}, {Crowley}, {del Peloso}, {del Pozo}, {Drimmel}, {Esquej}, {Fienga},
  {Fraile}, {Gai}, {Garcia-Reinaldos}, {Guerra}, {Hambly}, {Hauser},
  {Jan{\ss}en}, {Jordan}, {Kostrzewa-Rutkowska}, {Lattanzi}, {Liao}, {Licata},
  {Lister}, {L{\"o}ffler}, {Marchant}, {Masip}, {Mignard}, {Mints}, {Molina},
  {Mora}, {Morbidelli}, {Murphy}, {Pagani}, {Panuzzo}, {Pe{\~n}alosa Esteller},
  {Poggio}, {Re Fiorentin}, {Riva}, {Sagrist{\`a} Sell{\'e}s}, {Sanchez
  Gimenez}, {Sarasso}, {Sciacca}, {Siddiqui}, {Smart}, {Souami}, {Spagna},
  {Steele}, {Taris}, {Utrilla}, {van Reeven}, \& {Vecchiato}}]{Lindegren2021}
{Lindegren}, L., {Klioner}, S.~A., {Hern{\'a}ndez}, J., {et~al.} 2021, \aap,
  649, A2

\bibitem[{{Lopez}(2017)}]{Lopez17}
{Lopez}, E.~D. 2017, \mnras, 472, 245

\bibitem[{{Lundkvist} {et~al.}(2016){Lundkvist}, {Kjeldsen}, {Albrecht},
  {Davies}, {Basu}, {Huber}, {Justesen}, {Karoff}, {Silva Aguirre}, {van
  Eylen}, {Vang}, {Arentoft}, {Barclay}, {Bedding}, {Campante}, {Chaplin},
  {Christensen-Dalsgaard}, {Elsworth}, {Gilliland}, {Handberg}, {Hekker},
  {Kawaler}, {Lund}, {Metcalfe}, {Miglio}, {Rowe}, {Stello}, {Tingley}, \&
  {White}}]{Lundkvist16}
{Lundkvist}, M.~S., {Kjeldsen}, H., {Albrecht}, S., {et~al.} 2016, Nature
  Communications, 7, 11201

\bibitem[{{Luque} \& {Pall{\'e}}(2022)}]{Luque2022}
{Luque}, R. \& {Pall{\'e}}, E. 2022, Science, 377, 1211

\bibitem[{{Luque} {et~al.}(2019{\natexlab{a}}){Luque}, {Pall{\'e}},
  {Kossakowski}, {Dreizler}, {Kemmer}, {Espinoza}, {Burt},
  {Anglada-Escud{\'e}}, {B{\'e}jar}, {Caballero}, {Collins}, {Collins},
  {Cort{\'e}s-Contreras}, {D{\'\i}ez-Alonso}, {Feng}, {Hatzes}, {Hellier},
  {Henning}, {Jeffers}, {Kaltenegger}, {K{\"u}rster}, {Madden},
  {Molaverdikhani}, {Montes}, {Narita}, {Nowak}, {Ofir}, {Oshagh},
  {Parviainen}, {Quirrenbach}, {Reffert}, {Reiners},
  {Rodr{\'\i}guez-L{\'o}pez}, {Schlecker}, {Stock}, {Trifonov}, {Winn},
  {Zapatero Osorio}, {Zechmeister}, {Amado}, {Anderson}, {Batalha}, {Bauer},
  {Bluhm}, {Burke}, {Butler}, {Caldwell}, {Chen}, {Crane}, {Dragomir},
  {Dressing}, {Dynes}, {Jenkins}, {Kaminski}, {Klahr}, {Kotani}, {Lafarga},
  {Latham}, {Lewin}, {McDermott}, {Monta{\~n}{\'e}s-Rodr{\'\i}guez}, {Morales},
  {Murgas}, {Nagel}, {Pedraz}, {Ribas}, {Ricker}, {Rowden}, {Seager},
  {Shectman}, {Tamura}, {Teske}, {Twicken}, {Vanderspeck}, {Wang}, \&
  {Wohler}}]{GJ357}
{Luque}, R., {Pall{\'e}}, E., {Kossakowski}, D., {et~al.} 2019{\natexlab{a}},
  \aap, 628, A39

\bibitem[{{Luque} {et~al.}(2019{\natexlab{b}}){Luque}, {Pall{\'e}},
  {Kossakowski}, {Dreizler}, {Kemmer}, {Espinoza}, {Burt},
  {Anglada-Escud{\'e}}, {B{\'e}jar}, {Caballero}, {Collins}, {Collins},
  {Cort{\'e}s-Contreras}, {D{\'\i}ez-Alonso}, {Feng}, {Hatzes}, {Hellier},
  {Henning}, {Jeffers}, {Kaltenegger}, {K{\"u}rster}, {Madden},
  {Molaverdikhani}, {Montes}, {Narita}, {Nowak}, {Ofir}, {Oshagh},
  {Parviainen}, {Quirrenbach}, {Reffert}, {Reiners},
  {Rodr{\'\i}guez-L{\'o}pez}, {Schlecker}, {Stock}, {Trifonov}, {Winn},
  {Zapatero Osorio}, {Zechmeister}, {Amado}, {Anderson}, {Batalha}, {Bauer},
  {Bluhm}, {Burke}, {Butler}, {Caldwell}, {Chen}, {Crane}, {Dragomir},
  {Dressing}, {Dynes}, {Jenkins}, {Kaminski}, {Klahr}, {Kotani}, {Lafarga},
  {Latham}, {Lewin}, {McDermott}, {Monta{\~n}{\'e}s-Rodr{\'\i}guez}, {Morales},
  {Murgas}, {Nagel}, {Pedraz}, {Ribas}, {Ricker}, {Rowden}, {Seager},
  {Shectman}, {Tamura}, {Teske}, {Twicken}, {Vanderspeck}, {Wang}, \&
  {Wohler}}]{Luque2019}
{Luque}, R., {Pall{\'e}}, E., {Kossakowski}, D., {et~al.} 2019{\natexlab{b}},
  \aap, 628, A39

\bibitem[{{Mann} {et~al.}(2015){Mann}, {Feiden}, {Gaidos}, {Boyajian}, \& {von
  Braun}}]{Mann2015}
{Mann}, A.~W., {Feiden}, G.~A., {Gaidos}, E., {Boyajian}, T., \& {von Braun},
  K. 2015, \apj, 804, 64

\bibitem[{{Mansfield} {et~al.}(2019){Mansfield}, {Kite}, {Hu}, {Koll}, {Malik},
  {Bean}, \& {Kempton}}]{Mansfield2019}
{Mansfield}, M., {Kite}, E.~S., {Hu}, R., {et~al.} 2019, \apj, 886, 141

\bibitem[{{Marfil} {et~al.}(2021){Marfil}, {Tabernero}, {Montes}, {Caballero},
  {Lazaro}, {Gonzalez Hernandez}, {Nagel}, {Passegger}, {Schweitzer}, {Ribas},
  {Reiners}, {Quirrenbach}, {Amado}, {Cifuentes}, {Cortes-Contreras},
  {Dreizler}, {Duque-Arribas}, {Galadi-Enriquez}, {Henning}, {Jeffers},
  {Kaminski}, {Kurster}, {Lafarga}, {Lopez-Gallifa}, {Morales}, {Shan}, \&
  {Zechmeister}}]{Marfil2021}
{Marfil}, E., {Tabernero}, H.~M., {Montes}, D., {et~al.} 2021, arXiv e-prints,
  arXiv:2110.07329

\bibitem[{{Mazeh} {et~al.}(2016){Mazeh}, {Holczer}, \& {Faigler}}]{Mazeh2016}
{Mazeh}, T., {Holczer}, T., \& {Faigler}, S. 2016, \aap, 589, A75

\bibitem[{{McCully} {et~al.}(2018){McCully}, {Volgenau}, {Harbeck}, {Lister},
  {Saunders}, {Turner}, {Siiverd}, \& {Bowman}}]{mccully18}
{McCully}, C., {Volgenau}, N.~H., {Harbeck}, D.-R., {et~al.} 2018, in Society
  of Photo-Optical Instrumentation Engineers (SPIE) Conference Series, Vol.
  10707, Software and Cyberinfrastructure for Astronomy V, ed. J.~C. {Guzman}
  \& J.~{Ibsen}, 107070K

\bibitem[{{McDonald} {et~al.}(2019){McDonald}, {Kreidberg}, \&
  {Lopez}}]{McDonald2019}
{McDonald}, G.~D., {Kreidberg}, L., \& {Lopez}, E. 2019, \apj, 876, 22

\bibitem[{Mishra {et~al.}(2021)Mishra, Alibert, Leleu, Emsenhuber, Mordasini,
  Burn, Udry, \& Benz}]{Mishra2021b}
Mishra, L., Alibert, Y., Leleu, A., {et~al.} 2021, A\&A, 656, A74

\bibitem[{{Molaverdikhani} {et~al.}(2019{\natexlab{a}}){Molaverdikhani},
  {Henning}, \& {Molli{\`e}re}}]{molaverdikhani19b}
{Molaverdikhani}, K., {Henning}, T., \& {Molli{\`e}re}, P. 2019{\natexlab{a}},
  \apj, 883, 194

\bibitem[{{Molaverdikhani} {et~al.}(2019{\natexlab{b}}){Molaverdikhani},
  {Henning}, \& {Molli{\`e}re}}]{molaverdikhani19a}
{Molaverdikhani}, K., {Henning}, T., \& {Molli{\`e}re}, P. 2019{\natexlab{b}},
  \apj, 873, 32

\bibitem[{{Molaverdikhani} {et~al.}(2020){Molaverdikhani}, {Henning}, \&
  {Molli{\`e}re}}]{molaverdikhani20}
{Molaverdikhani}, K., {Henning}, T., \& {Molli{\`e}re}, P. 2020, \apj, 899, 53

\bibitem[{{Molli{\`e}re} {et~al.}(2019){Molli{\`e}re}, {Wardenier}, {van
  Boekel}, {Henning}, {Molaverdikhani}, \& {Snellen}}]{molliere19}
{Molli{\`e}re}, P., {Wardenier}, J.~P., {van Boekel}, R., {et~al.} 2019, \aap,
  627, A67

\bibitem[{{Montes} {et~al.}(2001){Montes}, {L{\'o}pez-Santiago}, {G{\'a}lvez},
  {Fern{\'a}ndez-Figueroa}, {De Castro}, \& {Cornide}}]{Montes2001}
{Montes}, D., {L{\'o}pez-Santiago}, J., {G{\'a}lvez}, M.~C., {et~al.} 2001,
  \mnras, 328, 45

\bibitem[{{Morello} {et~al.}(2021){Morello}, {Zingales}, {Martin-Lagarde},
  {Gastaud}, \& {Lagage}}]{morello21}
{Morello}, G., {Zingales}, T., {Martin-Lagarde}, M., {Gastaud}, R., \&
  {Lagage}, P.-O. 2021, \aj, 161, 174

\bibitem[{{Narita} {et~al.}(2019){Narita}, {Fukui}, {Kusakabe}, {Watanabe},
  {Palle}, {Parviainen}, {Monta{\~n}{\'e}s-Rodr{\'\i}guez}, {Murgas},
  {Monelli}, {Aguiar}, {Perez Prieto}, {Oscoz}, {de Leon}, {Mori}, {Tamura},
  {Yamamuro}, {B{\'e}jar}, {Crouzet}, {Hidalgo}, {Klagyivik}, {Luque}, \&
  {Nishiumi}}]{Narita2019}
{Narita}, N., {Fukui}, A., {Kusakabe}, N., {et~al.} 2019, Journal of
  Astronomical Telescopes, Instruments, and Systems, 5, 015001

\bibitem[{{Nidever} {et~al.}(2002){Nidever}, {Marcy}, {Butler}, {Fischer}, \&
  {Vogt}}]{Nidever2002}
{Nidever}, D.~L., {Marcy}, G.~W., {Butler}, R.~P., {Fischer}, D.~A., \& {Vogt},
  S.~S. 2002, \apjs, 141, 503

\bibitem[{{Nowak} {et~al.}(2020){Nowak}, {Luque}, {Parviainen}, {Pall{\'e}},
  {Molaverdikhani}, {B{\'e}jar}, {Lillo-Box}, {Rodr{\'\i}guez-L{\'o}pez},
  {Caballero}, {Zechmeister}, {Passegger}, {Cifuentes}, {Schweitzer}, {Narita},
  {Cale}, {Espinoza}, {Murgas}, {Hidalgo}, {Zapatero Osorio}, {Pozuelos},
  {Aceituno}, {Amado}, {Barkaoui}, {Barrado}, {Bauer}, {Benkhaldoun},
  {Caldwell}, {Casasayas Barris}, {Chaturvedi}, {Chen}, {Collins}, {Collins},
  {Cort{\'e}s-Contreras}, {Crossfield}, {de Le{\'o}n}, {D{\'\i}ez Alonso},
  {Dreizler}, {El Mufti}, {Esparza-Borges}, {Essack}, {Fukui}, {Gaidos},
  {Gillon}, {Gonzales}, {Guerra}, {Hatzes}, {Henning}, {Herrero}, {Hesse},
  {Hirano}, {Howell}, {Jeffers}, {Jehin}, {Jenkins}, {Kaminski}, {Kemmer},
  {Kielkopf}, {Kossakowski}, {Kotani}, {K{\"u}rster}, {Lafarga}, {Latham},
  {Law}, {Lissauer}, {Lodieu}, {Madrigal-Aguado}, {Mann}, {Massey}, {Matson},
  {Matthews}, {Monta{\~n}{\'e}s-Rodr{\'\i}guez}, {Montes}, {Morales}, {Mori},
  {Nagel}, {Oshagh}, {Pedraz}, {Plavchan}, {Pollacco}, {Quirrenbach},
  {Reffert}, {Reiners}, {Ribas}, {Ricker}, {Rose}, {Schlecker}, {Schlieder},
  {Seager}, {Stangret}, {Stock}, {Tamura}, {Tanner}, {Teske}, {Trifonov},
  {Twicken}, {Vanderspek}, {Watanabe}, {Wittrock}, {Ziegler}, \&
  {Zohrabi}}]{LTT3780-1}
{Nowak}, G., {Luque}, R., {Parviainen}, H., {et~al.} 2020, \aap, 642, A173

\bibitem[{{Orell-Miquel} {et~al.}(2022){Orell-Miquel}, {Murgas}, {Pall{\'e}},
  {Lamp{\'o}n}, {L{\'o}pez-Puertas}, {Sanz-Forcada}, {Nagel}, {Kaminski},
  {Casasayas-Barris}, {Nortmann}, {Luque}, {Molaverdikhani}, {Sedaghati},
  {Caballero}, {Amado}, {Bergond}, {Czesla}, {Hatzes}, {Henning},
  {Khalafinejad}, {Montes}, {Morello}, {Quirrenbach}, {Reiners}, {Ribas},
  {S{\'a}nchez-L{\'o}pez}, {Schweitzer}, {Stangret}, {Yan}, \& {Zapatero
  Osorio}}]{Orell2022}
{Orell-Miquel}, J., {Murgas}, F., {Pall{\'e}}, E., {et~al.} 2022, \aap, 659,
  A55

\bibitem[{{Osborn} {et~al.}(2021){Osborn}, {Armstrong}, {Cale}, {Brahm},
  {Wittenmyer}, {Dai}, {Crossfield}, {Bryant}, {Adibekyan}, {Cloutier},
  {Collins}, {Delgado Mena}, {Fridlund}, {Hellier}, {Howell}, {King},
  {Lillo-Box}, {Otegi}, {Sousa}, {Stassun}, {Matthews}, {Ziegler}, {Ricker},
  {Vanderspek}, {Latham}, {Seager}, {Winn}, {Jenkins}, {Acton}, {Addison},
  {Anderson}, {Ballard}, {Barrado}, {Barros}, {Batalha}, {Bayliss}, {Barclay},
  {Benneke}, {Berberian}, {Bouchy}, {Bowler}, {Brice{\~n}o}, {Burke},
  {Burleigh}, {Casewell}, {Ciardi}, {Collins}, {Cooke}, {Demangeon},
  {D{\'\i}az}, {Dorn}, {Dragomir}, {Dressing}, {Dumusque}, {Espinoza},
  {Figueira}, {Fulton}, {Furlan}, {Gaidos}, {Geneser}, {Gill}, {Goad},
  {Gonzales}, {Gorjian}, {G{\"u}nther}, {Helled}, {Henderson}, {Henning},
  {Hogan}, {Hojjatpanah}, {Horner}, {Howard}, {Hoyer}, {Huber}, {Isaacson},
  {Jenkins}, {Jensen}, {Jord{\'a}n}, {Kane}, {Kidwell}, {Kielkopf}, {Law},
  {Lendl}, {Lund}, {Matson}, {Mann}, {McCormac}, {Mengel}, {Morales},
  {Nielsen}, {Okumura}, {Osborn}, {Petigura}, {Plavchan}, {Pollacco},
  {Quintana}, {Raynard}, {Robertson}, {Rose}, {Roy}, {Reefe}, {Santerne},
  {Santos}, {Sarkis}, {Schlieder}, {Schwarz}, {Scott}, {Shporer}, {Smith},
  {Stibbard}, {Stockdale}, {Str{\o}m}, {Twicken}, {Tan}, {Tanner}, {Teske},
  {Tilbrook}, {Tinney}, {Udry}, {Villase{\~n}or}, {Vines}, {Wang}, {Weiss},
  {West}, {Wheatley}, {Wright}, {Zhang}, \& {Zohrabi}}]{osborn2021}
{Osborn}, A., {Armstrong}, D.~J., {Cale}, B., {et~al.} 2021, \mnras, 507, 2782

\bibitem[{{Owen} \& {Lai}(2018)}]{OwenLAi18}
{Owen}, J.~E. \& {Lai}, D. 2018, \mnras, 479, 5012

\bibitem[{{Palle} {et~al.}(2020){Palle}, {Nortmann}, {Casasayas-Barris},
  {Lamp{\'o}n}, {L{\'o}pez-Puertas}, {Caballero}, {Sanz-Forcada}, {Lara},
  {Nagel}, {Yan}, {Alonso-Floriano}, {Amado}, {Chen}, {Cifuentes},
  {Cort{\'e}s-Contreras}, {Czesla}, {Molaverdikhani}, {Montes}, {Passegger},
  {Quirrenbach}, {Reiners}, {Ribas}, {S{\'a}nchez-L{\'o}pez}, {Schweitzer},
  {Stangret}, {Zapatero Osorio}, \& {Zechmeister}}]{Palle2020}
{Palle}, E., {Nortmann}, L., {Casasayas-Barris}, N., {et~al.} 2020, \aap, 638,
  A61

\bibitem[{Parviainen(2015)}]{Parviainen2015}
Parviainen, H. 2015, MNRAS, 450, 3233

\bibitem[{Parviainen(2020)}]{Parviainen2020b}
Parviainen, H. 2020, Monthly Notices of the Royal Astronomical Society, 499,
  1633

\bibitem[{Parviainen \& Aigrain(2015)}]{Parviainen2015b}
Parviainen, H. \& Aigrain, S. 2015, MNRAS, 453, 3822

\bibitem[{Parviainen \& Korth(2020)}]{Parviainen2020a}
Parviainen, H. \& Korth, J. 2020, Monthly Notices of the Royal Astronomical
  Society [\eprint[arXiv]{2009.09965}]

\bibitem[{{Parviainen} {et~al.}(2019){Parviainen}, {Tingley}, {Deeg}, {Palle},
  {Alonso}, {Montanes Rodriguez}, {Murgas}, {Narita}, {Fukui}, {Watanabe},
  {Kusakabe}, {Tamura}, {Nishiumi}, {Prieto-Arranz}, {Klagyivik}, {B{\'e}jar},
  {Crouzet}, {Mori}, {Hidalgo Soto}, {Casasayas Barris}, \&
  {Luque}}]{Parviainen2019}
{Parviainen}, H., {Tingley}, B., {Deeg}, H.~J., {et~al.} 2019, \aap, 630, A89

\bibitem[{{Passegger} {et~al.}(2019{\natexlab{a}}){Passegger}, {Schweitzer},
  {Shulyak}, {Nagel}, {Hauschildt}, {Reiners}, {Amado}, {Caballero},
  {Cort{\'e}s-Contreras}, {Dom{\'\i}nguez-Fern{\'a}ndez}, {Quirrenbach},
  {Ribas}, {Azzaro}, {Anglada-Escud{\'e}}, {Bauer}, {B{\'e}jar}, {Dreizler},
  {Guenther}, {Henning}, {Jeffers}, {Kaminski}, {K{\"u}rster}, {Lafarga},
  {Mart{\'\i}n}, {Montes}, {Morales}, {Schmitt}, \&
  {Zechmeister}}]{Passegger2019}
{Passegger}, V.~M., {Schweitzer}, A., {Shulyak}, D., {et~al.}
  2019{\natexlab{a}}, \aap, 627, A161

\bibitem[{{Passegger} {et~al.}(2019{\natexlab{b}}){Passegger}, {Schweitzer},
  {Shulyak}, {Nagel}, {Hauschildt}, {Reiners}, {Amado}, {Caballero},
  {Cort{\'e}s-Contreras}, {Dom{\'\i}nguez-Fern{\'a}ndez}, {Quirrenbach},
  {Ribas}, {Azzaro}, {Anglada-Escud{\'e}}, {Bauer}, {B{\'e}jar}, {Dreizler},
  {Guenther}, {Henning}, {Jeffers}, {Kaminski}, {K{\"u}rster}, {Lafarga},
  {Mart{\'\i}n}, {Montes}, {Morales}, {Schmitt}, \&
  {Zechmeister}}]{2019A&A...627A.161P}
{Passegger}, V.~M., {Schweitzer}, A., {Shulyak}, D., {et~al.}
  2019{\natexlab{b}}, \aap, 627, A161

\bibitem[{{Perdelwitz} {et~al.}(2021){Perdelwitz}, {Mittag}, {Tal-Or},
  {Schmitt}, {Caballero}, {Jeffers}, {Reiners}, {Schweitzer}, {Trifonov},
  {Ribas}, {Quirrenbach}, {Amado}, {Seifert}, {Cifuentes},
  {Cort{\'e}s-Contreras}, {Montes}, {Revilla}, \&
  {Skrzypinski}}]{Perdelwitz2021}
{Perdelwitz}, V., {Mittag}, M., {Tal-Or}, L., {et~al.} 2021, \aap, 652, A116

\bibitem[{{P{\'e}rez-Torres} {et~al.}(2021){P{\'e}rez-Torres}, {G{\'o}mez},
  {Ortiz}, {Leto}, {Anglada}, {G{\'o}mez}, {Rodr{\'\i}guez}, {Trigilio},
  {Amado}, {Alberdi}, {Anglada-Escud{\'e}}, {Osorio}, {Umana}, {Berdi{\~n}as},
  {L{\'o}pez-Gonz{\'a}lez}, {Morales}, {Rodr{\'\i}guez-L{\'o}pez}, \&
  {Chibueze}}]{PerezTorres2021}
{P{\'e}rez-Torres}, M., {G{\'o}mez}, J.~F., {Ortiz}, J.~L., {et~al.} 2021,
  \aap, 645, A77

\bibitem[{{Quirrenbach} {et~al.}(2014){Quirrenbach}, {Amado}, {Caballero},
  {Mundt}, {Reiners}, {Ribas}, {Seifert}, {Abril}, {Aceituno},
  {Alonso-Floriano}, {Ammler-von Eiff}, {Antona Jim{\'e}nez},
  {Anwand-Heerwart}, {Azzaro}, {Bauer}, {Barrado}, {Becerril}, {B{\'e}jar},
  {Ben{\'{\i}}tez}, {Berdi{\~n}as}, {C{\'a}rdenas}, {Casal}, {Claret},
  {Colom{\'e}}, {Cort{\'e}s-Contreras}, {Czesla}, {Doellinger}, {Dreizler},
  {Feiz}, {Fern{\'a}ndez}, {Galad{\'{\i}}}, {G{\'a}lvez-Ortiz},
  {Garc{\'{\i}}a-Piquer}, {Garc{\'{\i}}a-Vargas}, {Garrido}, {Gesa}, {G{\'o}mez
  Galera}, {Gonz{\'a}lez {\'A}lvarez}, {Gonz{\'a}lez Hern{\'a}ndez},
  {Gr{\"o}zinger}, {Gu{\`a}rdia}, {Guenther}, {de Guindos},
  {Guti{\'e}rrez-Soto}, {Hagen}, {Hatzes}, {Hauschildt}, {Helmling}, {Henning},
  {Hermann}, {Hern{\'a}ndez Casta{\~n}o}, {Herrero}, {Hidalgo}, {Holgado},
  {Huber}, {Huber}, {Jeffers}, {Joergens}, {de Juan}, {Kehr}, {Klein},
  {K{\"u}rster}, {Lamert}, {Lalitha}, {Laun}, {Lemke}, {Lenzen}, {L{\'o}pez del
  Fresno}, {L{\'o}pez Mart{\'{\i}}}, {L{\'o}pez-Santiago}, {Mall}, {Mandel},
  {Mart{\'{\i}}n}, {Mart{\'{\i}}n-Ruiz}, {Mart{\'{\i}}nez-Rodr{\'{\i}}guez},
  {Marvin}, {Mathar}, {Mirabet}, {Montes}, {Morales Mu{\~n}oz}, {Moya},
  {Naranjo}, {Ofir}, {Oreiro}, {Pall{\'e}}, {Panduro}, {Passegger},
  {P{\'e}rez-Calpena}, {P{\'e}rez Medialdea}, {Perger}, {Pluto}, {Ram{\'o}n},
  {Rebolo}, {Redondo}, {Reffert}, {Reinhardt}, {Rhode}, {Rix}, {Rodler},
  {Rodr{\'{\i}}guez}, {Rodr{\'{\i}}guez-L{\'o}pez},
  {Rodr{\'{\i}}guez-P{\'e}rez}, {Rohloff}, {Rosich}, {S{\'a}nchez-Blanco},
  {S{\'a}nchez Carrasco}, {Sanz-Forcada}, {Sarmiento}, {Sch{\"a}fer},
  {Schiller}, {Schmidt}, {Schmitt}, {Solano}, {Stahl}, {Storz}, {St{\"u}rmer},
  {Su{\'a}rez}, {Ulbrich}, {Veredas}, {Wagner}, {Winkler}, {Zapatero Osorio},
  {Zechmeister}, {Abell{\'a}n de Paco}, {Anglada-Escud{\'e}}, {del Burgo},
  {Klutsch}, {Lizon}, {L{\'o}pez-Morales}, {Morales}, {Perryman}, {Tulloch}, \&
  {Xu}}]{CARMENES}
{Quirrenbach}, A., {Amado}, P.~J., {Caballero}, J.~A., {et~al.} 2014, in
  \procspie, Vol. 9147, Ground-based and Airborne Instrumentation for Astronomy
  V, 91471F

\bibitem[{{Quirrenbach} {et~al.}(2022){Quirrenbach}, {Passegger}, {Trifonov},
  {Amado}, {Caballero}, {Reiners}, {Ribas}, {Aceituno}, {Bejar}, {Chaturvedi},
  {Gonzalez-Cuesta}, {Henning}, {Herrero}, {Kaminski}, {Kuerster}, {Lalitha},
  {Lodieu}, {Lopez-Gonzalez}, {Montes}, {Palle}, {Perger}, {Pollacco},
  {Reffert}, {Rodriguez}, {Rodriguez Lopez}, {Shan}, {Tal-Or}, {Zapatero
  Osorio}, \& {Zechmeister}}]{Quirrenbach2022}
{Quirrenbach}, A., {Passegger}, V.~M., {Trifonov}, T., {et~al.} 2022, arXiv
  e-prints, arXiv:2203.16504

\bibitem[{{Rayner} {et~al.}(2009){Rayner}, {Cushing}, \& {Vacca}}]{Rayner2009}
{Rayner}, J.~T., {Cushing}, M.~C., \& {Vacca}, W.~D. 2009, \apjs, 185, 289

\bibitem[{{Reid} {et~al.}(1995){Reid}, {Hawley}, \& {Gizis}}]{Reid1995}
{Reid}, I.~N., {Hawley}, S.~L., \& {Gizis}, J.~E. 1995, \aj, 110, 1838

\bibitem[{{Reiners} {et~al.}(2022){Reiners}, {Shulyak}, {K{\"a}pyl{\"a}},
  {Ribas}, {Nagel}, {Zechmeister}, {Caballero}, {Shan}, {Fuhrmeister},
  {Quirrenbach}, {Amado}, {Montes}, {Jeffers}, {Azzaro}, {B{\'e}jar},
  {Chaturvedi}, {Henning}, {K{\"u}rster}, \& {Pall{\'e}}}]{Reiners2022}
{Reiners}, A., {Shulyak}, D., {K{\"a}pyl{\"a}}, P.~J., {et~al.} 2022, arXiv
  e-prints, arXiv:2204.00342

\bibitem[{{Reiners} {et~al.}(2018){Reiners}, {Zechmeister}, {Caballero},
  {Ribas}, {Morales}, {Jeffers}, {Sch{\"o}fer}, {Tal-Or}, {Quirrenbach},
  {Amado}, {Kaminski}, {Seifert}, {Abril}, {Aceituno}, {Alonso-Floriano},
  {Ammler-von Eiff}, {Antona}, {Anglada-Escud{\'e}}, {Anwand-Heerwart},
  {Arroyo-Torres}, {Azzaro}, {Baroch}, {Barrado}, {Bauer}, {Becerril},
  {B{\'e}jar}, {Ben{\'\i}tez}, {Berdinas}, {Bergond}, {Bl{\"u}mcke},
  {Brinkm{\"o}ller}, {del Burgo}, {Cano}, {C{\'a}rdenas V{\'a}zquez}, {Casal},
  {Cifuentes}, {Claret}, {Colom{\'e}}, {Cort{\'e}s-Contreras}, {Czesla},
  {D{\'\i}ez-Alonso}, {Dreizler}, {Feiz}, {Fern{\'a}ndez}, {Ferro},
  {Fuhrmeister}, {Galad{\'\i}-Enr{\'\i}quez}, {Garcia-Piquer}, {Garc{\'\i}a
  Vargas}, {Gesa}, {G{\'o}mez Galera}, {Gonz{\'a}lez Hern{\'a}ndez},
  {Gonz{\'a}lez-Peinado}, {Gr{\"o}zinger}, {Grohnert}, {Gu{\`a}rdia},
  {Guenther}, {Guijarro}, {de Guindos}, {Guti{\'e}rrez-Soto}, {Hagen},
  {Hatzes}, {Hauschildt}, {Hedrosa}, {Helmling}, {Henning}, {Hermelo},
  {Hern{\'a}ndez Arab{\'\i}}, {Hern{\'a}ndez Casta{\~n}o}, {Hern{\'a}ndez
  Hernando}, {Herrero}, {Huber}, {Huke}, {Johnson}, {de Juan}, {Kim}, {Klein},
  {Kl{\"u}ter}, {Klutsch}, {K{\"u}rster}, {Lafarga}, {Lamert}, {Lamp{\'o}n},
  {Lara}, {Laun}, {Lemke}, {Lenzen}, {Launhardt}, {L{\'o}pez del Fresno},
  {L{\'o}pez-Gonz{\'a}lez}, {L{\'o}pez-Puertas}, {L{\'o}pez Salas},
  {L{\'o}pez-Santiago}, {Luque}, {Mag{\'a}n Madinabeitia}, {Mall}, {Mancini},
  {Mandel}, {Marfil}, {Mar{\'\i}n Molina}, {Maroto Fern{\'a}ndez},
  {Mart{\'\i}n}, {Mart{\'\i}n-Ruiz}, {Marvin}, {Mathar}, {Mirabet}, {Montes},
  {Moreno-Raya}, {Moya}, {Mundt}, {Nagel}, {Naranjo}, {Nortmann}, {Nowak},
  {Ofir}, {Oreiro}, {Pall{\'e}}, {Panduro}, {Pascual}, {Passegger}, {Pavlov},
  {Pedraz}, {P{\'e}rez-Calpena}, {P{\'e}rez Medialdea}, {Perger}, {Perryman},
  {Pluto}, {Rabaza}, {Ram{\'o}n}, {Rebolo}, {Redondo}, {Reffert}, {Reinhart},
  {Rhode}, {Rix}, {Rodler}, {Rodr{\'\i}guez}, {Rodr{\'\i}guez-L{\'o}pez},
  {Rodr{\'\i}guez Trinidad}, {Rohloff}, {Rosich}, {Sadegi},
  {S{\'a}nchez-Blanco}, {S{\'a}nchez Carrasco}, {S{\'a}nchez-L{\'o}pez},
  {Sanz-Forcada}, {Sarkis}, {Sarmiento}, {Sch{\"a}fer}, {Schmitt}, {Schiller},
  {Schweitzer}, {Solano}, {Stahl}, {Strachan}, {St{\"u}rmer}, {Su{\'a}rez},
  {Tabernero}, {Tala}, {Trifonov}, {Tulloch}, {Ulbrich}, {Veredas}, {Vico
  Linares}, {Vilardell}, {Wagner}, {Winkler}, {Wolthoff}, {Xu}, {Yan}, \&
  {Zapatero Osorio}}]{Reiners2018}
{Reiners}, A., {Zechmeister}, M., {Caballero}, J.~A., {et~al.} 2018, \aap, 612,
  A49

\bibitem[{{Ricker} {et~al.}(2015){Ricker}, {Winn}, {Vanderspek}, {Latham},
  {Bakos}, {Bean}, {Berta-Thompson}, {Brown}, {Buchhave}, {Butler}, {Butler},
  {Chaplin}, {Charbonneau}, {Christensen-Dalsgaard}, {Clampin}, {Deming},
  {Doty}, {De Lee}, {Dressing}, {Dunham}, {Endl}, {Fressin}, {Ge}, {Henning},
  {Holman}, {Howard}, {Ida}, {Jenkins}, {Jernigan}, {Johnson}, {Kaltenegger},
  {Kawai}, {Kjeldsen}, {Laughlin}, {Levine}, {Lin}, {Lissauer}, {MacQueen},
  {Marcy}, {McCullough}, {Morton}, {Narita}, {Paegert}, {Palle}, {Pepe},
  {Pepper}, {Quirrenbach}, {Rinehart}, {Sasselov}, {Sato}, {Seager},
  {Sozzetti}, {Stassun}, {Sullivan}, {Szentgyorgyi}, {Torres}, {Udry}, \&
  {Villasenor}}]{Ricker2015}
{Ricker}, G.~R., {Winn}, J.~N., {Vanderspek}, R., {et~al.} 2015, Journal of
  Astronomical Telescopes, Instruments, and Systems, 1, 014003

\bibitem[{{Sahu} {et~al.}(2006){Sahu}, {Casertano}, {Bond}, {Valenti}, {Ed
  Smith}, {Minniti}, {Zoccali}, {Livio}, {Panagia}, {Piskunov}, {Brown},
  {Brown}, {Renzini}, {Rich}, {Clarkson}, \& {Lubow}}]{Sahu2006}
{Sahu}, K.~C., {Casertano}, S., {Bond}, H.~E., {et~al.} 2006, \nat, 443, 534

\bibitem[{{Sanchis-Ojeda} {et~al.}(2014){Sanchis-Ojeda}, {Rappaport}, {Winn},
  {Kotson}, {Levine}, \& {El Mellah}}]{Sanchis-Ojeda14}
{Sanchis-Ojeda}, R., {Rappaport}, S., {Winn}, J.~N., {et~al.} 2014, \apj, 787,
  47

\bibitem[{{Sanz-Forcada} {et~al.}(2011){Sanz-Forcada}, {Micela}, {Ribas},
  {Pollock}, {Eiroa}, {Velasco}, {Solano}, \&
  {Garc{\'{\i}}a-{\'A}lvarez}}]{san11}
{Sanz-Forcada}, J., {Micela}, G., {Ribas}, I., {et~al.} 2011, \aap, 532, A6+

\bibitem[{Schlecker {et~al.}(2022)Schlecker, Burn, Sabotta, Seifert, Henning,
  Emsenhuber, Mordasini, Reffert, Shan, \& Klahr}]{Schlecker2022}
Schlecker, M., Burn, R., Sabotta, S., {et~al.} 2022, RV-detected planets around
  M dwarfs: Challenges for core accretion models

\bibitem[{Schlecker {et~al.}(2021{\natexlab{a}})Schlecker, Mordasini,
  Emsenhuber, Klahr, Henning, Burn, Alibert, \& Benz}]{Schlecker2021}
Schlecker, M., Mordasini, C., Emsenhuber, A., {et~al.} 2021{\natexlab{a}},
  A\&A, 656, A71

\bibitem[{Schlecker {et~al.}(2021{\natexlab{b}})Schlecker, Pham, Burn, Alibert,
  Mordasini, Emsenhuber, Klahr, Henning, \& Mishra}]{Schlecker2021b}
Schlecker, M., Pham, D., Burn, R., {et~al.} 2021{\natexlab{b}}, A\&A, 656, A73

\bibitem[{{Sch{\"o}fer} {et~al.}(2019){Sch{\"o}fer}, {Jeffers}, {Reiners},
  {Shulyak}, {Fuhrmeister}, {Johnson}, {Zechmeister}, {Ribas}, {Quirrenbach},
  {Amado}, {Caballero}, {Anglada-Escud{\'e}}, {Bauer}, {B{\'e}jar},
  {Cort{\'e}s-Contreras}, {Dreizler}, {Guenther}, {Kaminski}, {K{\"u}rster},
  {Lafarga}, {Montes}, {Morales}, {Pedraz}, \& {Tal-Or}}]{Schoefer2019}
{Sch{\"o}fer}, P., {Jeffers}, S.~V., {Reiners}, A., {et~al.} 2019, \aap, 623,
  A44

\bibitem[{{Schweitzer} {et~al.}(2019){Schweitzer}, {Passegger}, {Cifuentes},
  {B{\'e}jar}, {Cort{\'e}s-Contreras}, {Caballero}, {del Burgo}, {Czesla},
  {K{\"u}rster}, {Montes}, {Zapatero Osorio}, {Ribas}, {Reiners},
  {Quirrenbach}, {Amado}, {Aceituno}, {Anglada-Escud{\'e}}, {Bauer},
  {Dreizler}, {Jeffers}, {Guenther}, {Henning}, {Kaminski}, {Lafarga},
  {Marfil}, {Morales}, {Schmitt}, {Seifert}, {Solano}, {Tabernero}, \&
  {Zechmeister}}]{Schweitzer2019}
{Schweitzer}, A., {Passegger}, V.~M., {Cifuentes}, C., {et~al.} 2019, \aap,
  625, A68

\bibitem[{{Seifahrt} {et~al.}(2016){Seifahrt}, {Bean}, {St{\"u}rmer}, {Gers},
  {Grobler}, {Reed}, \& {Jones}}]{Seifahrt2016}
{Seifahrt}, A., {Bean}, J.~L., {St{\"u}rmer}, J., {et~al.} 2016, in Society of
  Photo-Optical Instrumentation Engineers (SPIE) Conference Series, Vol. 9908,
  Ground-based and Airborne Instrumentation for Astronomy VI, ed. C.~J.
  {Evans}, L.~{Simard}, \& H.~{Takami}, 990818

\bibitem[{{Seifahrt} {et~al.}(2020){Seifahrt}, {Bean}, {St{\"u}rmer}, {Kasper},
  {Gers}, {Schwab}, {Zechmeister}, {Stef{\'a}nsson}, {Montet}, {Dos Santos},
  {Peck}, {White}, \& {Tapia}}]{Seifahrt2020}
{Seifahrt}, A., {Bean}, J.~L., {St{\"u}rmer}, J., {et~al.} 2020, in Society of
  Photo-Optical Instrumentation Engineers (SPIE) Conference Series, Vol. 11447,
  Society of Photo-Optical Instrumentation Engineers (SPIE) Conference Series,
  114471F

\bibitem[{{Seifahrt} {et~al.}(2018){Seifahrt}, {St{\"u}rmer}, {Bean}, \&
  {Schwab}}]{Seifahrt18}
{Seifahrt}, A., {St{\"u}rmer}, J., {Bean}, J.~L., \& {Schwab}, C. 2018, in
  Society of Photo-Optical Instrumentation Engineers (SPIE) Conference Series,
  Vol. 10702, Ground-based and Airborne Instrumentation for Astronomy VII, ed.
  C.~J. {Evans}, L.~{Simard}, \& H.~{Takami}, 107026D

\bibitem[{{Shporer} {et~al.}(2020){Shporer}, {Collins}, {Astudillo-Defru},
  {Irwin}, {Bonfils}, {Collins}, {Matthews}, {Winters}, {Anderson},
  {Armstrong}, {Charbonneau}, {Cloutier}, {Daylan}, {Gan}, {G{\"u}nther},
  {Hellier}, {Horne}, {Huang}, {Jensen}, {Kielkopf}, {Palle}, {Sefako},
  {Stassun}, {Tan}, {Vanderburg}, {Ricker}, {Latham}, {Vanderspek}, {Seager},
  {Winn}, {Jenkins}, {Colon}, {Dressing}, {L{\'e}epine}, {Muirhead}, {Rose},
  {Twicken}, \& {Villasenor}}]{shporer2020}
{Shporer}, A., {Collins}, K.~A., {Astudillo-Defru}, N., {et~al.} 2020, \apjl,
  890, L7

\bibitem[{{Skrutskie} {et~al.}(2006){Skrutskie}, {Cutri}, {Stiening},
  {Weinberg}, {Schneider}, {Carpenter}, {Beichman}, {Capps}, {Chester},
  {Elias}, {Huchra}, {Liebert}, {Lonsdale}, {Monet}, {Price}, {Seitzer},
  {Jarrett}, {Kirkpatrick}, {Gizis}, {Howard}, {Evans}, {Fowler}, {Fullmer},
  {Hurt}, {Light}, {Kopan}, {Marsh}, {McCallon}, {Tam}, {Van Dyk}, \&
  {Wheelock}}]{Skrutskie2006}
{Skrutskie}, M.~F., {Cutri}, R.~M., {Stiening}, R., {et~al.} 2006, \aj, 131,
  1163

\bibitem[{{Smith} {et~al.}(2012){Smith}, {Stumpe}, {Van Cleve}, {Jenkins},
  {Barclay}, {Fanelli}, {Girouard}, {Kolodziejczak}, {McCauliff}, {Morris}, \&
  {Twicken}}]{2012PASP..124.1000S}
{Smith}, J.~C., {Stumpe}, M.~C., {Van Cleve}, J.~E., {et~al.} 2012, \pasp, 124,
  1000

\bibitem[{{Snellen} {et~al.}(2013){Snellen}, {de Kok}, {le Poole}, {Brogi}, \&
  {Birkby}}]{Snellen2013}
{Snellen}, I.~A.~G., {de Kok}, R.~J., {le Poole}, R., {Brogi}, M., \& {Birkby},
  J. 2013, \apj, 764, 182

\bibitem[{{Soto} {et~al.}(2021){Soto}, {Anglada-Escud{\'e}}, {Dreizler},
  {Molaverdikhani}, {Kemmer}, {Rodr{\'\i}guez-L{\'o}pez}, {Lillo-Box},
  {Pall{\'e}}, {Espinoza}, {Caballero}, {Quirrenbach}, {Ribas}, {Reiners},
  {Narita}, {Hirano}, {Amado}, {B{\'e}jar}, {Bluhm}, {Burke}, {Caldwell},
  {Charbonneau}, {Cloutier}, {Collins}, {Cort{\'e}s-Contreras}, {Girardin},
  {Guerra}, {Harakawa}, {Hatzes}, {Irwin}, {Jenkins}, {Jensen}, {Kawauchi},
  {Kotani}, {Kudo}, {Kunimoto}, {Kuzuhara}, {Latham}, {Montes}, {Morales},
  {Mori}, {Nelson}, {Omiya}, {Pedraz}, {Passegger}, {Rackham}, {Rudat},
  {Schlieder}, {Sch{\"o}fer}, {Schweitzer}, {Selezneva}, {Stockdale}, {Tamura},
  {Trifonov}, {Vanderspek}, \& {Watanabe}}]{LHS1478}
{Soto}, M.~G., {Anglada-Escud{\'e}}, G., {Dreizler}, S., {et~al.} 2021, \aap,
  649, A144

\bibitem[{{Soubiran} {et~al.}(2018){Soubiran}, {Jasniewicz}, {Chemin},
  {Zurbach}, {Brouillet}, {Panuzzo}, {Sartoretti}, {Katz}, {Le Campion},
  {Marchal}, {Hestroffer}, {Th{\'e}venin}, {Crifo}, {Udry}, {Cropper},
  {Seabroke}, {Viala}, {Benson}, {Blomme}, {Jean-Antoine}, {Huckle}, {Smith},
  {Baker}, {Damerdji}, {Dolding}, {Fr{\'e}mat}, {Gosset}, {Guerrier}, {Guy},
  {Haigron}, {Jan{\ss}en}, {Plum}, {Fabre}, {Lasne}, {Pailler}, {Panem},
  {Riclet}, {Royer}, {Tauran}, {Zwitter}, {Gueguen}, \& {Turon}}]{Soubiran2018}
{Soubiran}, C., {Jasniewicz}, G., {Chemin}, L., {et~al.} 2018, \aap, 616, A7

\bibitem[{{Southworth}(2011)}]{Southworth2011}
{Southworth}, J. 2011, \mnras, 417, 2166

\bibitem[{{Strand} \& {Hall}(1951)}]{Strand1951}
{Strand}, K.~A. \& {Hall}, R.~G., J. 1951, \aj, 56, 106

\bibitem[{{Stumpe} {et~al.}(2014){Stumpe}, {Smith}, {Catanzarite}, {Van Cleve},
  {Jenkins}, {Twicken}, \& {Girouard}}]{Stumpe2014PASP..126..100S}
{Stumpe}, M.~C., {Smith}, J.~C., {Catanzarite}, J.~H., {et~al.} 2014, \pasp,
  126, 100

\bibitem[{{Stumpe} {et~al.}(2012){Stumpe}, {Smith}, {Van Cleve}, {Twicken},
  {Barclay}, {Fanelli}, {Girouard}, {Jenkins}, {Kolodziejczak}, {McCauliff}, \&
  {Morris}}]{Stumpe2012PASP..124..985S}
{Stumpe}, M.~C., {Smith}, J.~C., {Van Cleve}, J.~E., {et~al.} 2012, \pasp, 124,
  985

\bibitem[{{Swain} {et~al.}(2021){Swain}, {Estrela}, {Roudier}, {Sotin},
  {Rimmer}, {Valio}, {West}, {Pearson}, {Huber-Feely}, \& {Zellem}}]{Swain2021}
{Swain}, M.~R., {Estrela}, R., {Roudier}, G.~M., {et~al.} 2021, \aj, 161, 213

\bibitem[{{Szab{\'o}} \& {Kiss}(2011)}]{Szabo2011}
{Szab{\'o}}, G.~M. \& {Kiss}, L.~L. 2011, \apjl, 727, L44

\bibitem[{{Tal-Or} {et~al.}(2018){Tal-Or}, {Zechmeister}, {Reiners}, {Jeffers},
  {Sch{\"o}fer}, {Quirrenbach}, {Amado}, {Ribas}, {Caballero}, {Aceituno},
  {Bauer}, {B{\'e}jar}, {Czesla}, {Dreizler}, {Fuhrmeister}, {Hatzes},
  {Johnson}, {K{\"u}rster}, {Lafarga}, {Montes}, {Morales}, {Reffert},
  {Sadegi}, {Seifert}, \& {Shulyak}}]{Lev2018}
{Tal-Or}, L., {Zechmeister}, M., {Reiners}, A., {et~al.} 2018, \aap, 614, A122

\bibitem[{{Trifonov} {et~al.}(2021){Trifonov}, {Caballero}, {Morales},
  {Seifahrt}, {Ribas}, {Reiners}, {Bean}, {Luque}, {Parviainen}, {Pall{\'e}},
  {Stock}, {Zechmeister}, {Amado}, {Anglada-Escud{\'e}}, {Azzaro}, {Barclay},
  {B{\'e}jar}, {Bluhm}, {Casasayas-Barris}, {Cifuentes}, {Collins}, {Collins},
  {Cort{\'e}s-Contreras}, {de Leon}, {Dreizler}, {Dressing}, {Esparza-Borges},
  {Espinoza}, {Fausnaugh}, {Fukui}, {Hatzes}, {Hellier}, {Henning}, {Henze},
  {Herrero}, {Jeffers}, {Jenkins}, {Jensen}, {Kaminski}, {Kasper},
  {Kossakowski}, {K{\"u}rster}, {Lafarga}, {Latham}, {Mann}, {Molaverdikhani},
  {Montes}, {Montet}, {Murgas}, {Narita}, {Oshagh}, {Passegger}, {Pollacco},
  {Quinn}, {Quirrenbach}, {Ricker}, {Rodr{\'\i}guez L{\'o}pez}, {Sanz-Forcada},
  {Schwarz}, {Schweitzer}, {Seager}, {Shporer}, {Stangret}, {St{\"u}rmer},
  {Tan}, {Tenenbaum}, {Twicken}, {Vanderspek}, \& {Winn}}]{trifonov2021}
{Trifonov}, T., {Caballero}, J.~A., {Morales}, J.~C., {et~al.} 2021, Science,
  371, 1038

\bibitem[{{Trifonov} {et~al.}(2020){Trifonov}, {Tal-Or}, {Zechmeister},
  {Kaminski}, {Zucker}, \& {Mazeh}}]{Trifonov2020}
{Trifonov}, T., {Tal-Or}, L., {Zechmeister}, M., {et~al.} 2020, \aap, 636, A74

\bibitem[{{Turbet} {et~al.}(2020){Turbet}, {Bolmont}, {Ehrenreich}, {Gratier},
  {Leconte}, {Selsis}, {Hara}, \& {Lovis}}]{Turbet2020}
{Turbet}, M., {Bolmont}, E., {Ehrenreich}, D., {et~al.} 2020, \aap, 638, A41

\bibitem[{{Turnpenney} {et~al.}(2018){Turnpenney}, {Nichols}, {Wynn}, \&
  {Burleigh}}]{Turnpenney2018}
{Turnpenney}, S., {Nichols}, J.~D., {Wynn}, G.~A., \& {Burleigh}, M.~R. 2018,
  \apj, 854, 72

\bibitem[{{Valsecchi} {et~al.}(2014){Valsecchi}, {Rasio}, \&
  {Steffen}}]{valsecchi2014}
{Valsecchi}, F., {Rasio}, F.~A., \& {Steffen}, J.~H. 2014, \apjl, 793, L3

\bibitem[{{Vedantham} {et~al.}(2020){Vedantham}, {Callingham}, {Shimwell},
  {Tasse}, {Pope}, {Bedell}, {Snellen}, {Best}, {Hardcastle}, {Haverkorn},
  {Mechev}, {O'Sullivan}, {R{\"o}ttgering}, \& {White}}]{Vedantham2020}
{Vedantham}, H.~K., {Callingham}, J.~R., {Shimwell}, T.~W., {et~al.} 2020,
  Nature Astronomy, 4, 577

\bibitem[{{Vogt}(1992)}]{Vogt1992}
{Vogt}, S.~S. 1992, in European Southern Observatory Conference and Workshop
  Proceedings, Vol.~40, European Southern Observatory Conference and Workshop
  Proceedings, 223

\bibitem[{{Vogt} {et~al.}(1994){Vogt}, {Allen}, {Bigelow}, {Bresee}, {Brown},
  {Cantrall}, {Conrad}, {Couture}, {Delaney}, {Epps}, {Hilyard}, {Hilyard},
  {Horn}, {Jern}, {Kanto}, {Keane}, {Kibrick}, {Lewis}, {Osborne},
  {Pardeilhan}, {Pfister}, {Ricketts}, {Robinson}, {Stover}, {Tucker}, {Ward},
  \& {Wei}}]{Vogt1994}
{Vogt}, S.~S., {Allen}, S.~L., {Bigelow}, B.~C., {et~al.} 1994, in Society of
  Photo-Optical Instrumentation Engineers (SPIE) Conference Series, Vol. 2198,
  Instrumentation in Astronomy VIII, ed. D.~L. {Crawford} \& E.~R. {Craine},
  362

\bibitem[{{Wagman}(1967)}]{Wagman1967}
{Wagman}, N.~E. 1967, \aj, 72, 957

\bibitem[{{Wilson}(1953)}]{Wilson1953}
{Wilson}, R.~E. 1953, Carnegie Institute Washington D.C. Publication, 0

\bibitem[{{Winters} {et~al.}(2022){Winters}, {Cloutier}, {Medina}, {Irwin},
  {Charbonneau}, {Astudillo-Defru}, {Bonfils}, {Howard}, {Isaacson}, {Bean},
  {Seifahrt}, {Teske}, {Eastman}, {Twicken}, {Collins}, {Jensen}, {Quinn},
  {Payne}, {Kristiansen}, {Spencer}, {Vanderburg}, {Zechmeister}, {Weiss},
  {Wang}, {Wang}, {Udry}, {Terentev}, {St{\"u}rmer}, {Stef{\'a}nsson},
  {Shporer}, {Shectman}, {Sefako}, {Schwengeler}, {Schwarz}, {Scarsdale},
  {Rubenzahl}, {Roy}, {Rosenthal}, {Robertson}, {Petigura}, {Pepe},
  {Omohundro}, {Murphy}, {Murgas}, {Mo{\v{c}}nik}, {Montet}, {Mennickent},
  {Mayo}, {Massey}, {Lubin}, {Lovis}, {Lewin}, {Kasper}, {Kane}, {Jenkins},
  {Huber}, {Horne}, {Hill}, {Gorrini}, {Giacalone}, {Fulton}, {Forveille},
  {Figueira}, {Fetherolf}, {Dressing}, {D{\'\i}az}, {Delfosse}, {Dalba}, {Dai},
  {Cort{\'e}s}, {Crossfield}, {Crane}, {Conti}, {Collins}, {Chontos}, {Butler},
  {Brown}, {Brady}, {Behmard}, {Beard}, {Batalha}, \& {Almenara}}]{winters2022}
{Winters}, J.~G., {Cloutier}, R., {Medina}, A.~A., {et~al.} 2022, \aj, 163, 168

\bibitem[{{Wright} {et~al.}(2004){Wright}, {Marcy}, {Butler}, \&
  {Vogt}}]{Wright2004}
{Wright}, J.~T., {Marcy}, G.~W., {Butler}, R.~P., \& {Vogt}, S.~S. 2004, \apjs,
  152, 261

\bibitem[{{Wright} {et~al.}(2011){Wright}, {Drake}, {Mamajek}, \&
  {Henry}}]{wri11}
{Wright}, N.~J., {Drake}, J.~J., {Mamajek}, E.~E., \& {Henry}, G.~W. 2011,
  \apj, 743, 48

\bibitem[{{Zacharias} {et~al.}(2013){Zacharias}, {Finch}, {Girard}, {Henden},
  {Bartlett}, {Monet}, \& {Zacharias}}]{Zacharias2013}
{Zacharias}, N., {Finch}, C.~T., {Girard}, T.~M., {et~al.} 2013, \aj, 145, 44

\bibitem[{{Zechmeister} {et~al.}(2014){Zechmeister}, {Anglada-Escud{\'e}}, \&
  {Reiners}}]{FOX_extraction}
{Zechmeister}, M., {Anglada-Escud{\'e}}, G., \& {Reiners}, A. 2014, \aap, 561,
  A59

\bibitem[{{Zechmeister} \& {K{\"u}rster}(2009{\natexlab{a}})}]{Zechmeister2009}
{Zechmeister}, M. \& {K{\"u}rster}, M. 2009{\natexlab{a}}, \aap, 496, 577

\bibitem[{{Zechmeister} \& {K{\"u}rster}(2009{\natexlab{b}})}]{GLS_paper}
{Zechmeister}, M. \& {K{\"u}rster}, M. 2009{\natexlab{b}}, \aap, 496, 577

\bibitem[{{Zechmeister} {et~al.}(2018){Zechmeister}, {Reiners}, {Amado},
  {Azzaro}, {Bauer}, {B{\'e}jar}, {Caballero}, {Guenther}, {Hagen}, {Jeffers},
  {Kaminski}, {K{\"u}rster}, {Launhardt}, {Montes}, {Morales}, {Quirrenbach},
  {Reffert}, {Ribas}, {Seifert}, {Tal-Or}, \& {Wolthoff}}]{SERVAL}
{Zechmeister}, M., {Reiners}, A., {Amado}, P.~J., {et~al.} 2018, \aap, 609, A12

\bibitem[{{Zeng} {et~al.}(2019){Zeng}, {Jacobsen}, {Sasselov}, {Petaev},
  {Vanderburg}, {Lopez-Morales}, {Perez-Mercader}, {Mattsson}, {Li}, {Heising},
  {Bonomo}, {Damasso}, {Berger}, {Cao}, {Levi}, \& {Wordsworth}}]{Zeng2019}
{Zeng}, L., {Jacobsen}, S.~B., {Sasselov}, D.~D., {et~al.} 2019, Proceedings of
  the National Academy of Science, 116, 9723

\end{thebibliography}


\newpage 

\appendix

\section{Additional stellar parameters}

\begin{table}[!ht]
\centering
\small
\caption{Multiband photometry of Gl~806$^a$.} 
\label{tab:phot}
\begin{tabular}{lcr}
\hline
\hline
\noalign{\smallskip}
Band & Magnitude & Reference \\
    & [mag] & \\
\noalign{\smallskip}
\hline
\noalign{\smallskip}
$B_T$   & $12.370\pm0.154$      & TYC \\
$B$     & $12.495\pm0.010$      & UCAC4 \\
$g'$    & $11.610\pm0.010$      & UCAC4 \\
$G_{BP}$& $11.0059\pm0.0029$    & {\it Gaia} EDR3 \\
$V_T$   & $10.974\pm0.061$      & TYC \\
$V$     & $10.704\pm0.010$      & UCAC4 \\
$r'$    & $10.137\pm0.010$      & UCAC4 \\
$G$     & $ 9.8287\pm0.0028$    & {\it Gaia} EDR3 \\
$i'$    & $ 9.127\pm0.010$      & UCAC4 \\
$G_{RP}$& $ 8.7507\pm0.0038$    & {\it Gaia} EDR3 \\
$J$     & $7.329\pm0.018$       & 2MASS \\
$H$     & $6.769\pm0.023$       & 2MASS \\
$K_s$   & $6.533\pm0.016$       & 2MASS \\
$W1$    & $6.409\pm0.075$       & AllWISE \\
$W2$    & $6.169\pm0.027$       & AllWISE \\
$W3$    & $6.239\pm0.016$       & AllWISE \\
$W4$    & $6.296\pm0.052$       & AllWISE \\
\noalign{\smallskip}
\hline
\end{tabular}
\tablebib{
TYC: Tycho-2, \citet{Hog2000};
2MASS: Two Micron All-Sky Survey, \citet{Skrutskie2006};
UCAC4: The Fourth US Naval Observatory CCD Astrograph Catalog, \citet{Zacharias2013};
AllWISE: Wide-field Infrared Survey Explorer, \citet{Cutri2014};
{\em Gaia} EDR3: \citet{2018A&A...616A...1G}.
}
\end{table}


\section{RV data}

\begin{table*}
\centering
\small
\caption{CARMENES Radial velocity measurements and spectroscopic activity indicators for GJ 806 from optical spectra.} 
\label{tab:RV_table}
\begin{tabular}{l c c c c c c c}
\hline
\hline
        \noalign{\smallskip}
        BJD & RV & CRX & dLW & H$\alpha$ & Na~{\sc i}~D$_1$ & Na~{\sc i}~D$_2$ & Ca~{\sc ii}~IRT \\
         & (m\,s$^{-1}$) & (m\,s$^{-1}$\,Np$^{-1}$) & (m$^{2}$\,s$^{-2}$) &  (m\,s$^{-1}$) & (m\,s$^{-1}$) & (m\,s$^{-1}$) & (m\,s$^{-1}$) \\
        \noalign{\smallskip}
        \hline
        \noalign{\smallskip}
2457501.67419 & $1.5\pm2.4$  & $35.0\pm12.0$  & $-9.2\pm1.4$  & $0.8246\pm0.0016$  & $0.1616\pm0.003$  & $0.1704\pm0.0031$  & $0.559\pm0.0014$  \\ 
2457535.64118 & $0.9\pm1.6$  & $40.0\pm17.0$  & $-14.1\pm2.0$  & $0.8234\pm0.002$  & $0.1638\pm0.0038$  & $0.1716\pm0.0039$  & $0.5593\pm0.0017$  \\ 
2457564.63377 & $-1.7\pm1.4$  & $31.0\pm12.0$  & $3.9\pm1.7$  & $0.8244\pm0.0012$  & $0.1869\pm0.0017$  & $0.1819\pm0.0017$  & $0.5741\pm0.0012$  \\ 
2457587.47999 & $-0.4\pm1.5$  & $24.0\pm10.0$  & $1.4\pm1.2$  & $0.8176\pm0.0011$  & $0.1732\pm0.0018$  & $0.1743\pm0.0018$  & $0.56998\pm0.0009$  \\ 
2457605.4934 & $2.2\pm1.5$  & $17.0\pm14.0$  & $-11.1\pm1.7$  & $0.8346\pm0.0019$  & $0.1719\pm0.0035$  & $0.1718\pm0.0035$  & $0.5685\pm0.0016$  \\ 
2457617.45895 & $-0.5\pm1.7$  & $-18.0\pm15.0$  & $-18.0\pm1.8$  & $0.8196\pm0.0024$  & $0.15\pm0.0047$  & $0.1664\pm0.0048$  & $0.5583\pm0.0022$  \\ 
2457652.49537 & $-9.3\pm1.6$  & $8.0\pm15.0$  & $-10.0\pm1.8$  & $0.829\pm0.0019$  & $0.1638\pm0.0033$  & $0.1753\pm0.0033$  & $0.562\pm0.0018$  \\ 
2457655.36104 & $3.4\pm2.7$  & $-18.0\pm27.0$  & $-24.6\pm3.9$  & $0.8269\pm0.0034$  & $0.1766\pm0.0083$  & $0.1671\pm0.0085$  & $0.5651\pm0.0029$  \\ 
2457678.30743 & $-9.2\pm1.4$  & $9.0\pm12.0$  & $-11.7\pm2.8$  & $0.826\pm0.0019$  & $0.157\pm0.0035$  & $0.1585\pm0.0036$  & $0.5712\pm0.0017$  \\ 
2457709.38972 & $0.3\pm1.4$  & $-4.0\pm12.0$  & $-0.1\pm1.6$  & $0.8215\pm0.0016$  & $0.1999\pm0.0029$  & $0.1872\pm0.0029$  & $0.5699\pm0.0014$  \\ 
2457754.29392 & $1.1\pm1.1$  & $32.0\pm10.0$  & $-3.0\pm1.9$  & $0.8252\pm0.0014$  & $0.1829\pm0.0029$  & $0.1707\pm0.0028$  & $0.5735\pm0.0013$  \\ 
2458093.39094 & $-0.3\pm1.1$  & $3.9\pm9.4$  & $8.14\pm0.83$  & $0.82167\pm0.00095$  & $0.187\pm0.0018$  & $0.1881\pm0.0017$  & $0.56646\pm0.00083$  \\ 
2458205.69105 & $3.9\pm1.2$  & $-102.0\pm75.0$  & $-22.8\pm8.3$  & $0.812\pm0.01$  & $0.103\pm0.044$  & $0.155\pm0.042$  & $0.5633\pm0.0085$  \\ 
2458263.52974 & $-3.5\pm2.0$  & $-2.0\pm8.8$  & $5.3\pm0.98$  & $0.816\pm0.0011$  & $0.167\pm0.0016$  & $0.1774\pm0.0016$  & $0.5648\pm0.00098$  \\ 
2458295.60007 & $3.1\pm1.3$  & $12.0\pm13.0$  & $4.5\pm1.4$  & $0.8298\pm0.0012$  & $0.1939\pm0.0018$  & $0.184\pm0.0018$  & $0.5634\pm0.0011$  \\ 
2458322.48587 & $-2.4\pm1.2$  & $2.0\pm11.0$  & $6.3\pm1.5$  & $0.8425\pm0.0011$  & $0.1912\pm0.0017$  & $0.1868\pm0.0017$  & $0.57471\pm0.00099$  \\ 
2458349.41466 & $-1.9\pm1.2$  & $10.5\pm9.1$  & $4.2\pm1.1$  & $0.8201\pm0.0011$  & $0.1768\pm0.0017$  & $0.1762\pm0.0017$  & $0.56799\pm0.00098$  \\ 
2458392.39999 & $-0.5\pm1.2$  & $-9.0\pm12.0$  & $3.5\pm1.5$  & $0.8364\pm0.0011$  & $0.1783\pm0.0017$  & $0.1848\pm0.0017$  & $0.57408\pm0.00098$  \\ 
2458537.7464 & $-2.2\pm1.1$  & $11.0\pm10.0$  & $5.8\pm1.2$  & $0.834\pm0.0011$  & $0.1766\pm0.0016$  & $0.1879\pm0.0016$  & $0.56599\pm0.001$  \\ 
2458619.65315 & $-7.6\pm1.2$  & $1.9\pm8.8$  & $17.8\pm2.1$  & $0.8361\pm0.0011$  & $0.1988\pm0.0018$  & $0.1964\pm0.0018$  & $0.5758\pm0.001$  \\ 
2458644.60548 & $-0.0\pm1.7$  & $15.0\pm12.0$  & $8.7\pm1.7$  & $0.8239\pm0.0012$  & $0.1893\pm0.0018$  & $0.1826\pm0.0018$  & $0.5712\pm0.001$  \\ 
2458664.51735 & $8.0\pm1.2$  & $5.0\pm15.0$  & $-5.0\pm2.0$  & $0.8213\pm0.0021$  & $0.1684\pm0.0041$  & $0.1729\pm0.0041$  & $0.5693\pm0.0018$  \\ 
2458678.54766 & $2.3\pm2.0$  & $29.0\pm10.0$  & $6.9\pm1.2$  & $0.8134\pm0.0011$  & $0.1814\pm0.0018$  & $0.1784\pm0.0017$  & $0.5682\pm0.001$  \\ 
2458681.49083 & $0.2\pm1.1$  & $-4.0\pm11.0$  & $-2.0\pm1.5$  & $0.8323\pm0.0015$  & $0.175\pm0.0026$  & $0.1773\pm0.0026$  & $0.5729\pm0.0013$  \\ 
2458686.64815 & $-4.0\pm1.9$  & $-22.1\pm7.7$  & $7.9\pm1.3$  & $0.8189\pm0.0011$  & $0.177\pm0.0017$  & $0.1805\pm0.0017$  & $0.5684\pm0.001$  \\ 
2458690.44247 & $11.7\pm1.3$  & $-6.6\pm9.2$  & $7.7\pm1.5$  & $0.8248\pm0.0011$  & $0.1867\pm0.0016$  & $0.1863\pm0.0016$  & $0.57308\pm0.00098$  \\ 
2458694.38266 & $0.5\pm1.3$  & $11.0\pm11.0$  & $5.3\pm1.9$  & $0.8376\pm0.0011$  & $0.1935\pm0.0018$  & $0.1935\pm0.0018$  & $0.5702\pm0.001$  \\ 
2458697.61517 & $5.9\pm1.3$  & $-9.0\pm11.0$  & $6.9\pm1.8$  & $0.8486\pm0.0011$  & $0.1986\pm0.0017$  & $0.1933\pm0.0017$  & $0.57839\pm0.00096$  \\ 
2458700.46186 & $-4.1\pm1.1$  & $18.2\pm9.0$  & $6.4\pm1.7$  & $0.846\pm0.0011$  & $0.1986\pm0.0017$  & $0.1953\pm0.0016$  & $0.57662\pm0.00098$  \\ 
2458706.46897 & $-3.5\pm1.5$  & $5.2\pm7.6$  & $6.8\pm1.5$  & $0.8436\pm0.0011$  & $0.1887\pm0.0017$  & $0.1878\pm0.0017$  & $0.57566\pm0.00098$  \\ 
2458710.42108 & $-2.8\pm1.2$  & $-5.0\pm11.0$  & $7.2\pm1.2$  & $0.8169\pm0.0011$  & $0.1726\pm0.0016$  & $0.1784\pm0.0016$  & $0.55978\pm0.00097$  \\ 
2458710.54438 & $-0.4\pm1.3$  & $-10.7\pm9.3$  & $7.07\pm0.98$  & $0.8328\pm0.0011$  & $0.1822\pm0.0017$  & $0.1896\pm0.0017$  & $0.57106\pm0.00099$  \\ 
2458714.37356 & $-3.7\pm1.6$  & $13.0\pm10.0$  & $7.5\pm1.3$  & $0.8153\pm0.001$  & $0.1733\pm0.0015$  & $0.1795\pm0.0015$  & $0.56709\pm0.00097$  \\ 
2458904.74432 & $-9.4\pm1.7$  & $-11.0\pm11.0$  & $6.5\pm1.1$  & $0.829\pm0.0011$  & $0.1767\pm0.0018$  & $0.1797\pm0.0018$  & $0.5702\pm0.001$  \\ 
2458921.68833 & $6.1\pm1.4$  & $21.0\pm15.0$  & $38.9\pm2.4$  & $0.8356\pm0.0016$  & $0.1893\pm0.0029$  & $0.1911\pm0.0029$  & $0.5758\pm0.0014$  \\ 
2459064.49425 & $-3.4\pm1.1$  & $-4.0\pm11.0$  & $4.2\pm1.4$  & $0.8567\pm0.0011$  & $0.1818\pm0.0018$  & $0.1903\pm0.0018$  & $0.5757\pm0.001$  \\ 
2459065.52916 & $-5.0\pm1.2$  & $-12.7\pm7.5$  & $5.4\pm1.2$  & $0.8249\pm0.0011$  & $0.1697\pm0.0016$  & $0.1801\pm0.0016$  & $0.56627\pm0.00095$  \\ 
2459103.41552 & $3.4\pm1.5$  & $-2.8\pm8.4$  & $1.8\pm1.4$  & $0.8235\pm0.0011$  & $0.1687\pm0.0016$  & $0.1766\pm0.0017$  & $0.55952\pm0.00097$  \\ 
2459147.29728 & $0.2\pm1.2$  & $-2.2\pm9.6$  & $1.3\pm1.2$  & $0.82597\pm0.00099$  & $0.1614\pm0.0014$  & $0.1778\pm0.0014$  & $0.56499\pm0.00088$  \\ 
2459414.61833 & $8.3\pm1.3$  & $-1.6\pm9.2$  & $2.8\pm1.2$  & $0.82827\pm0.001$  & $0.179\pm0.0015$  & $0.19\pm0.0015$  & $0.56804\pm0.00092$  \\ 
2459425.53878 & $0.3\pm1.5$  & $25.5\pm9.2$  & $-3.3\pm1.4$  & $0.8283\pm0.001$  & $0.1539\pm0.0015$  & $0.1671\pm0.0015$  & $0.56217\pm0.00092$  \\ 
2459469.40333 & $5.5\pm1.5$  & $-1.0\pm12.0$  & $-6.0\pm1.3$  & $0.8232\pm0.0014$  & $0.1617\pm0.0022$  & $0.1683\pm0.0021$  & $0.5564\pm0.0012$  \\ 
2459490.43845 & $-8.7\pm1.7$  & $-10.0\pm12.0$  & $4.1\pm1.4$  & $0.82253\pm0.00097$  & $0.1765\pm0.0014$  & $0.1797\pm0.0014$  & $0.57058\pm0.00088$  \\ 
2459495.31045 & $6.3\pm1.3$  & $17.0\pm11.0$  & $-12.3\pm1.8$  & $0.8348\pm0.0014$  & $0.1583\pm0.0022$  & $0.1656\pm0.0022$  & $0.5644\pm0.0013$  \\ 
2459495.40711 & $6.2\pm1.4$  & $13.0\pm9.5$  & $1.0\pm1.1$  & $0.82395\pm0.00082$  & $0.1612\pm0.0011$  & $0.1691\pm0.0011$  & $0.56218\pm0.00074$  \\ 
2459495.56347 & $4.2\pm1.5$  & $10.0\pm11.0$  & $-0.2\pm1.0$  & $0.82128\pm0.00088$  & $0.1642\pm0.0012$  & $0.1681\pm0.0012$  & $0.56283\pm0.00081$  \\ 
2459501.32239 & $-2.0\pm1.5$  & $-3.0\pm13.0$  & $-6.3\pm1.3$  & $0.8265\pm0.001$  & $0.1737\pm0.0019$  & $0.1764\pm0.0019$  & $0.565\pm0.00094$  \\ 
2459501.41235 & $-2.8\pm1.4$  & $5.5\pm9.2$  & $-0.8\pm1.2$  & $0.83275\pm0.001$  & $0.169\pm0.0014$  & $0.1797\pm0.0015$  & $0.56613\pm0.00088$  \\ 
2459501.49202 & $-1.1\pm1.5$  & $8.0\pm8.0$  & $2.5\pm1.1$  & $0.82788\pm0.00087$  & $0.1762\pm0.0012$  & $0.1808\pm0.0012$  & $0.56605\pm0.0008$  \\ 
2459502.29233 & $-7.1\pm1.2$  & $21.8\pm9.3$  & $-4.0\pm1.1$  & $0.8299\pm0.0011$  & $0.1766\pm0.0018$  & $0.1778\pm0.0018$  & $0.56618\pm0.00097$  \\ 
2459503.29509 & $-9.8\pm1.6$  & $2.6\pm7.5$  & $4.43\pm0.98$  & $0.82592\pm0.0009$  & $0.173\pm0.0013$  & $0.1776\pm0.0013$  & $0.56558\pm0.0008$  \\ 
2459503.39991 & $-7.5\pm1.6$  & $-3.0\pm8.5$  & $2.2\pm1.2$  & $0.82809\pm0.00098$  & $0.1723\pm0.0014$  & $0.1766\pm0.0014$  & $0.56707\pm0.00086$  \\ 
2459503.48331 & $-4.4\pm1.6$  & $-1.8\pm8.7$  & $2.4\pm1.2$  & $0.82656\pm0.00096$  & $0.1715\pm0.0013$  & $0.1792\pm0.0014$  & $0.56895\pm0.00089$  \\ 
2459505.29227 & $-1.2\pm2.0$  & $-11.8\pm8.3$  & $-4.2\pm1.6$  & $0.8296\pm0.0013$  & $0.186\pm0.0021$  & $0.1892\pm0.0021$  & $0.5674\pm0.0011$  \\ 
2459505.38371 & $-1.0\pm1.9$  & $13.9\pm9.5$  & $0.2\pm1.4$  & $0.8206\pm0.001$  & $0.175\pm0.0015$  & $0.1874\pm0.0016$  & $0.56845\pm0.00089$  \\ 
2459505.45737 & $3.8\pm2.0$  & $16.4\pm7.7$  & $2.2\pm1.1$  & $0.82318\pm0.00092$  & $0.1821\pm0.0013$  & $0.1937\pm0.0013$  & $0.56131\pm0.00084$  \\ 
2459506.28726 & $4.1\pm1.6$  & $-26.0\pm8.7$  & $0.7\pm1.1$  & $0.8189\pm0.00098$  & $0.1779\pm0.0015$  & $0.1869\pm0.0015$  & $0.55876\pm0.00091$  \\ 
2459506.43675 & $6.1\pm1.6$  & $6.8\pm9.1$  & $2.8\pm1.2$  & $0.83573\pm0.00093$  & $0.1788\pm0.0014$  & $0.1919\pm0.0014$  & $0.56093\pm0.00082$  \\ 
2459506.49778 & $8.0\pm1.6$  & $0.8\pm7.8$  & $3.5\pm1.2$  & $0.83057\pm0.0009$  & $0.1753\pm0.0013$  & $0.1857\pm0.0013$  & $0.55936\pm0.00083$  \\ 
2459509.3356 & $8.0\pm1.6$  & $-2.5\pm9.0$  & $1.3\pm1.3$  & $0.83074\pm0.00096$  & $0.1849\pm0.0016$  & $0.1908\pm0.0016$  & $0.56743\pm0.00086$  \\ 
2459509.41194 & $7.9\pm1.5$  & $-19.0\pm12.0$  & $-3.7\pm1.1$  & $0.8343\pm0.0012$  & $0.1809\pm0.0019$  & $0.1887\pm0.0019$  & $0.5682\pm0.0011$  \\ 
2459509.4965 & $5.7\pm1.6$  & $-14.0\pm9.1$  & $3.2\pm1.2$  & $0.82631\pm0.00091$  & $0.1826\pm0.0013$  & $0.1948\pm0.0013$  & $0.56783\pm0.00083$  \\ 
2459512.36055 & $-6.2\pm1.9$  & $-16.0\pm12.0$  & $-1.7\pm1.3$  & $0.8299\pm0.0012$  & $0.2013\pm0.0021$  & $0.2001\pm0.002$  & $0.5609\pm0.0011$  \\ 
2459512.44856 & $-0.2\pm1.9$  & $-30.8\pm8.2$  & $-2.4\pm1.3$  & $0.8271\pm0.0012$  & $0.172\pm0.0018$  & $0.1757\pm0.0018$  & $0.5606\pm0.0011$  \\ 
2459512.53415 & $1.0\pm2.2$  & $-10.7\pm8.9$  & $-5.0\pm1.1$  & $0.8281\pm0.0011$  & $0.1831\pm0.0018$  & $0.1809\pm0.0018$  & $0.5607\pm0.001$  \\ 
2459514.28808 & $-2.9\pm3.0$  & $-45.0\pm15.0$  & $-8.7\pm1.5$  & $0.8321\pm0.0015$  & $0.2521\pm0.004$  & $0.2208\pm0.0038$  & $0.5598\pm0.0013$  \\ 
2459516.38193 & $-9.3\pm2.9$  & $-9.0\pm17.0$  & $-19.0\pm2.5$  & $0.8291\pm0.0024$  & $0.2102\pm0.005$  & $0.1894\pm0.005$  & $0.5583\pm0.002$  \\ 
    \hline
\end{tabular}
\end{table*}


\newpage
\section{TESS individual transit light curves}

\begin{figure*}
\includegraphics[width=0.8\linewidth]{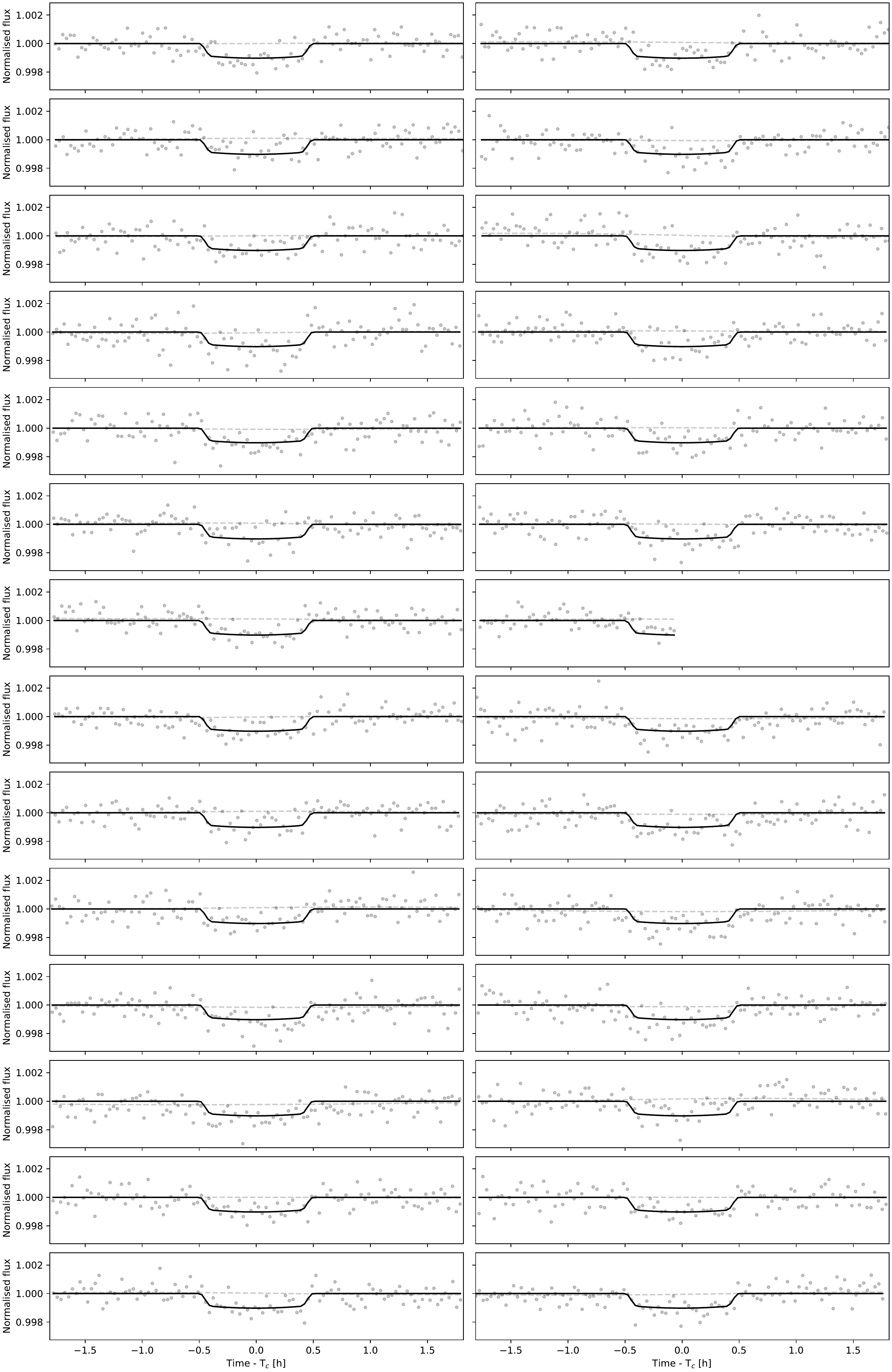}
\caption{All individual transits from TESS}
\label{fig:indiv_transits}
\end{figure*}


\newpage





\begin{figure*}
\includegraphics[width=1\linewidth]{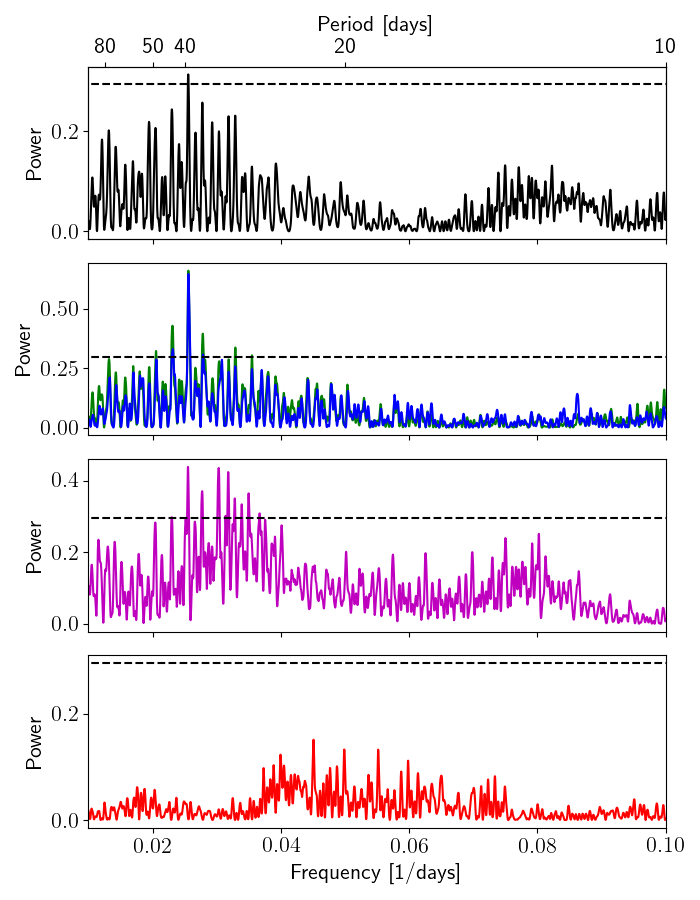}
\caption{GLS of GJ 806 for different detrended time series:
The black line (top panel) denotes the time series of pEW(H$\alpha$), the blue and the green line (second panel from top)
the bluest and the middle line of the \ion{Ca}{ii} IRT, respectively, the violet (third panel from top)
line the TiO index, and the red line (bottom panel) the window function. In all panels, the dashed horizontal
line marks the FAP of 0.01 above which we rate signals as significant.}
\label{fig:rothalpha}
\end{figure*}




\newpage
\section{Joint fit additional material}

\begin{table*}
\caption[width=\textwidth]{
\label{tab:Priors_TABLE3}
Parameter prior functions used in the models showed in Table\,\ref{tab:Semiampl}. The prior labels $\mathcal{F}$, $\mathcal{U}$, $\mathcal{N}$, $\mathcal{J}$ represent fixed, uniform, normal and Jeffreys distributions, respectively. Central time of transit ($t_0$) units are BJD\,$-$\,2457000.
}
\centering

\begin{tabular}{llr}
\hline \hline
\vspace{-0.25cm} \\
Parameter &  Prior & Description \vspace{0.05cm}\\
\hline
\vspace{-0.25cm} \\

\multicolumn{3}{c}{\textit{ Planet parameters }} \vspace{0.15cm} \\

$P_b$ [d] & $\mathcal{N}(0.926323,10^{-5})$ & Period of planet b  \vspace{0.05cm}\\
$t_{0,b}$\,$^{(a)}$ & $\mathcal{N}(2445.5737,0.0005)$ & Central time of transit of planet b \vspace{0.05cm}\\
$K_b$ [$\mathrm{m\,s^{-1}}$] & $\mathcal{U}(0.0,20.0)$ & RV semi-amplitude of planet b \vspace{0.05cm}\\
$e_b$ & $\mathcal{F}(0.0)$ & Eccentricity of planet b \vspace{0.05cm}\\
$\omega_b$ & $\mathcal{F}(90.0)$ & Argument of periastron of planet b \vspace{0.15cm}\\

$P_c$ [d] & $\mathcal{U}(5,7)$ & Period of planet c  \vspace{0.05cm}\\
$t_{0,c}$\,$^{(a)}$ & $\mathcal{U}(2420.5,2424.0)$ & Central time of transit of planet c \vspace{0.05cm}\\
$K_c$ [$\mathrm{m\,s^{-1}}$] & $\mathcal{U}(0.0,20.0)$ & RV semi-amplitude of planet c \vspace{0.05cm}\\
$e_c$ & $\mathcal{F}(0.0)$ & Eccentricity of planet c \vspace{0.05cm}\\
$\omega_c$ & $\mathcal{F}(90.0)$ & Argument of periastron of planet c \vspace{0.15cm}\\

$P_{13d}$ [d] & $\mathcal{U}(12,14)$ & Period of 13\,d signal as a planet  \vspace{0.05cm}\\
$t_{0,13d}$\,$^{(a)}$ & $\mathcal{U}(2414.0,2420.0)$ & Central time of transit of 13\,d signal \vspace{0.05cm}\\
$K_{13d}$ [$\mathrm{m\,s^{-1}}$] & $\mathcal{U}(0.0,20.0)$ & RV semi-amplitude of 13\,d signal \vspace{0.05cm}\\
$e_{13d}$ & $\mathcal{F}(0.0)$ & Eccentricity of 13\,d signal \vspace{0.05cm}\\
$\omega_{13d}$ & $\mathcal{F}(90.0)$ & Argument of periastron of 13\,d signal \vspace{0.15cm}\\

\multicolumn{3}{c}{\textit{ Quasi-periodic GP parameters }} \vspace{0.15cm} \\

P$_{\mathrm{rot}}$ [d] & $\mathcal{U}(12,14)$ & GP rotational period  \vspace{0.05cm}\\
B$_{GP}$ [$\mathrm{m\,s^{-1}}$] & $\mathcal{J}(0.1,100)$ & GP hyperparameter B  \vspace{0.05cm}\\
C$_{GP}$ [$\mathrm{m\,s^{-1}}$] & $\mathcal{J}(10^{-10},100)$ & GP hyperparameter C  \vspace{0.05cm}\\
L$_{GP}$ [d] & $\mathcal{J}(10^{-10},100)$ & GP hyperparameter L  \vspace{0.15cm}\\

\multicolumn{3}{c}{\textit{ Exponential Sinus Squared GP parameters }} \vspace{0.15cm} \\

P$_{\mathrm{rot}}$ [d] & $\mathcal{U}(12,14)$ & GP rotational period  \vspace{0.05cm}\\
$\sigma_{GP}$ [$\mathrm{m\,s^{-1}}$] & $\mathcal{J}(0.1,100)$ & GP hyperparameter $\sigma$  \vspace{0.05cm}\\
$\alpha_{GP}$ [d$^{-2}$] & $\mathcal{J}(10^{-10},100)$ & GP hyperparameter $\alpha$  \vspace{0.05cm}\\
$\gamma_{GP}$ & $\mathcal{J}(10^{-3},100)$ & GP hyperparameter $\gamma$  \vspace{0.15cm}\\

\multicolumn{3}{c}{\textit{ Instrument parameters }} \vspace{0.15cm} \\

$\gamma_{CARMENES}$ [$\mathrm{m\,s^{-1}}$] & $\mathcal{U}(-10.0,10.0)$ & Systemic velocity for CARMENES \vspace{0.05cm}\\
$\sigma_{CARMENES}$ [$\mathrm{m\,s^{-1}}$] & $\mathcal{J}(0.1,10)$ & Extra jitter term for CARMENES \vspace{0.05cm}\\

$\gamma_{HIRES}$ [$\mathrm{m\,s^{-1}}$] & $\mathcal{U}(-10.0,10.0)$ & Systemic velocity for HIRES \vspace{0.05cm}\\
$\sigma_{HIRES}$ [$\mathrm{m\,s^{-1}}$] & $\mathcal{J}(0.1,10)$ & Extra jitter term for HIRES \vspace{0.05cm}\\

$\gamma_{MAROON-X_{red}}$ [$\mathrm{m\,s^{-1}}$] & $\mathcal{U}(-10.0,10.0)$ & Systemic velocity for MAROON-X red arm \vspace{0.05cm}\\
$\sigma_{MAROON-X_{red}}$ [$\mathrm{m\,s^{-1}}$] & $\mathcal{J}(0.1,10)$ & Extra jitter term for MAROON-X red arm \vspace{0.05cm}\\

$\gamma_{MAROON-X_{blue}}$ [$\mathrm{m\,s^{-1}}$] & $\mathcal{U}(-10.0,10.0)$ & Systemic velocity for MAROON-X blue arm \vspace{0.05cm}\\
$\sigma_{MAROON-X_{blue}}$ [$\mathrm{m\,s^{-1}}$] & $\mathcal{J}(0.1,10)$ & Extra jitter term for MAROON-X blue arm \vspace{0.05cm}\\
\hline
\hline
\end{tabular}
\tablefoot{ $^{(a)}$ Central time of transit ($t_0$) units are BJD\,$-$\,2457000. }
\end{table*}

\begin{figure*}[ht!]
\includegraphics[width=1\linewidth]{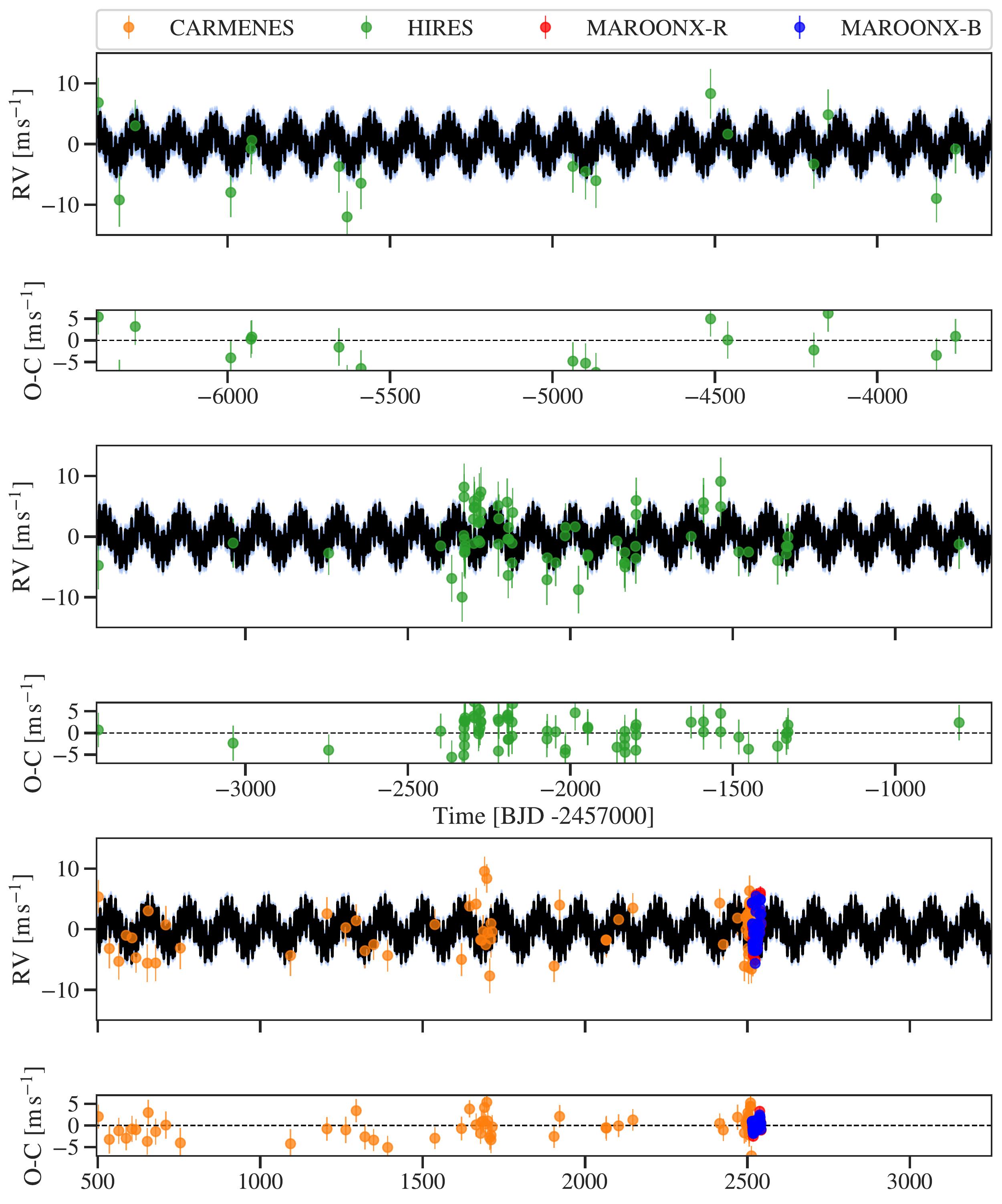}
\caption{RV results from the joint fit. CARMENES (orange), HIRES (green), MAROON-X red channel (red) and blue channel (blue) along with the best-fit model (black line) and the $3\sigma$ confidence interval (shaded light blue area). 
\label{fig: RV + MODEL JF}}
\end{figure*}

\begin{figure*}[ht!]
\includegraphics[width=1\linewidth]{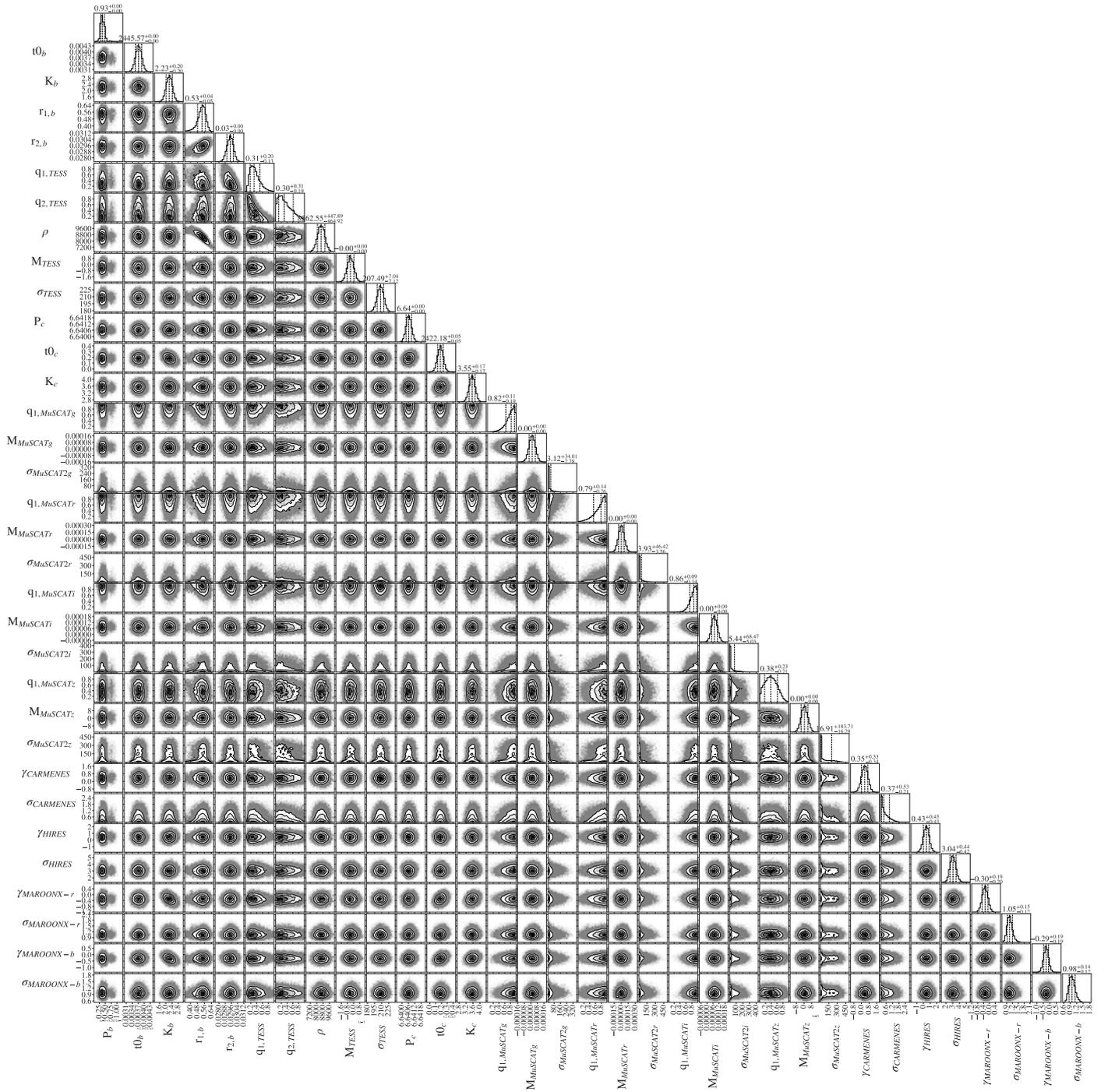}
\caption{Corner plot for the orbital parameters fitted with \texttt{juliet} using the joint fit model. The plot makes use of \texttt{corner.py} package (\citealp{CORNER_PLOT}). 
\label{fig: JF CORNER PLOT}}
\end{figure*}

\begin{table*}
\caption[width=\textwidth]{
\label{table - Joint Fit PRIORS}
Parameter prior functions used in the \texttt{juliet} joint fit model for GJ~806. The prior labels $\mathcal{F}$, $\mathcal{U}$, $\mathcal{N}$, $\mathcal{J}$ represent fixed, uniform, normal and Jeffreys distributions, respectively.
The parametrization for $(p,b)$ using $(r_1,r_2)$ \citep{Espinoza2018} and $(q_1,q_2)$ quadratic limb darkening \citep{Kipping2013} are both explained in Section\,\ref{sec:jfit}. Central time of transit ($t_0$) units are BJD\,$-$\,2457000.
}
\centering

\begin{tabular}{llr}
\hline \hline
\vspace{-0.25cm} \\
Parameter &  Prior & Description \vspace{0.05cm}\\
\hline
\vspace{-0.25cm} \\
\multicolumn{3}{c}{\textit{ Stellar parameters }} \vspace{0.15cm} \\
$\rho_{\star}$ [kg\,m$^{-3}$] & $\mathcal{N}(8200,500)$ & Stellar density \vspace{0.05cm}\\
\vspace{-0.25cm} \\

\multicolumn{3}{c}{\textit{ Planet parameters }} \vspace{0.15cm} \\

$P_b$ [d] & $\mathcal{N}(0.9263,0.001)$ & Period of planet b  \vspace{0.05cm}\\
$t_{0,b}$\,$^{(a)}$ & $\mathcal{N}(2445.57,0.1)$ & Central time of transit of planet b \vspace{0.05cm}\\
$K_b$ [$\mathrm{m\,s^{-1}}$] & $\mathcal{U}(0.0,20.0)$ & RV semi-amplitude of planet b \vspace{0.05cm}\\
$e_b$ & $\mathcal{F}(0.0)$ & Eccentricity of planet b \vspace{0.05cm}\\
$\omega_b$ & $\mathcal{F}(90.0)$ & Argument of periastron of planet b \vspace{0.05cm}\\
$r_{1,b}$ & $\mathcal{U}(0.0,1.0)$ & Parametrization for $p$ and $b$ for planet b \vspace{0.05cm}\\
$r_{2,b}$ & $\mathcal{U}(0.0,1.0)$ & Parametrization for $p$ and $b$ for planet b \vspace{0.05cm} \\

$P_c$ [d] & $\mathcal{N}(6.6,0.1)$ & Period of planet c  \vspace{0.05cm}\\
$t_{0,c}$\,$^{(a)}$ & $\mathcal{U}(2420.5,2424.0)$ & Central time of transit of planet c \vspace{0.05cm}\\
$K_c$ [$\mathrm{m\,s^{-1}}$] & $\mathcal{U}(0.0,20.0)$ & RV semi-amplitude of planet c \vspace{0.05cm}\\
$e_c$ & $\mathcal{F}(0.0)$ & Eccentricity of planet c \vspace{0.05cm}\\
$\omega_c$ & $\mathcal{F}(90.0)$ & Argument of periastron of planet c \vspace{0.05cm}\\

\multicolumn{3}{c}{\textit{ Photometry parameters }} \vspace{0.15cm} \\

$D$ (ppm) & $\mathcal{F}(1.0)$ & Dilution factor for \textit{TESS} and MuSCAT2 \vspace{0.05cm}\\

$q_{1,\textit{TESS}}$ & $\mathcal{U}(0.0,1.0)$ & Quadratic limb darkening parametrization for \textit{TESS} \vspace{0.05cm} \\
$q_{2,\textit{TESS}}$ & $\mathcal{U}(0.0,1.0)$ & Quadratic limb darkening parametrization for \textit{TESS} \vspace{0.05cm} \\
$M_{\textit{TESS}}$ (ppm) & $\mathcal{N}(0.0,0.1)$ & Relative flux offset for \textit{TESS} \vspace{0.05cm}\\
$\sigma_{\textit{TESS}}$ (ppm) & $\mathcal{J}(0.1,10^3)$ & Extra jitter term for \textit{TESS} \vspace{0.05cm}\\

$q_{1,\textit{MuSCAT2-g}}$ & $\mathcal{U}(0.0,1.0)$ & Linear limb darkening parametrization for MuSCAT2-g \vspace{0.05cm} \\
$M_{\textit{MuSCAT2-g}}$ (ppm) & $\mathcal{N}(0.0,0.1)$ & Relative flux offset for MuSCAT2-g \vspace{0.05cm}\\
$\sigma_{\textit{MuSCAT2-g}}$ (ppm) & $\mathcal{J}(0.1,10^3)$ & Extra jitter term for MuSCAT2-g \vspace{0.05cm}\\

$q_{1,\textit{MuSCAT2-r}}$ & $\mathcal{U}(0.0,1.0)$ & Linear limb darkening parametrization for MuSCAT2-r \vspace{0.05cm} \\
$M_{\textit{MuSCAT2-r}}$ (ppm) & $\mathcal{N}(0.0,0.1)$ & Relative flux offset for MuSCAT2-r \vspace{0.05cm}\\
$\sigma_{\textit{MuSCAT2-r}}$ (ppm) & $\mathcal{J}(0.1,10^3)$ & Extra jitter term for MuSCAT2-r \vspace{0.05cm}\\

$q_{1,\textit{MuSCAT2-i}}$ & $\mathcal{U}(0.0,1.0)$ & Linear limb darkening parametrization for MuSCAT2-i \vspace{0.05cm} \\
$M_{\textit{MuSCAT2-i}}$ (ppm) & $\mathcal{N}(0.0,0.1)$ & Relative flux offset for MuSCAT2-i \vspace{0.05cm}\\
$\sigma_{\textit{MuSCAT2-i}}$ (ppm) & $\mathcal{J}(0.1,10^3)$ & Extra jitter term for MuSCAT2-i \vspace{0.05cm}\\

$q_{1,\textit{MuSCAT2-z}}$ & $\mathcal{U}(0.0,1.0)$ & Linear limb darkening parametrization for MuSCAT2-z \vspace{0.05cm} \\
$M_{\textit{MuSCAT2-z}}$ (ppm) & $\mathcal{N}(0.0,0.1)$ & Relative flux offset for MuSCAT2-z \vspace{0.05cm}\\
$\sigma_{\textit{MuSCAT2-z}}$ (ppm) & $\mathcal{J}(0.1,10^3)$ & Extra jitter term for MuSCAT2-z \vspace{0.25cm}\\

\multicolumn{3}{c}{\textit{ RV parameters }} \vspace{0.15cm} \\

$\gamma_{CARMENES}$ [$\mathrm{m\,s^{-1}}$] & $\mathcal{U}(-10.0,10.0)$ & Systemic velocity for CARMENES \vspace{0.05cm}\\
$\sigma_{CARMENES}$ [$\mathrm{m\,s^{-1}}$] & $\mathcal{J}(0.1,10)$ & Extra jitter term for CARMENES \vspace{0.05cm}\\

$\gamma_{HIRES}$ [$\mathrm{m\,s^{-1}}$] & $\mathcal{U}(-10.0,10.0)$ & Systemic velocity for HIRES \vspace{0.05cm}\\
$\sigma_{HIRES}$ [$\mathrm{m\,s^{-1}}$] & $\mathcal{J}(0.1,10)$ & Extra jitter term for HIRES \vspace{0.05cm}\\

$\gamma_{MAROON-X_{red}}$ [$\mathrm{m\,s^{-1}}$] & $\mathcal{U}(-10.0,10.0)$ & Systemic velocity for MAROON-X red arm \vspace{0.05cm}\\
$\sigma_{MAROON-X_{red}}$ [$\mathrm{m\,s^{-1}}$] & $\mathcal{J}(0.1,10)$ & Extra jitter term for MAROON-X red arm \vspace{0.05cm}\\

$\gamma_{MAROON-X_{blue}}$ [$\mathrm{m\,s^{-1}}$] & $\mathcal{U}(-10.0,10.0)$ & Systemic velocity for MAROON-X blue arm \vspace{0.05cm}\\
$\sigma_{MAROON-X_{blue}}$ [$\mathrm{m\,s^{-1}}$] & $\mathcal{J}(0.1,10)$ & Extra jitter term for MAROON-X blue arm \vspace{0.05cm}\\
\hline
\hline
\end{tabular}
\tablefoot{ $^{(a)}$ Central time of transit ($t_0$) units are BJD\,$-$\,2457000. }
\end{table*}

\end{document}